\documentclass[iop]{emulateapj-rtx4}
\usepackage{apjfonts}
\input psfig.tex

\newcommand{\B}{$m_{\rm F435W}$}
\newcommand{\V}{$m_{\rm F555W}$}
\newcommand{\I}{$m_{\rm F814W}$}
\newcommand{\BI}{$m_{\rm F435W} - m_{\rm F814W}$}

\newcommand{\VI}{$m_{\rm F555W} - m_{\rm F814W}$}

\newcommand{\FeH}{[Fe/H]}

\newcommand{\afe}{[$\alpha$/Fe]}
\newcommand{\kms}{km\,s$^{-1}$} 
\newcommand{\Msun}{M$_{\odot}$}

\newcommand{\picplace}[1]{\vbox{\hrule\@height 0.4pt\@width\hsize
\hbox to\hsize{\vrule\@width 0.4pt\@height#1\hfil
\vrule\@width 0.4pt\@height#1}\hrule\@height 0.4pt\@width\hsize}}

\slugcomment{Accepted for publication in ApJ}

\shorttitle{Properties of Extended MSTOs in LMC Star Clusters}
\shortauthors{Goudfrooij et al.}

\begin{document}

%% LaTeX will automatically break titles if they run longer than
%% one line. However, you may use \\ to force a line break if
%% you desire.

\title{Population Parameters of Intermediate-Age Star Clusters in the
  Large Magellanic Cloud. \\ II. New Insights from Extended
  Main Sequence Turnoffs in 7 Star Clusters\altaffilmark{1}}  

%% Use \author, \affil, and the \and command to format
%% author and affiliation information.
%% Note that \email has replaced the old \authoremail command
%% from AASTeX v4.0. You can use \email to mark an email address
%% anywhere in the paper, not just in the front matter.
%% As in the title, you can use \\ to force line breaks.

\author{Paul Goudfrooij\altaffilmark{2}, Thomas H. Puzia\altaffilmark{3,4},
  Vera Kozhurina-Platais\altaffilmark{2}, and Rupali Chandar\altaffilmark{5}}

\altaffiltext{1}{Based on observations with the NASA/ESA {\it Hubble
    Space Telescope}, obtained at the Space Telescope Science
  Institute, which is operated by the Association of Universities for
  Research in Astronomy, Inc., under NASA contract NAS5-26555} 
\altaffiltext{2}{Space Telescope Science Institute, 3700 San Martin
  Drive, Baltimore, MD 21218; goudfroo@stsci.edu, verap@stsci.edu} 
\altaffiltext{3}{Department of Astronomy and Astrophysics, Pontificia
  Universidad Cat\'olica de Chile, Av. Vicu\~{n}a Mackenna 4860, Macul
  7820436, Santiago, Chile; tpuzia@gmail.com} 
\altaffiltext{4}{Herzberg Institute of Astrophysics, 5071 West Saanich Road,
  Victoria, BC V9E 2E7, Canada}
\altaffiltext{5}{Department of Physics and Astronomy, The University of Toledo,
  2801 West Bancroft Street, Toledo, OH 43606; rupali.chandar@utoledo.edu} 

%\author{Paul Goudfrooij}
%\affil{Space Telescope Science Institute, 3700 San Martin Drive,
%  Baltimore, MD 21218} 
%\email{goudfroo@stsci.edu}

%\and 

%\author{Thomas H. Puzia} 
% \affil{Herzberg Institute of Astrophysics, 5071 West Saanich Road,
%  Victoria, BC V9E 2E7, Canada}
%\email{puziat@nrc.ca}

%\and

%\author{Vera Kozhurina-Platais}
%\affil{Space Telescope Science Institute, 3700 San Martin Drive,
%  Baltimore, MD 21218} 
%\email{verap@stsci.edu}

%\and

%\author{Rupali Chandar}
%\affil{Department of Physics and Astronomy, The University of Toledo,
%  2801 West Bancroft Street, Toledo, OH 43606} 
%\email{rupali.chandar@utoledo.edu}

%% Notice that each of these authors has alternate affiliations, which
%% are identified by the \altaffilmark after each name.  Specify alternate
%% affiliation information with \altaffiltext, with one command per each
%% affiliation.

%\altaffiltext{1}{Based on observations with the NASA/ESA {\it Hubble
%    Space Telescope}, obtained at the Space Telescope Science
%    Institute, which is operated by AURA, Inc., under NASA contract
%    NAS5--26555.} 
%\altaffiltext{2}{Society of Fellows, Harvard University.}

%% Mark off your abstract in the ``abstract'' environment. In the manuscript
%% style, abstract will output a Received/Accepted line after the
%% title and affiliation information. No date will appear since the author
%% does not have this information. The dates will be filled in by the
%% editorial office after submission.

\begin{abstract}
We discuss new photometry from high-resolution images of 7
intermediate-age (1\,--\,2~Gyr) star clusters in the Large Magellanic
Cloud taken with the Advanced Camera for Surveys on board the 
  Hubble Space Telescope. 
We fit color-magnitude diagrams (CMDs) with several
different sets of theoretical isochrones, and determine systematic
uncertainties for population parameters when derived using any one set of
isochrones. The cluster CMDs show several interesting features, including
extended main sequence turnoff (MSTO) regions, narrow red giant
branches, and clear sequences of unresolved binary stars. We show that
the extended MSTOs are not caused by photometric uncertainties, 
contamination by field stars, or the presence of binary stars.
Enhanced helium abundances in a fraction of cluster 
stars are also ruled out as the reason for the extended MSTOs.
Quantitative comparisons with simulations
indicate that the MSTO regions are better described by a spread in
ages than by a bimodal age distribution, although we can not formally rule out
the latter for the three lowest-mass clusters in our sample (which have masses
lower than $\sim 3\times10^4$ \Msun). 
This conclusion differs from that of some previous works which suggested that
the age distribution in massive clusters in our sample is bimodal.  
This suggests that any secondary star formation occurred in an
extended fashion rather than through short bursts. We discuss these results in
the context of the nature of multiple stellar populations in star clusters. 
\end{abstract}

%% Keywords should appear after the \end{abstract} command. The uncommented
%% example has been keyed in ApJ style. See the instructions to authors
%% for the journal to which you are submitting your paper to determine
%% what keyword punctuation is appropriate.


%% From the front matter, we move on to the body of the paper.
%% In the first two sections, notice the use of the natbib \citep
%% and \citet commands to identify citations.  The citations are
%% tied to the reference list via symbolic KEYs. The KEY corresponds
%% to the KEY in the \bibitem in the reference list below. We have
%% chosen the first three characters of the first author's name plus
%% the last two numeral of the year of publication as our KEY for
%% each reference.

\section{Introduction}              \label{s:intro}

An accurate knowledge of stellar populations of ``intermediate'' age 
($\approx1-3\times10^9$~yr) is important within the context of several
currently hot topics in astrophysics.  
Intermediate-age stars typically dominate the emission observed from
galaxies at high redshift \citep[e.g.,][]{vdwel+06}. Furthermore, star clusters 
in this age range are critical for testing predictions of the dynamical
evolution of star clusters  \citep[e.g.,][]{goud+07}, and for understanding
the evolution of intermediate-mass stars. The Large Magellanic Cloud (LMC)
hosts a rich system of intermediate-age star clusters. The first surveys
dedicated to studying properties of these clusters were based on integrated
colors \citep[e.g.,][hereafter SWB]{swb80}. These studies led to empirical and
homogeneous age scales such as the ``SWB parameter'' and the ``$S$ parameter''
\citep{elsfal85,elsfal88,gira+95,pess+08}, which describe the position of the
cluster in integrated-light color-color diagrams. While a number of more
recent studies of such intermediate-age clusters have determined ages more
directly by fitting the location of the main sequence turnoff (MSTO) region
with model isochrones in color-magnitude diagrams
\citep[CMD; e.g.,][]{bert+03,kerb+07,mucc+07,mack+08,milo+09}, such age
determinations are still rather sparse and somewhat dependent on the stellar
model being used. 
It is therefore important to obtain more CMD-based ages and metallicities of
intermediate-age clusters and to study any systematic uncertainties related to
the choice of any particular stellar model and its ingredients.  

Another current hot topic in astrophysics is that of multiple stellar
populations in globular clusters (GCs). The standard paradigm that GCs consist
of stars born at the same time out of the same material has
faced serious challenges over the last decade. It is now known that the most
massive GCs in our Galaxy such as $\omega$\,Cen and M\,54 host multiple red
giant branches (RGBs) due to populations with different [Fe/H]
\citep[e.g.,][]{sarlay95,lee+99,hilric00,vill+07,carr+10}. Slightly less
massive Galactic GCs such as NGC~2808, NGC~1851 and 47~Tuc show multiple
sub-giant branches (SGBs) and/or multiple main sequences, which are typically
interpreted as populations with different Helium abundance
\citep[e.g.,][]{piot+07,milo+08,ande+09}. While lower-mass Galactic GCs
typically do not show clear evidence for multiple populations from optical
broad-band photometry, spectroscopic surveys do show significant star-to-star
abundance variations in {\it light elements\/} such as C, N, O, F, and Na
(often dubbed ``Na-O anticorrelations'') within all Galactic GCs 
studied to date in sufficient detail (\citealt{carr+09}, and references
therein). Since these abundance variations have been found among 
RGB stars as well as main sequence (MS) stars within a given GC \citep{grat+04}, the
cause of the variations seems to be that secondary population(s) formed out of
material shed by an evolved population within the cluster.  While the chemical
processes involved in causing the light element abundance variations 
have largely been identified as proton-capture reactions in Hydrogen burning
at high temperature \citep[$\ga 40 \times 10^6$ K, see e.g.][]{grat+04}, the
old age of Galactic GCs has precluded a clear picture of the time scales and
hence the types of stars involved in the chemical enrichment of the
second-generation stars. Currently, the most popular candidates are {\it (i)\/}
intermediate-mass AGB stars ($3 \la {\cal{M}}/M_{\odot} \la 8$,
hereafter IM-AGB; e.g., \citealt{danven07} and references therein),
{\it (ii)\/} rapidly rotating massive stars 
(often abbreviated as FRMS; e.g.,
\citealt{decr+07}) and {\it (iii)\/} massive binary stars
\citep{demink+09}.  

Recently, deep CMDs from images taken with the Advanced Camera for Surveys
(ACS) aboard the Hubble Space Telescope (HST) provided conclusive evidence that
several massive intermediate-age star clusters in the Magellanic Clouds have 
extended and/or multiple main sequence turn-off (MSTO) regions 
\citep[][hereafter Paper I]{mack+08,glat+08,milo+09,goud+09}, in some cases
accompanied by composite red clumps \citep{gira+09,rube+11}. To date, these
properties have been interpreted in three main ways: {\it (i)\/} Bimodal age
distributions \citep{mack+08,milo+09}, {\it (ii)\/} age spreads of
200\,--\,500 Myr \citep{goud+09,mack+08,milo+09,gira+09,rube+10}, and
{\it (iii)\/} spreads in rotation velocity among turn-off stars (\citealt{basdem09}). 

In this second paper of our series we follow methods presented in
Paper I and conduct a detailed investigation of population parameters of 7
intermediate-age star clusters in the LMC. While color-magnitude
diagrams of clusters in this sample were already presented in \citet{mack+08}
and \citet{milo+09}, we employ different methods and present
additional analysis of the clusters.  
At the data reduction level this includes corrections for charge transfer
inefficiency that are specifically determined for the datasets used
  here. At the analysis and interpretation level, we investigate 
the impact of unresolved binary stars associated with one {\it and\/}
  two generations of stars on the MSTO region, 
we use different techniques for assessing background contamination, and we 
investigate how well different sets of isochrones 
fare when compared with the observations. The additional analysis
allows us to reveal new properties relevant to the assembly of these
intermediate-age star clusters and their evolutionary association with
multiple stellar populations in ancient Galactic globular clusters. In
a companion paper in this series (\citealt{goud+11b}, hereafter Paper III), we 
synthesize the results from the current paper along with new
radial distributions of cluster stars at different evolutionary phases and
dynamical considerations to constrain the origin of extended MSTO (hereafter
eMSTO) regions in intermediate-age star clusters in the LMC.  

The remainder of this paper is organized as follows. \S\,\ref{s:obs} presents
the observations, \S\,\ref{s:anal} discusses 
details of the stellar photometry 
and evaluates contamination by field stars.
\S\,\ref{s:isofits} presents our isochrone fitting analysis using
different sets of theoretical isochrones. 
We discuss the level of helium abundance enhancement of any subpopulation of
stars in these clusters in \S\,\ref{s:Helium}. 
In \S\,\ref{s:MSTOmorph} we analyze the morphology of the eMSTO
regions. \S\,\ref{s:disc} discusses the results in the context of the
nature of multiple stellar populations in star clusters in general, and
\S\,\ref{s:conc} presents our main conclusions.  

\section{Sample Selection and Observations} \label{s:obs}

Our sample consists of seven star clusters 
(NGC~1751, NGC~1783, NGC~1806, NGC~1846, NGC~1987, NGC~2108, and LW431)
in the LMC
with an 'SWB' parameter in the range IV{\sc b}~$-$~VI, translating to ages between
roughly 1.0 and 2.5 Gyr. Main properties of the clusters
are listed in Table~\ref{t:sample}, along with a journal of their
{\it HST\/} observations with the wide-field channel (WFC) of the
Advanced Camera for Surveys (ACS) instrument as part of {\it HST\/}
program GO-10595 (PI: P.\ Goudfrooij).  
We centered the target clusters on one of the two CCD chips of the ACS/WFC
camera so that the observations cover a fairly large radial
extent, enabling us to study variations of cluster properties with radius and
evaluate the properties of the field star population. Three exposures were
taken in each of the  
F435W, F555W, and F814W filters: Two long exposures of 340 s each and one
shorter exposure to avoid saturation of the brightest stars in the cluster 
(90 s, 40 s, and 8 s in F435W, F555W, and F814W respectively). The two long
exposures in each filter were spatially offset from each other by
3\farcs011 in a direction +85\fdg28 with respect to the positive X 
axis of the CCD array. This was done to move across the gap between the
two ACS/WFC CCD chips, as well as to simplify the identification and removal of
hot pixels. 

% Put Table 1 here
\begin{table}[tbh]
\begin{center}
\footnotesize
%\scriptsize
\caption{Main properties of the star clusters studied in this paper.}
 \label{t:sample}
\begin{tabular}{@{}lrccc@{}}
\multicolumn{3}{c}{~} \\ [-2.5ex]   
 \tableline \tableline
\multicolumn{3}{c}{~} \\ [-2.ex]                                                
\multicolumn{1}{c}{Cluster} & \multicolumn{1}{c}{Obs.\ Date} & $V$ & Ref. & 
 SWB \\
\multicolumn{1}{c}{(1)}     & \multicolumn{1}{c}{(2)}        & (3) & (4) & 
 (5) \\ [0.5ex] \tableline  
\multicolumn{3}{c}{~} \\ [-2.ex]              
NGC 1751 & Oct 18-19, 2006 & 11.67 $\pm$ 0.13 & 1 & VI    \\
NGC 1783 &    Jan 01, 2006 & 10.39 $\pm$ 0.03 & 1 & V     \\
NGC 1806 &    Sep 29, 2005 & 11.00 $\pm$ 0.05 & 1 & V     \\
NGC 1846 &    Jan 12, 2006 & 10.68 $\pm$ 0.20 & 1 & VI    \\
NGC 1987 &    Oct 18, 2006 & 11.74 $\pm$ 0.09 & 1 & {\sc IVb} \\
NGC 2108 &    Aug 22, 2006 & 12.32 $\pm$ 0.04 & 2\tablenotemark{a} & {\sc IVb} \\
  LW 431 & Nov 05-06, 2006 & 13.67 $\pm$ 0.04 & 2\tablenotemark{a} & VI    \\ [0.5ex] \tableline
\multicolumn{3}{c}{~} \\ [-2.5ex]              
\end{tabular}
\tablecomments{Column (1): Name of star cluster. (2): Date of {\it
    HST/ACS\/} observations. (3): Integrated $V$ magnitude. (4): Reference
  of $V$ magnitude. Ref. 1: \citet{goud+06}; Ref.\ 2: \citet{bica+96}. (5): SWB type
  from \citet{bica+96}.}  
\tablenotetext{a}{uncertainty only includes internal errors associated with
  measurements of cluster and one background aperture.}
\end{center}
\end{table}

\addtocounter{table}{1}

\section{Analysis}          \label{s:anal}

\subsection{Photometry}    \label{s:phot}

Stellar photometry was conducted using PSF fitting on the flatfielded ({\tt
  flt}) files produced by the {\it HST\/} calibration pipeline, using the
spatially variable ``effective PSF'' {\it (ePSF)} method described in
\citet{andkin00}, and tailored for ACS/WFC data by \citet{andkin06}. A detailed
description of the application of the ePSF method to ACS/WFC data is given
in \citet{ande+08a,ande+08b}.
We selected all stars with the {\it ePSF} parameter ``PSF fit quality'' $q <
0.5$  and ``isolation index'' of 5. The latter parameter selects stars that
have no brighter neighbors within a radius of 5 pixels. To further weed out
hot pixels, cosmic rays and spurious detections along diffraction spikes, the 
geometrically corrected positions among the three images per filter were
compared. We selected objects with coordinates matching within a tolerance of
0.2 pixels in either axis, which eliminated the hot pixels and cosmic ray hits
effectively. 

Corrections for imperfect charge transfer efficiency (CTE) of the
ACS/WFC CCDs were made specifically for the case of photometry from {\tt
    flt} files featuring varying exposure times following \citet{kozh+07}.   
The accuracy of our CTE corrections is such that the rms scatter of the
magnitude residuals is 0.02 mag at an instrumental magnitude of $-2$
(corresponding to \B\ = 23.8, \V\ = 24.4, and \I\ = 23.5  in magnitude units
relative to Vega). This uncertainty is smaller than the photometric
measurement errors at those magnitudes.  

Photometric incompleteness as functions of stellar brightness, color, and
position in the cluster was quantified by repeatedly adding small numbers of
artificial ePSFs to the images, covering the magnitude and color ranges found
in the CMDs.  The overall radial distribution of the inserted artificial stars
followed that of the stars in the image. 
We refer the reader to Paper I for further details of our photometric methods
and calibration. 

\subsection{Color-Magnitude Diagrams}  \label{s:CMDs}

\begin{figure*}[pt]
\centerline{\includegraphics[bb=45 341 552 491,clip,width=0.953\textwidth]{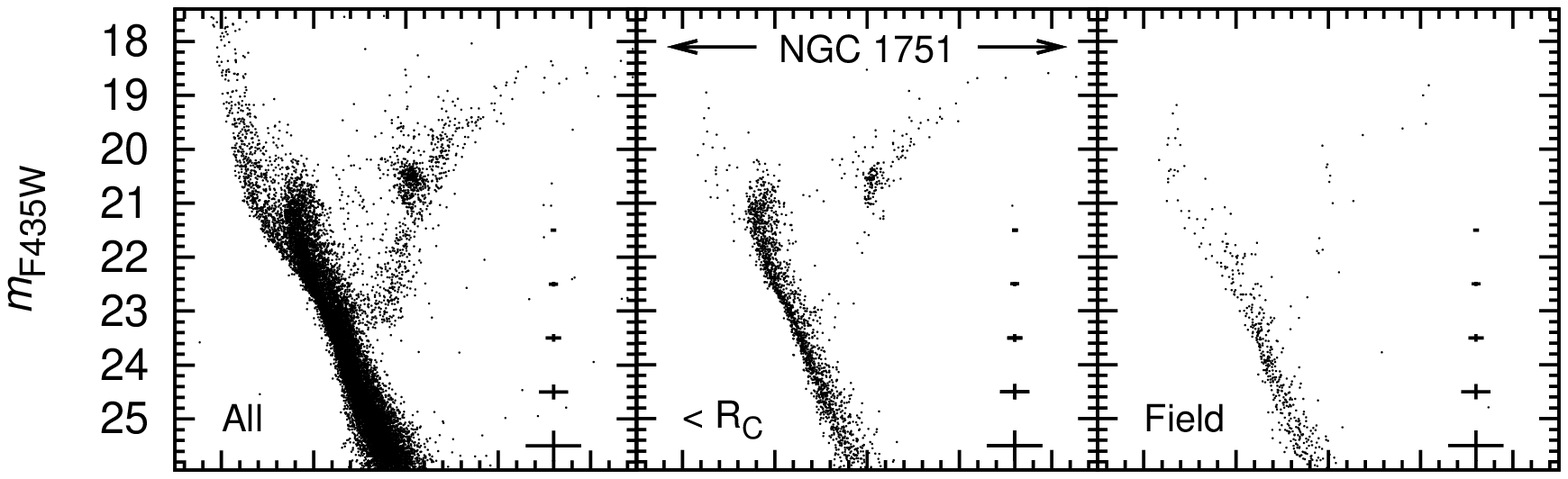}}
\centerline{\includegraphics[bb=51 341 548 490,clip,width=0.935\textwidth]{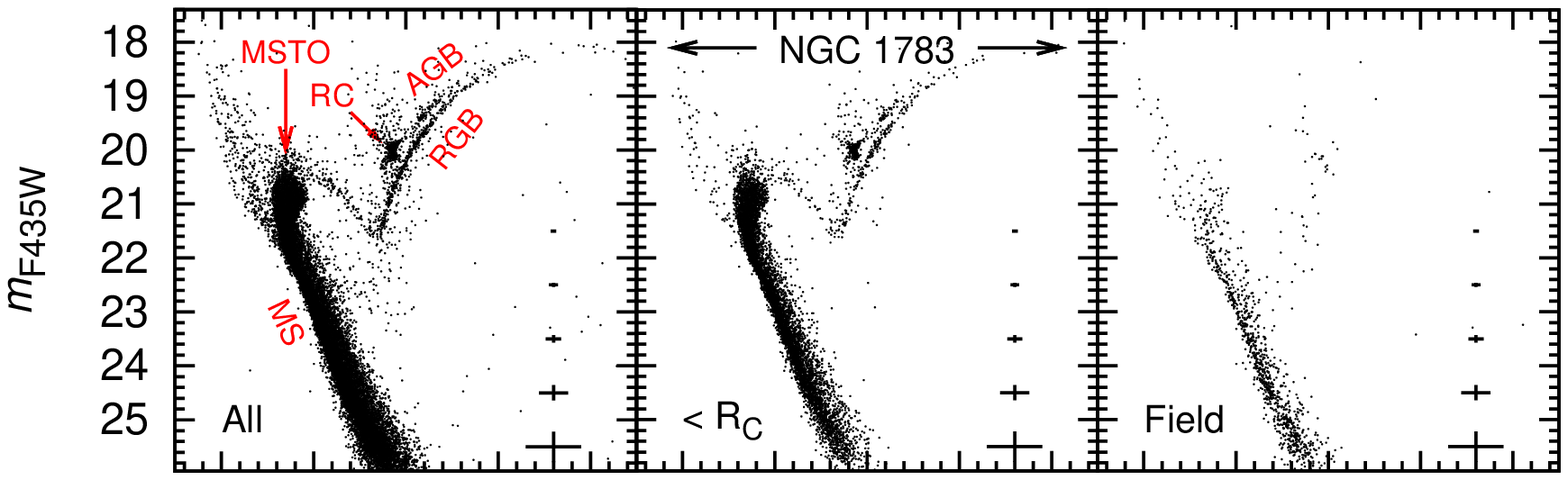}}
\centerline{\includegraphics[width=0.935\textwidth]{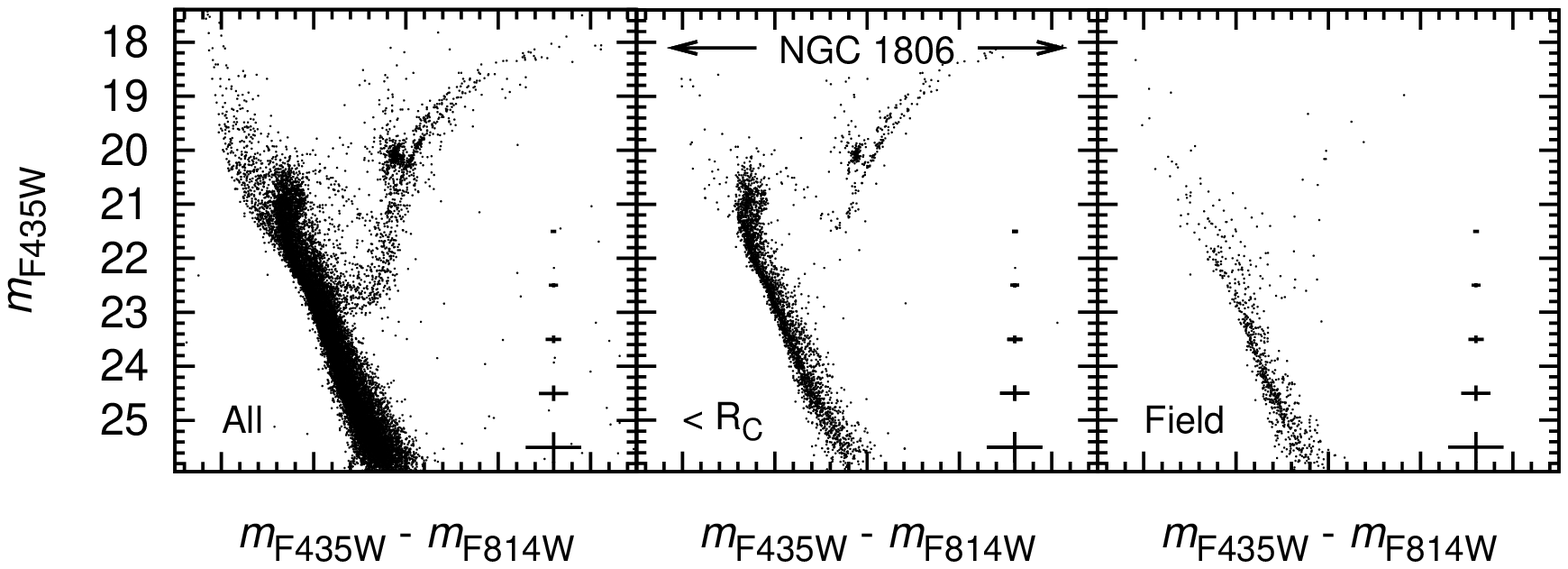}}
\caption{\B\ vs.\ \BI\ CMDs for the HST/ACS images of NGC~1751, NGC~1783, and
  NGC~1806 (we refer to Paper I for the CMDs of NGC~1846). The left panels show
  all detected sources that passed our selection criteria. The middle panels
  show all such sources within the King core radius of the star cluster in
  question. The right panels show all sources detected in areas near the
  corners of the image furthest away from the cluster center. These composite
  ``field'' regions have the same area as those shown in the middle panels.  
  The left panel for NGC~1783 shows labels for the MS, MSTO, RC, RGB, and
  AGB features in the CMD. 
\label{f:fullCMDs_1}}  
\end{figure*}

\begin{figure*}[pt]
\centerline{\includegraphics[bb=45 341 552 491,clip,width=0.95\textwidth]{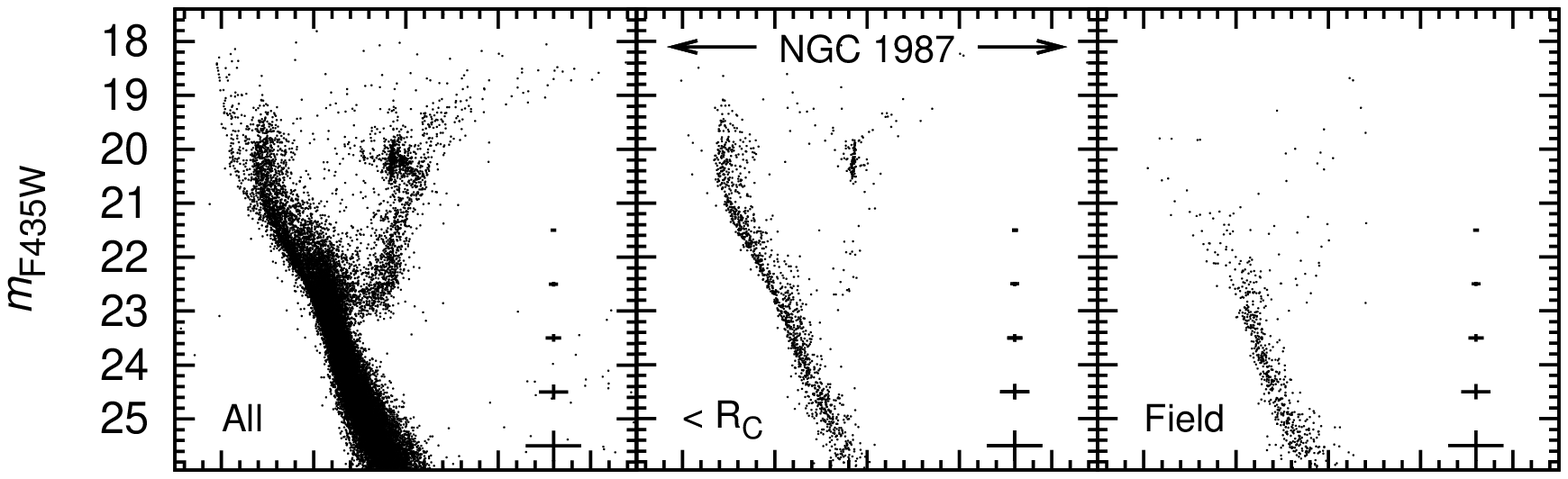}}
\centerline{\includegraphics[bb=45 341 552 491,clip,width=0.95\textwidth]{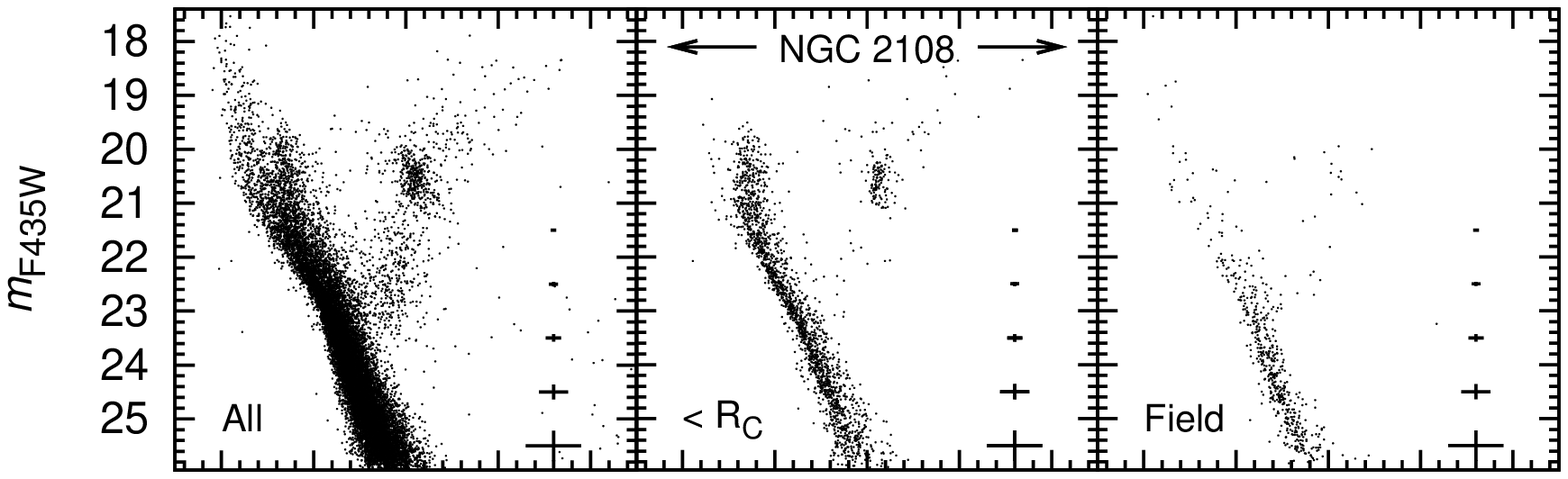}}
\centerline{\includegraphics[bb=45 298 552 491,clip,width=0.95\textwidth]{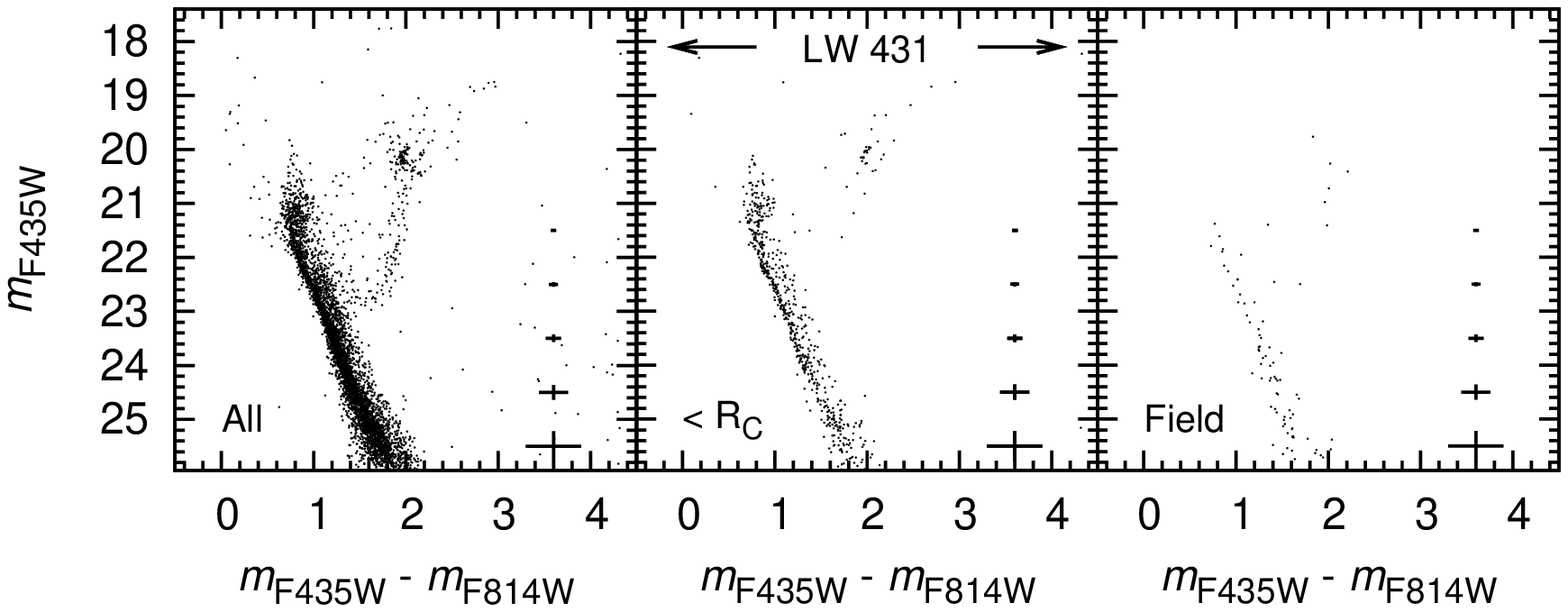}}
\caption{Same as Fig.~\ref{f:fullCMDs_1}, but now for NGC 1987, NGC 2108, and
  LW 431. 
\label{f:fullCMDs_2}}  
\end{figure*}

Color-magnitude diagrams (CMDs) for the star clusters in our sample were
created for both \B\ vs.\  \BI\ and \V\ vs.\ \VI. CMDs for all selected stars
are plotted in the left panels of Figs.\
\ref{f:fullCMDs_1}\,--\,\ref{f:fullCMDs_2}. 
To decrease and assess contamination by field stars, we also plot CMDs for 
all stars within one core radius based on fits to a single-mass King (1962)
model from the cluster centers 
(derived as described in \S\ \ref{s:rad_dist} below) in the middle panels,
and CMDs for stars located in regions near the corners of the {\it
  HST/ACS\/} image which are presumably dominated by
non-cluster stars,
in the right panels. The latter regions were chosen to cover the
same surface area as those in the middle panels.
In Figure~\ref{f:fullCMDs_1} we identify and label the following
evolved phases of stellar evolution: 
the MS, MSTO, red clump (hereafter RC\footnote{The red clump is sometimes
  called Helium clump since this is where stars undergo core Helium burning}),
the RGB, and the AGB.   

\subsubsection{General Comments}

Figs.\ \ref{f:fullCMDs_1}\,--\,\ref{f:fullCMDs_2} show that the star clusters
in our sample all contain a clear red 
clump as well as RGBs and AGBs with various extents and levels of 
population. 
Focusing on the relative F435W-band luminosities of the
MSTO and the RC as well as the extent of the RGB within the individual
clusters, it is clear that the clusters NGC~1987 and NGC~2108 are younger than
the others in our sample.  
The ages of the other clusters appear to be similar to one another. 
Cluster ages will be quantified in Section~4.
A comparison of the middle and right-hand panels in Figs.\ 
\ref{f:fullCMDs_1}\,--\,\ref{f:fullCMDs_2} also shows that the
  MSTO's seen in the star clusters (middle panels) are brighter than
  those seen in the right-hand panels, which are most likely dominated
  by field stars. 
The field population is typically dominated (in terms of star number
  density) by a 5\,--\,6 Gyr old population with a broad MSTO located
at \B~$\simeq$~23.0 and \V~$\simeq$~22.5 plus an associated evolved
RGB population. 
  The field population also includes stars younger
  than the clusters in our sample, as evidenced by the presence of MS
  stars that are brighter and bluer than the MSTOs of the star
  clusters.  

A comparison of the morphologies of the MSTO regions in all star clusters in our
sample with the local photometric uncertainties 
(see Figs.\ \ref{f:fullCMDs_1}\,--\,\ref{f:fullCMDs_2}) 
shows that the MSTO regions are significantly more extended than expected for
a coeval simple stellar population (SSP). 
  This is consistent with the findings of \citet{macbro07},
  \citet{mack+08}, and \citet{milo+09}. However, these authors
  interpreted the extended MSTO regions in NGC~1751, NGC~1783, NGC~1806,
  and/or NGC~1846 in their photometry as due to the presence of two distinct
  stellar populations of identical metallicity but different age.  
  Our photometry shows differences from those of Mackey et al.\ and
  Milone et al., which potentially affects the interpretation of how
  these clusters formed.  These differences are discussed further in
  \S\ \ref{s:disc}.    

\subsubsection{Differential Reddening}

\begin{figure}[tbh]
\centerline{\includegraphics[width=8.3cm]{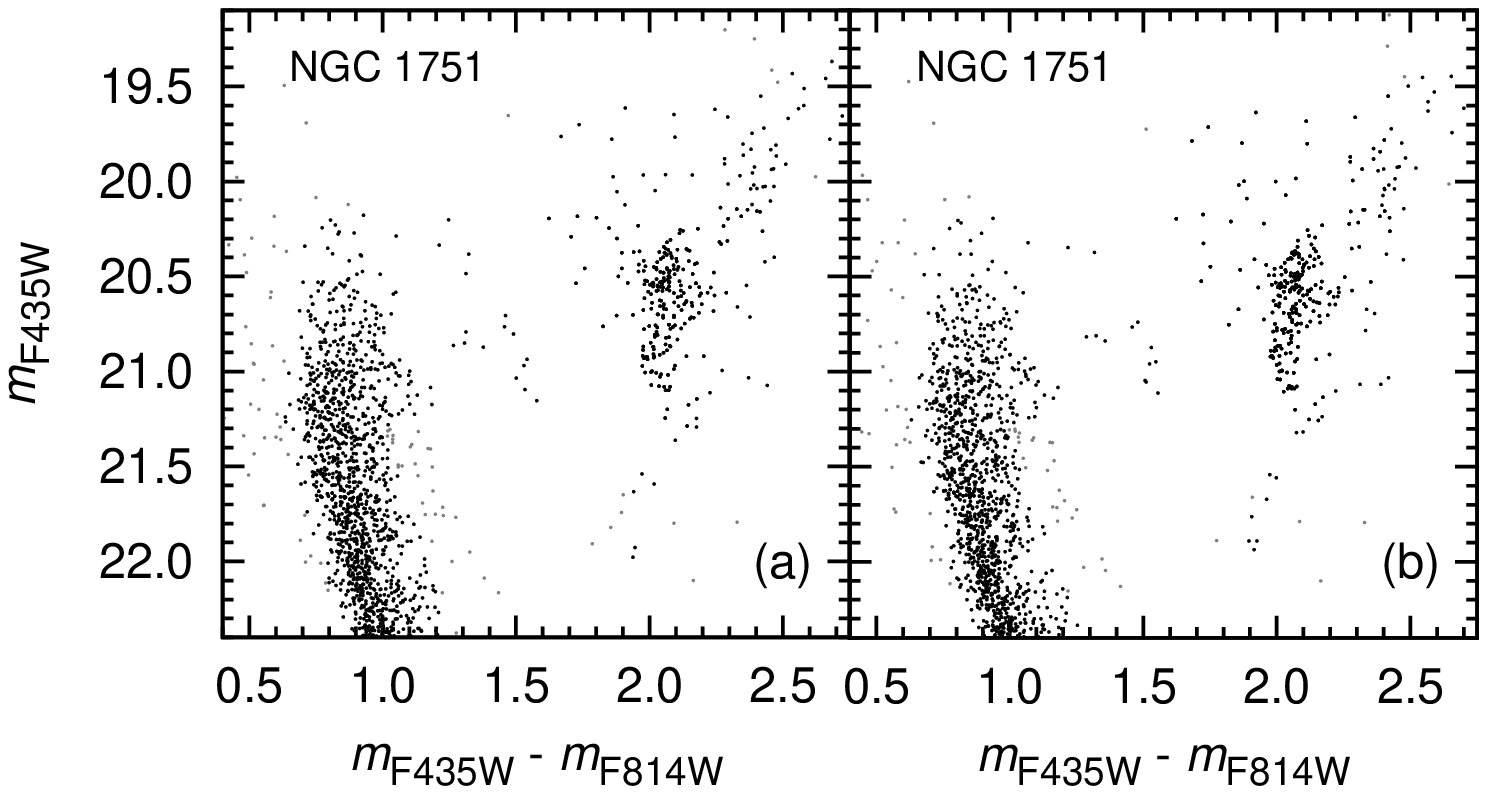}}
\caption{Portion of the \B\ vs.\ \BI\ CMD of NGC 1751. Panel (a): Observed
  CMD. Panel (b): After correction for differential reddening.  
\label{f:difred_1751}}
\end{figure}

\begin{figure}[tbh]
\centerline{\includegraphics[width=8.3cm]{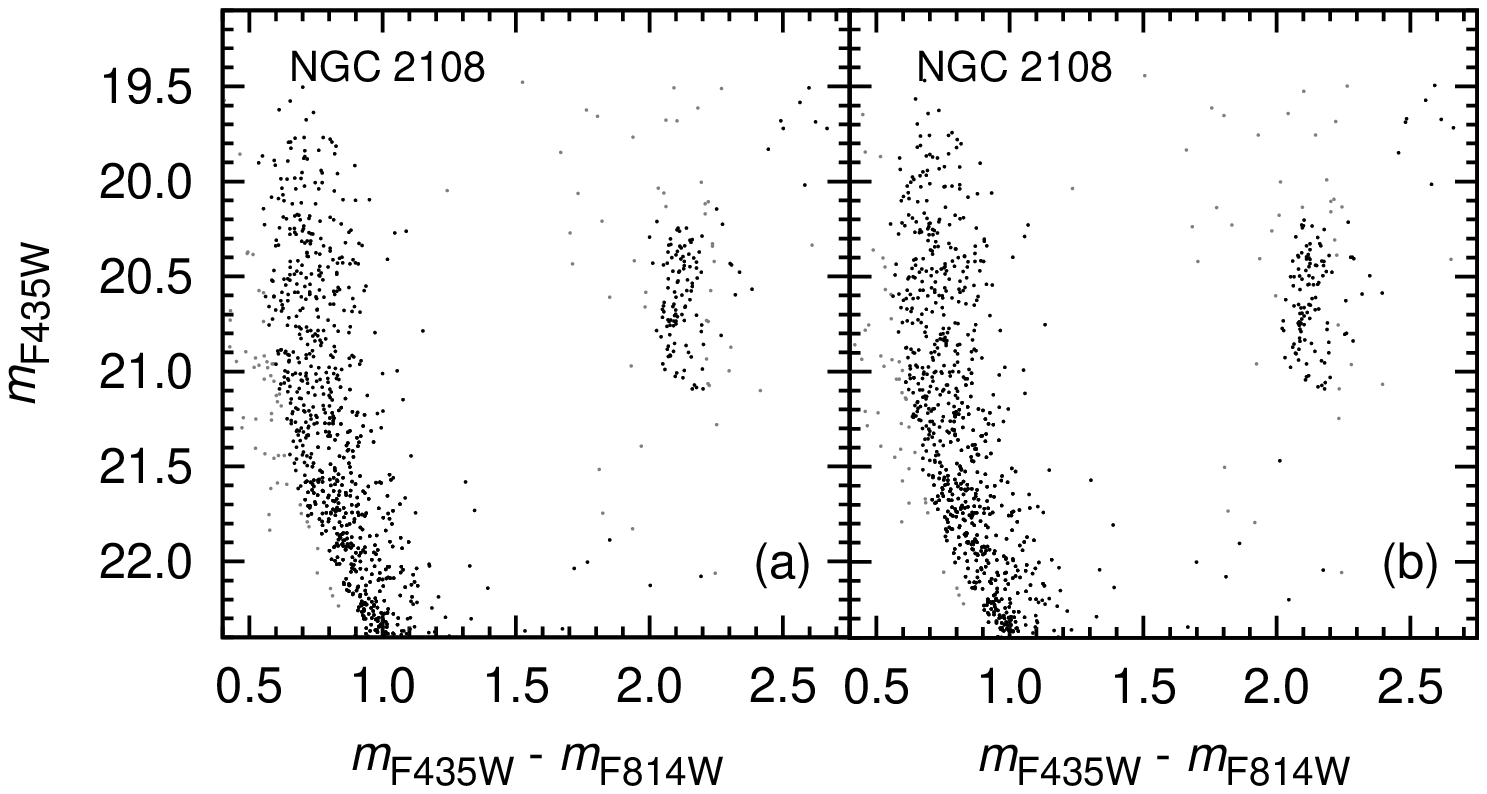}}
\caption{Same as Fig.\ \ref{f:difred_1751}, but now for NGC 2108.
\label{f:difred_2108}}
\end{figure}

The CMDs of most star clusters in our sample reveal compact RCs 
and/or well-defined RGBs and AGBs that have widths which are comparable to
the photometric uncertainties. Since the reddening vector is approximately
perpendicular to the RGB and AGB, this suggests that spatial variations in
reddening are negligible for most clusters in our sample. There are, however,
two exceptions: NGC~1751 and NGC~2108, both of which have a ``fuzzy'' RC and
RGB. To correct their CMDs for differential reddening, we follow a method
similar to that used by \citet{sara+07}. We divide the cluster field
into several subareas (the number of subareas depends on the total number of
stars found in the cluster). After defining a grid of magnitude and color
intervals along the MS below the MSTO region, a fiducial ridge line for the
MS is derived for all stars located within the King core radius
from running medians of star magnitudes and colors.  
We then measure the weighted mean \B, \V, and \I\ magnitudes of
stars within those grids for all subareas within the cluster. These
weighted mean magnitudes define the local reddening in each subarea relative
to the mean reddening within the King core radius of the cluster. These
reddening values are then used to correct the magnitudes of stars in all
subareas to a uniform reddening value. The effect of the correction for
differential reddening is shown in Figures \ref{f:difred_1751} and
\ref{f:difred_2108}. The improvement is particularly significant for NGC~1751
where the RC becomes more compact than before the correction.  
The total amplitude of differential reddening within the cluster
was found to be $\Delta\,E(m_{\rm   F435W}-m_{\rm F814W})$ = 0.14 and 0.10
for NGC~1751 and NGC~2108, respectively.  

\subsection{Radial Surface Number Density Distributions} \label{s:rad_dist} 

\begin{figure*}[tb]
\centerline{
\psfig{figure=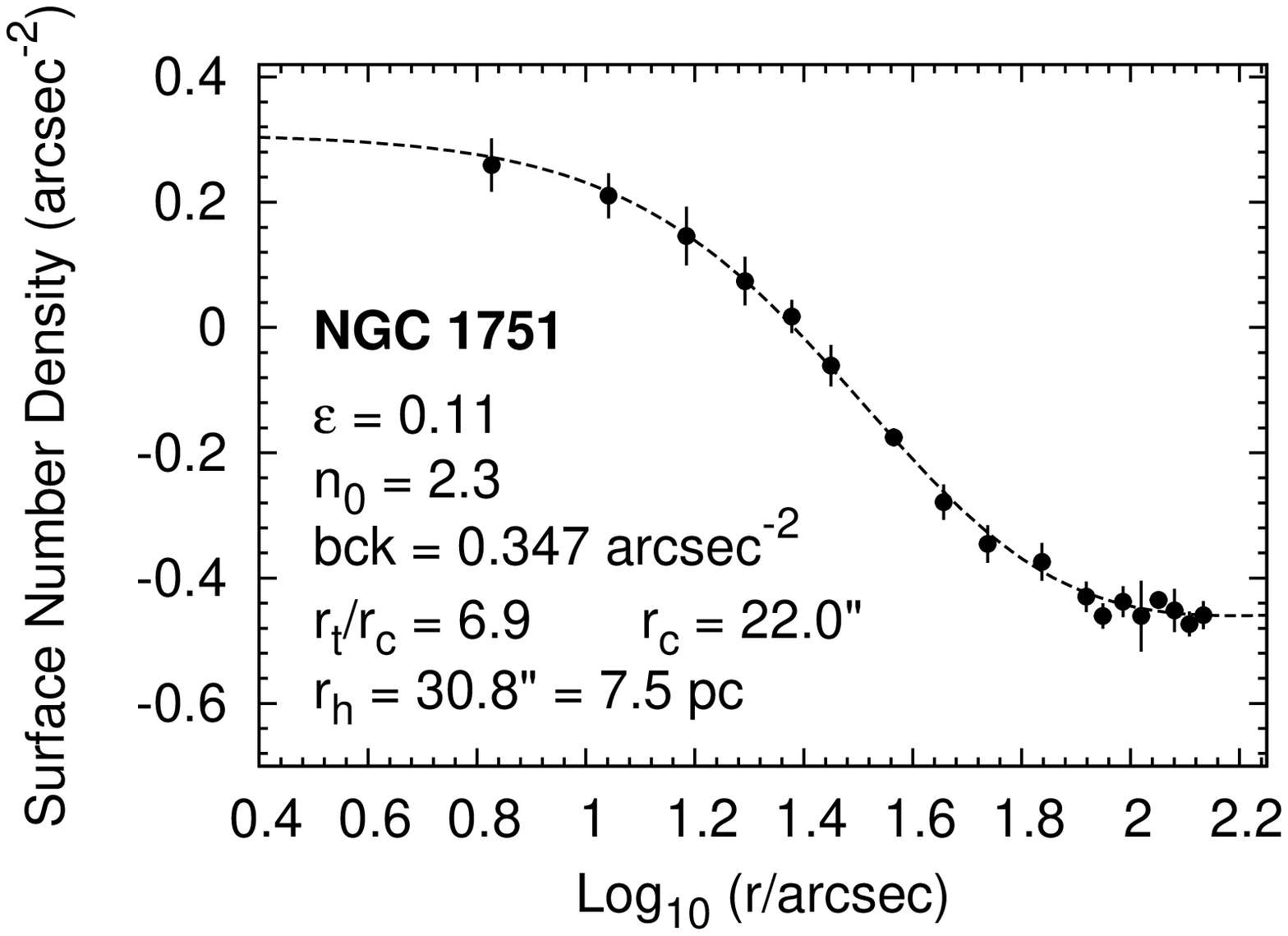,height=4.05cm,angle=-90}
\psfig{figure=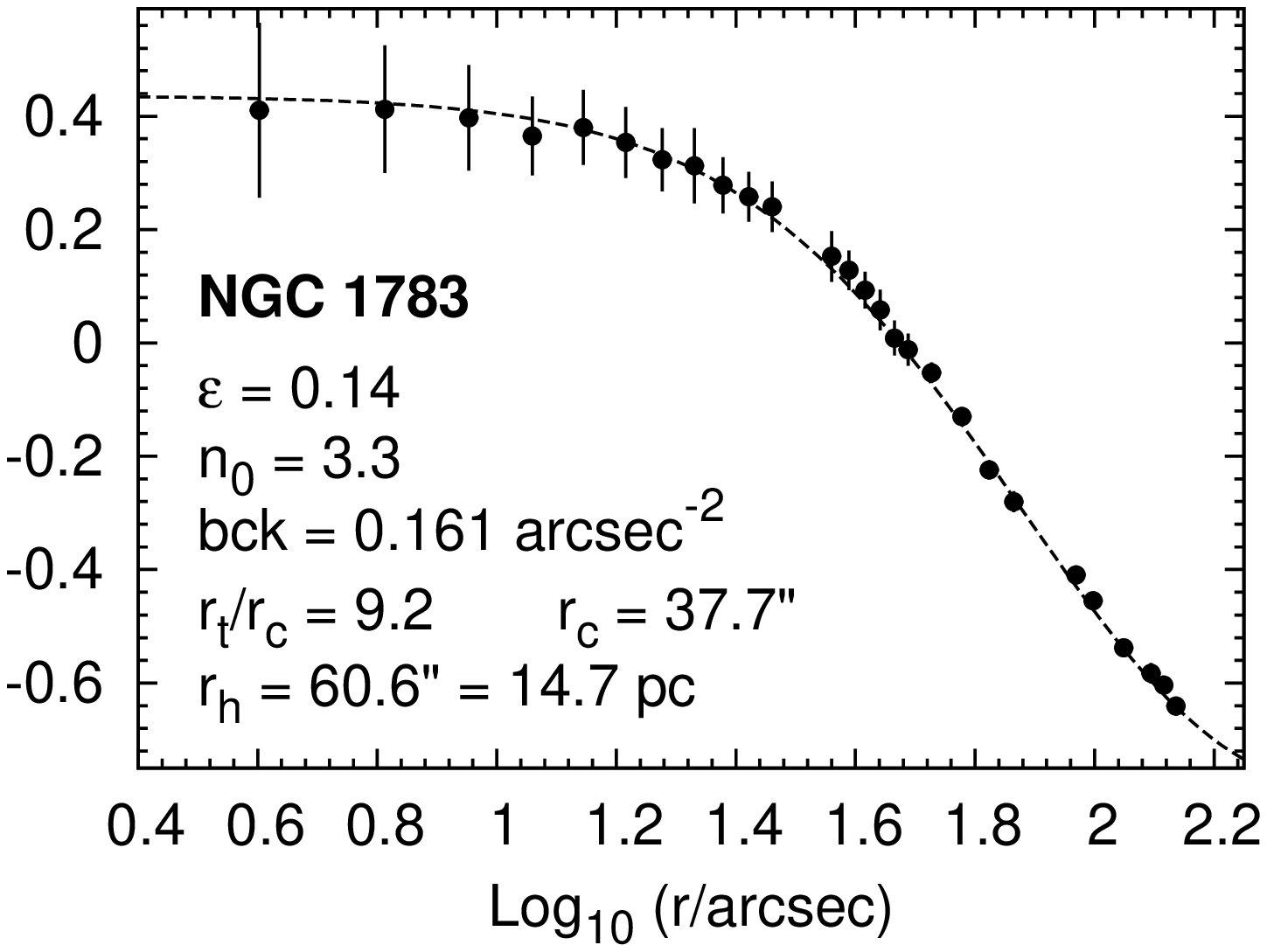,height=3.8cm,angle=-90}
\psfig{figure=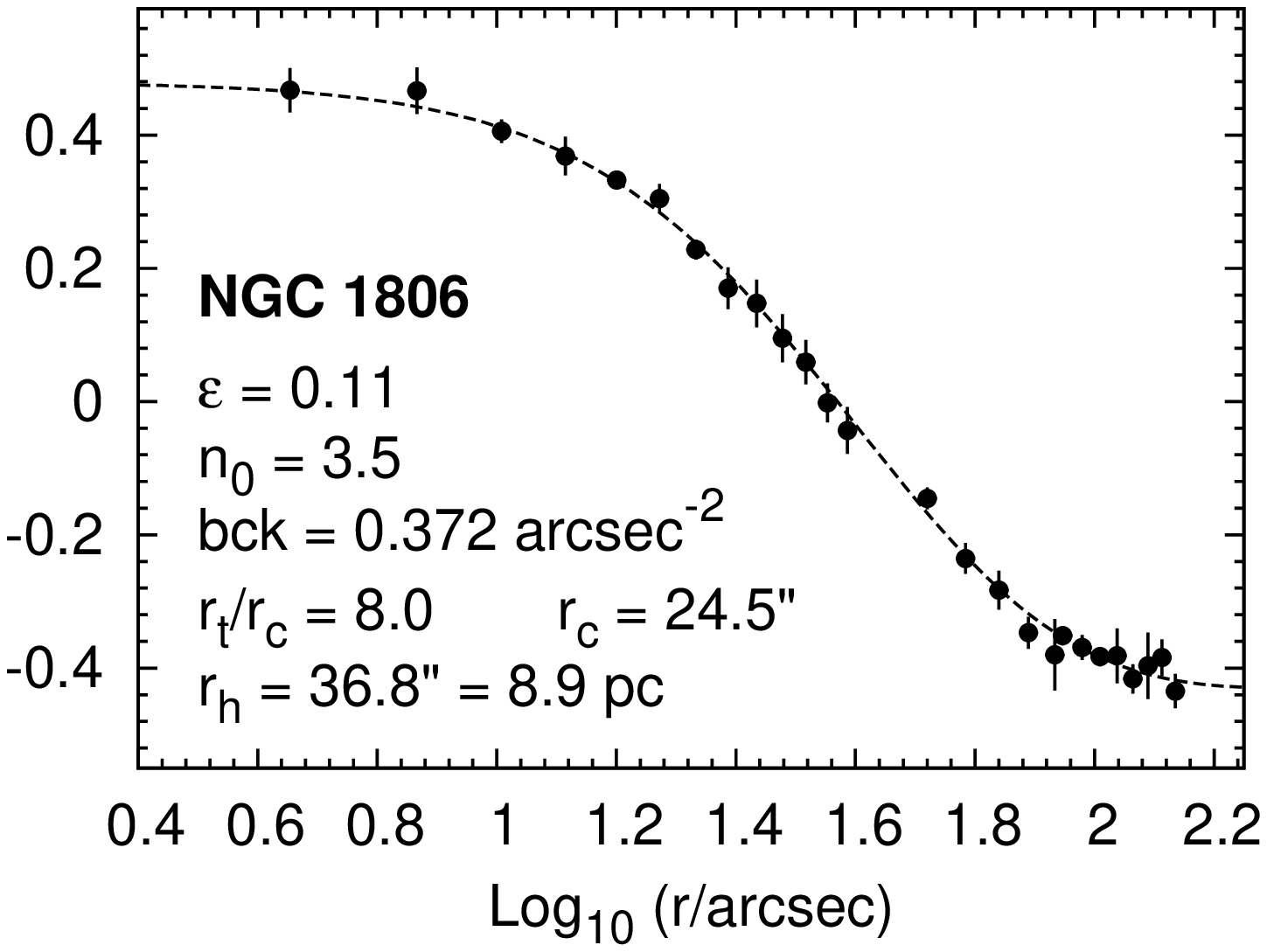,height=3.8cm,angle=-90}
}
\vspace*{0.5mm}
\centerline{
\psfig{figure=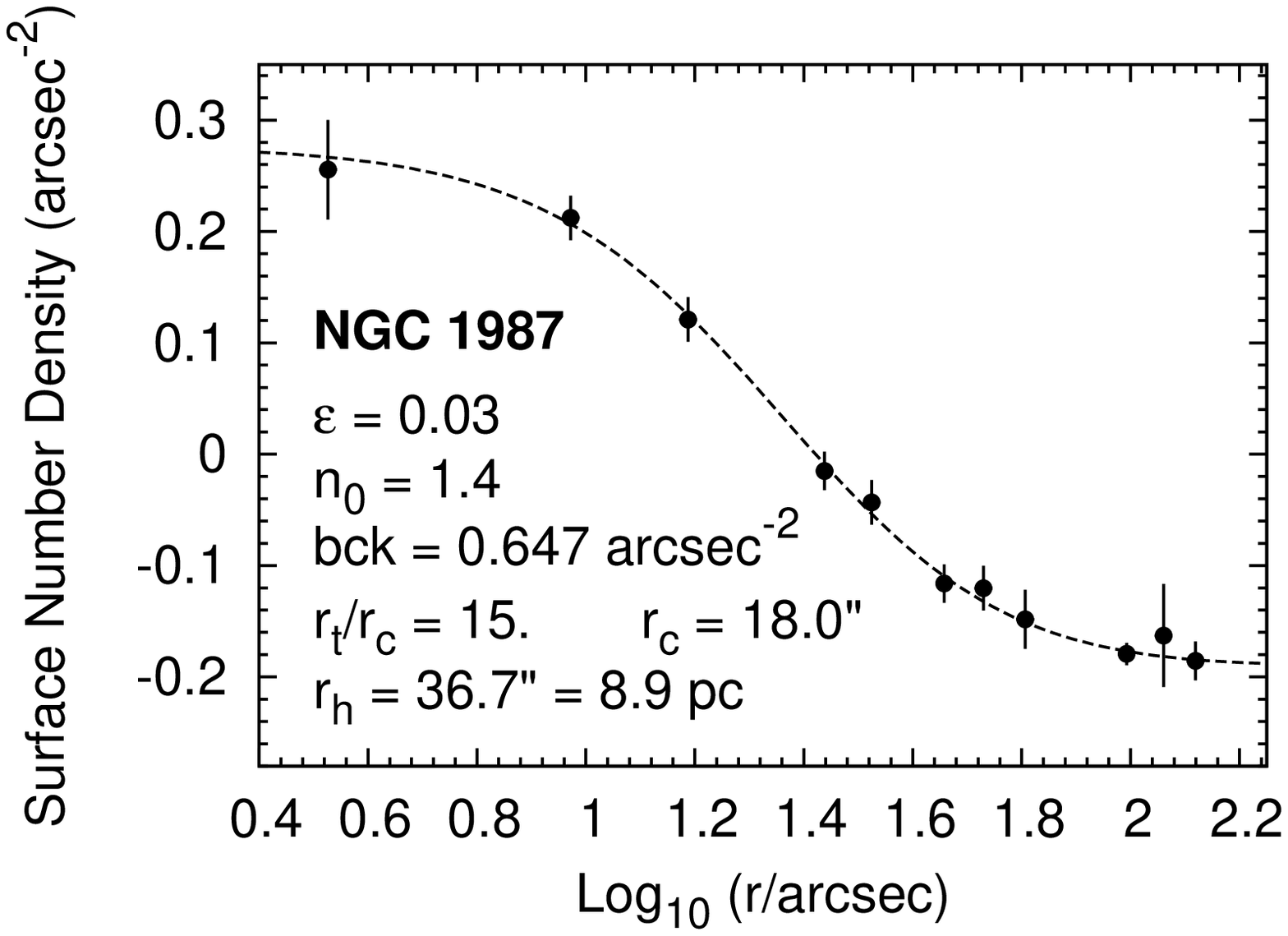,height=4.05cm,angle=-90}
\psfig{figure=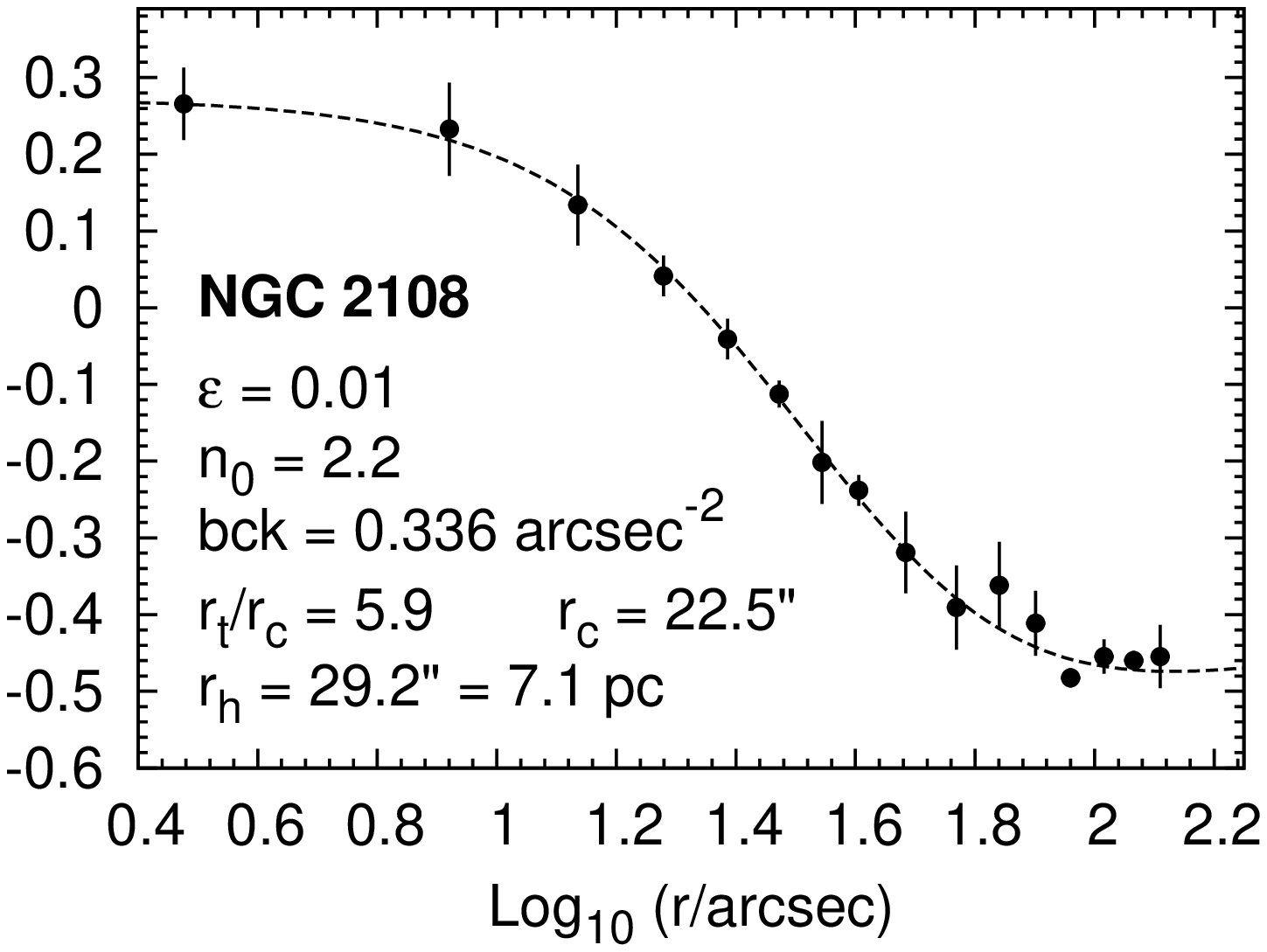,height=3.8cm,angle=-90}
\psfig{figure=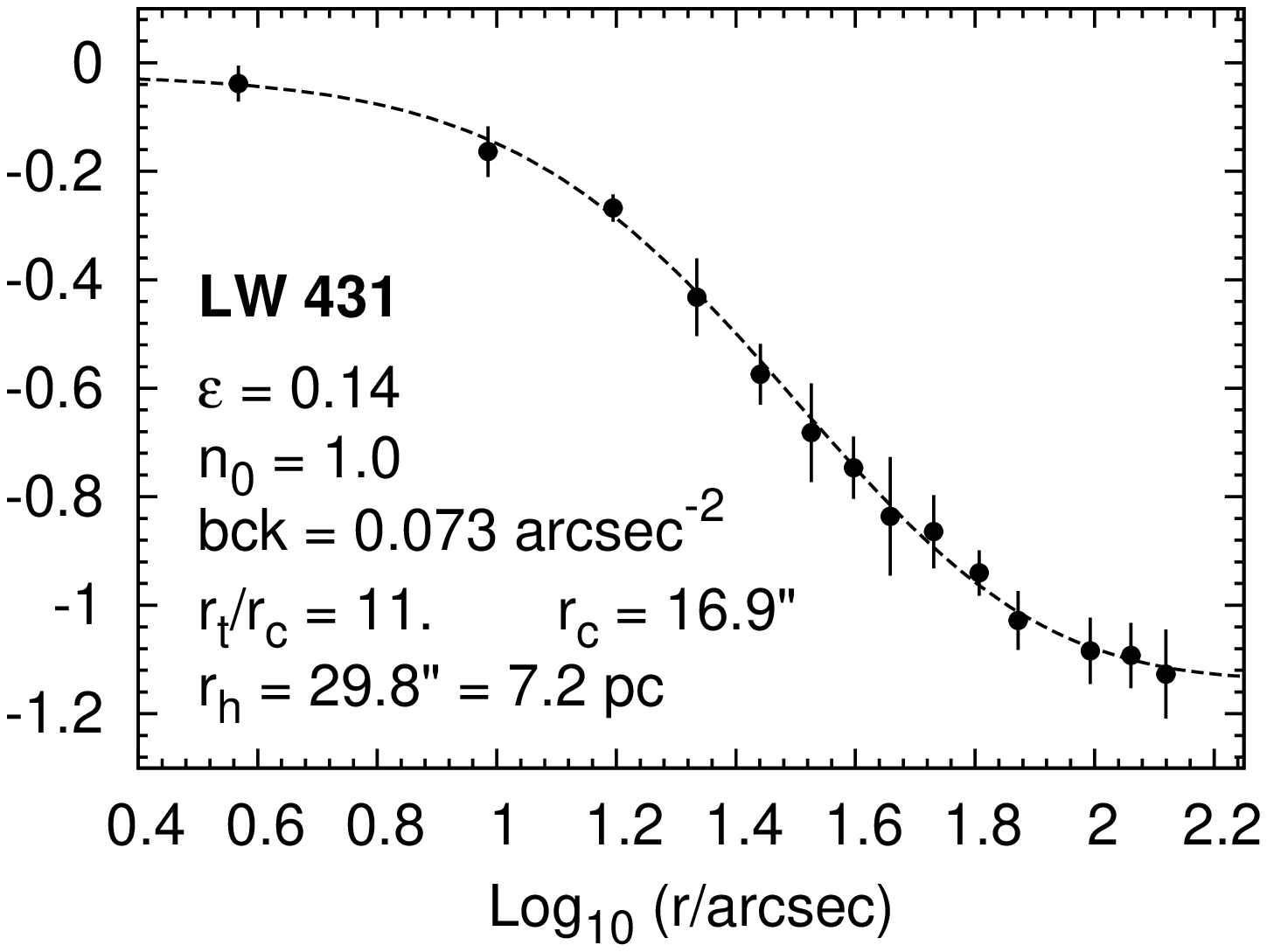,height=3.8cm,angle=-90}
}
\caption{Radial surface numbers density profiles of the star clusters in our
  sample. The points represent observed values. The dashed lines represent the
  best-fit King models (cf.\ Eq.\ \ref{eq:King}) whose parameters are shown in
  the legends. The names and ellipticities of the clusters are also
  shown in the legends.  
\label{f:numdensfits}}
\end{figure*}

We analyze the projected surface number density of stars in the sample star
clusters for two main reasons: {\it (i)\/} to determine regions in the CMD
that are strongly dominated by stars belonging to the cluster rather than to
the underlying field in the LMC, and {\it (ii)\/} to allow an
evaluation of dynamical properties of the clusters,
which can help shed light on the origin of the 
eMSTOs. 
The cluster centers were determined by fitting a 2-D gaussian
to an image constructed from the 
 completeness-corrected number density of
  stars brighter than the magnitude for which the compleneteness is 50\% in
  the central regions in a given star cluster. 
This image was constructed using a bin size of $50 \times 50$ pixels (i.e.,
2\farcs5 $\times$ 2\farcs5). 
Note that using the number density instead of surface
brightness avoids biases that can arise because of a few
bright stars near the center. The typical uncertainty of the centering
procedure was $\pm$\,5 pixels in either axis. 
The ellipticities of the clusters were derived by running the task {\tt
  ellipse} within {\sc iraf/stsdas}\footnote{STSDAS is a product of the Space
  Telescope Science Institute, which is operated by AURA for NASA} on the
number density images mentioned above. 
  Derived ellipticities stayed constant with radius to within the
  uncertainties. 
The area sampled by the ACS image was then divided into a number of centered,
concentric elliptical annuli. The number of such annuli
was chosen in an adaptive manner so as to include a minimum of 100 stars per
annulus. The surface number density was corrected for
incompleteness by dividing the number of stars by the average completeness
fraction in each annulus. For annuli with radii larger than
$\sim$\,850 pixels ($\hat{=}$ 42\farcs5), care was taken to account for the
limited azimuthal coverage of the cluster by the ACS/WFC image. 
  Specifically, we first constructed a parallelogram whose edges stay 50 
  pixels within the area exposed by all (three) F555W exposures of a given
  star cluster. We then constructed `elliptical pie slices' that subtend angle
  intervals which are radius-dependent in a way such that the angle subtended
  by the outer end of the pie slice fits fully within the parallelogram mentioned
  above. Stars were then counted within those pie slices, and the areas of
  each pie slice were evaluated to measure surface number densities. Error
  values were derived from Poisson statistics of the star number counts.
Radii are expressed in terms of the ``equivalent'' radius of the ellipse, 
$r = a\, \sqrt{1-\epsilon}$ where $a$ is the semimajor axis of the ellipse and
$\epsilon$ its ellipticity. 
The resulting radial surface number density profiles were fit with a
\citet{king62} model combined with a constant background level: 
\begin{equation}
n(r) = n_0 \: \left( \frac{1}{\sqrt{1 + (r/r_c)^2}} - \frac{1}{\sqrt{1+c^2}}
 \right)^2 \; + \; {\rm bkg} 
\label{eq:King}
\end{equation}
where $n_0$ is the central surface number density, $r_c$ is the core radius,
and $c \equiv (r_t/r_c)$ is the concentration index ($r_t$ being the tidal
radius). The best-fit King models were selected using a $\chi^2$
minimization routine that involves varying values of $c$. Fig.\
\ref{f:numdensfits} shows the best-fit models along with the individual
surface number density values for each star cluster in our sample (except
NGC~1846 for which we refer the reader to Paper~I). Radius values in arcsec 
were converted to parsecs using the distance moduli listed in
Table~\ref{t:bestisotab2}. 

\subsubsection{Selection of Cluster-Dominated Regions on the CMD}

We establish portions of the CMD that are strongly dominated
by cluster stars using the statistical method described in detail in 
Paper~I. Briefly, we compared star surface number 
densities in selection boxes on the CMD from two radial ranges: ``inner''
stars with $\log(r) \leq 1.5$ versus ``outer'' stars with $2.0 \leq \log(r)
\leq 2.2$ (with radius $r$ in arcsec). Selection boxes for which the
``inner/outer'' surface number density ratio (after completeness correction)
exceeded the value given by the best-fit King model to {\it all} stars in the
ACS image were tagged as dominated by cluster stars. 

Stars in regions on the CMD found to have less than 20\% contamination by 
field stars are shown with black dots in all CMDs shown in this paper,
whereas the other stars are shown with grey dots.

\section{Isochrone Fitting} \label{s:isofits}  

We fit isochrones to the CMDs of the star clusters in our sample 
to determine their age, metallicity, \afe\
ratios.
We use three sets of stellar models with predictions computed for the
ACS/WFC filter system: 
Padova isochrones \citep{marigo+08,gira+08}, Teramo isochrones
\citep[sometimes referred to as BaSTI isochrones;][]{piet+04,piet+06}, and
Dartmouth isochrones \citep{dott+08}.  

{\bf Padova isochrones:} We use the default models which involve scaled solar
abundance ratios (i.e., \afe\ = 0.0) and which include some degree of
convective overshooting \citep[see][]{gira+00}.
 The Padova isochrones are calculated through the thermally pulsing
 AGB (TP-AGB) stage of stellar evolution. 
Using the web interface of the Padova
team\footnote{\tt http://stev.oapd.inaf.it/cgi-bin/cmd}, we construct a grid of
isochrones that covers the ages $0.3 \leq \tau [{\rm Gyr}] \leq 3.0$ (where
$\tau$ is the age) with a step of $\Delta \tau = 0.05$ Gyr and metallicities
$Z$ = 0.001, 0.002, 0.004, 0.006, 0.008, 0.01, 0.02, and 0.03. 

{\bf Teramo isochrones:}  We use that team's web
site\footnote{\tt http://albione.oa-teramo.inaf.it} to construct grids of
isochrones that cover the same ages and metallicities as for
the Padova models mentioned above, except that $Z = 0.006$ is not available. 
We use the Teramo isochrones with \afe\ = 0.0  which are calculated through
the TP-AGB phase and which include prescriptions for convective
overshooting. 

{\bf Dartmouth isochrones:} We use the full grid available from their web
site\footnote{\tt http://stellar.dartmouth.edu/$\sim$models/complete.html} which
covers the ages $0.25 \leq \tau [{\rm Gyr}] \leq 1.0$ with $\Delta\tau = 0.05$ Gyr and
$1.0 < \tau [{\rm Gyr}] \leq 5.0$ with $\Delta\tau = 0.25$ Gyr, metallicities
[Fe/H] = $-$2.5, $-$2.0, $-$1.5, $-$1.0, $-$0.5, 0.0, +0.3, and +0.5, and \afe\
= $-$0.2, 0.0, +0.2, +0.4, +0.6, and +0.8. Isochrones with a finer grid in 
Age and/or [Fe/H] were also created around the initial best-fit values found for
the star clusters in our sample (see next Section), using the interpolation
routine made available through the Dartmouth team web site. 
The Dartmouth isochrones terminate at the He flash, so that the HB
and AGB sequences are not included.  
Finally, we use Dartmouth isochrones featuring enhanced Helium abundance ($Y$
= 0.33) to assess the impact of enhanced He on the MSTO region.
This topic is addressed in \S\,\ref{s:Helium}.   

\subsection{Fitting Method} \label{s:fitmeth}

The isochrone fitting was performed using the method described in
detail in Paper I. We provide a less comprehensive description here, and
concentrate most on parts of the procedure that are additions to the steps
described in Paper I. We start the isochrone fitting by using parameters that
involve pairs of fiducial points on the CMD that are: {\it (i)\/} relatively
easy to measure or determine from both the data and the isochrone tables, {\it
  (ii)\/} sensitive to at least one population parameter such as age or
metallicity, and {\it (iii) independent\/} of the distance and foreground
reddening of the cluster.  

In the case of clusters with a well-defined RGB bump (i.e., NGC~1751,
NGC~1783, NGC~1806, and NGC~1846), we use the following parameters (cf.\ Paper~I):   
\begin{enumerate}
\item The difference in magnitude between the MSTO and the 
RGB bump\footnote{The RGB bump marks the time in stellar
  evolution when the outward moving hydrogen-burning shell encounters the
  base of the convective envelope. At this point, fresh hydrogen fuels the
  fusion processes in the shell which becomes hotter and fainter for a
  short period, causing stars to pile up in the CMD \citep[e.g.,][]{fusi+90}.},
called
  $\Delta B_{\rm RGBB}^{\rm MSTO}$ and $\Delta V_{\rm RGBB}^{\rm
    MSTO}$ in the $B$ and $V$ filters, respectively. 
  The MSTO is defined as the point where a polynomial fit
  to the stars (or the isochrones) near the turn-off is vertical in the
  CMD. 
We define the location of the RGB bump in the isochrones as the average
  magnitude and color of isochrone RGB entries between the two
  masses at which the magnitudes and colors ``turn around'' in direction on
  the CMD with increasing stellar mass. 
\item The difference in color between the MSTO and the RGB bump, referred to as
  $\Delta (B-I)_{\rm RGBB}^{\rm MSTO}$ and $\Delta (V-I)_{\rm RGBB}^{\rm MSTO}$. 
\item The slope of the RGB. This was evaluated using the (mean) color of the RGB
  stars at two fiducial magnitudes, namely at $m_{\rm RGBB} + 1$ and $m_{\rm
    RGBB} - 0.75$. The former magnitude was chosen to represent a point
  intermediate between the RGB bump and the lower end of the RGB; the latter
  magnitude was chosen to avoid issues related to confusing RGB with AGB stars
  on the CMD.   
  The mean colors of the RGB stars were derived from the CMD
  by means of a polynomial fit to the RGB star positions in the CMD.
The predicted colors
were derived from a linear interpolation
  between isochrone table entries. 
\end{enumerate}
The main reason for using the RGB bump as a prime parameter in this context is
that all three isochrone families can be used this way. However, in the case of
clusters for which the location of the RGB bump is not well constrained
from the observations (i.e., NGC~1987, NGC~2108, and LW~431), we replace
parameters (1) and (2) above  with the following:   
\begin{itemize}
\item[1a.] The difference in magnitude between the MSTO and the red
  clump (RC), named 
  $\Delta B_{\rm RC}^{\rm MSTO}$ and $\Delta V_{\rm RC}^{\rm
   MSTO}$ in the $B$ and $V$ filters, respectively. 
%  The RC marks the core helium-burning phase in stellar evolution for
%  stars with a significant amount of residual hydrogen envelope
%  material that have effective temperatures cooler than stars on the
%  instability strip. This phase is often referred to as the red
%  horizontal branch \citep[e.g.][]{gira+98}.
%  For the location of the RC on the CMD w
We simply calculate the mean observed
  magnitude and color of stars in a box centered on the RC by eye. For the
  isochrones, we define the ``mean'' location of the RC as follows. After
  identifying the start of the RC in the Padova and Teramo isochrone
  tables, isochrone magnitudes and colors are recorded up to the point
  where the difference in color between two subsequent isochrone entries
  becomes $\ge 3\,\sigma$ larger (redder) than the average color
  accumulated from the isochrone entries in the RC recorded up to that
  point. This procedure was empirically verified to yield the appropriate
  end point of the RC. Weighted magnitudes and colors for the RC are then
  derived from the recorded isochrone entries. To simulate the distribution of
  stars in the RC of a star cluster, weight factors are assigned during the
  latter operation by using a \citet{salp55} mass function.  
\item[2a.] The difference in color between the MSTO and the RC, named 
  $\Delta (B-I)_{\rm RC}^{\rm MSTO}$ and $\Delta (V-I)_{\rm RC}^{\rm MSTO}$. 
\end{itemize} 

\begin{figure*}[tp]
\epsscale{0.6}
\plotone{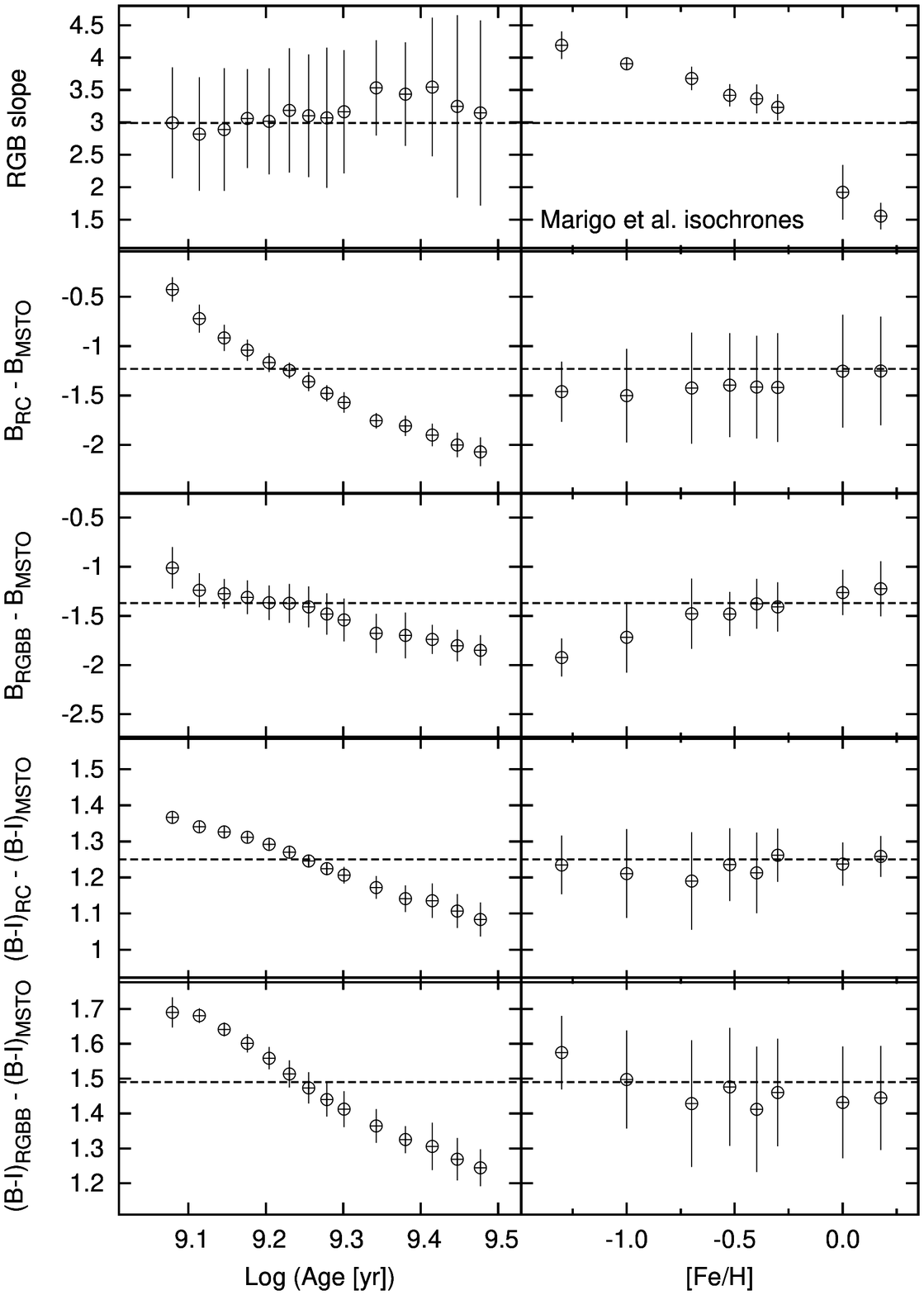}
\caption{Dependence of five distance-- and reddening-independent parameters
  involving fiducial points on the \B\ vs.\
  \BI\ CMD on age and metallicity.  Predictions from the Padova
  isochrones \citep{marigo+08} are shown. Top panels: Slope of the RGB as defined in \S\
  \ref{s:fitmeth}. 
  Second row of panels: $B_{\rm RC} - B_{\rm MSTO}$. 
  Third row of panels: $B_{\rm RGBB} - B_{\rm MSTO}$. 
  Fourth row of panels: $(B-I)_{\rm RC} - (B-I)_{\rm MSTO}$. 
  Bottom panels: $(B-I)_{\rm RGBB} - (B-I)_{\rm MSTO}$. 
  Error bars in the left and
  right panels reflect the variation of the parameter values among the
  isochrones with different metallicities and ages, respectively. The dashed
  lines in each panel represent the measurements of these parameters
  for NGC 1806. 
\label{f:agemetplot_girardi}}    
\end{figure*}

\begin{figure*}[tp]
\epsscale{0.6}
\plotone{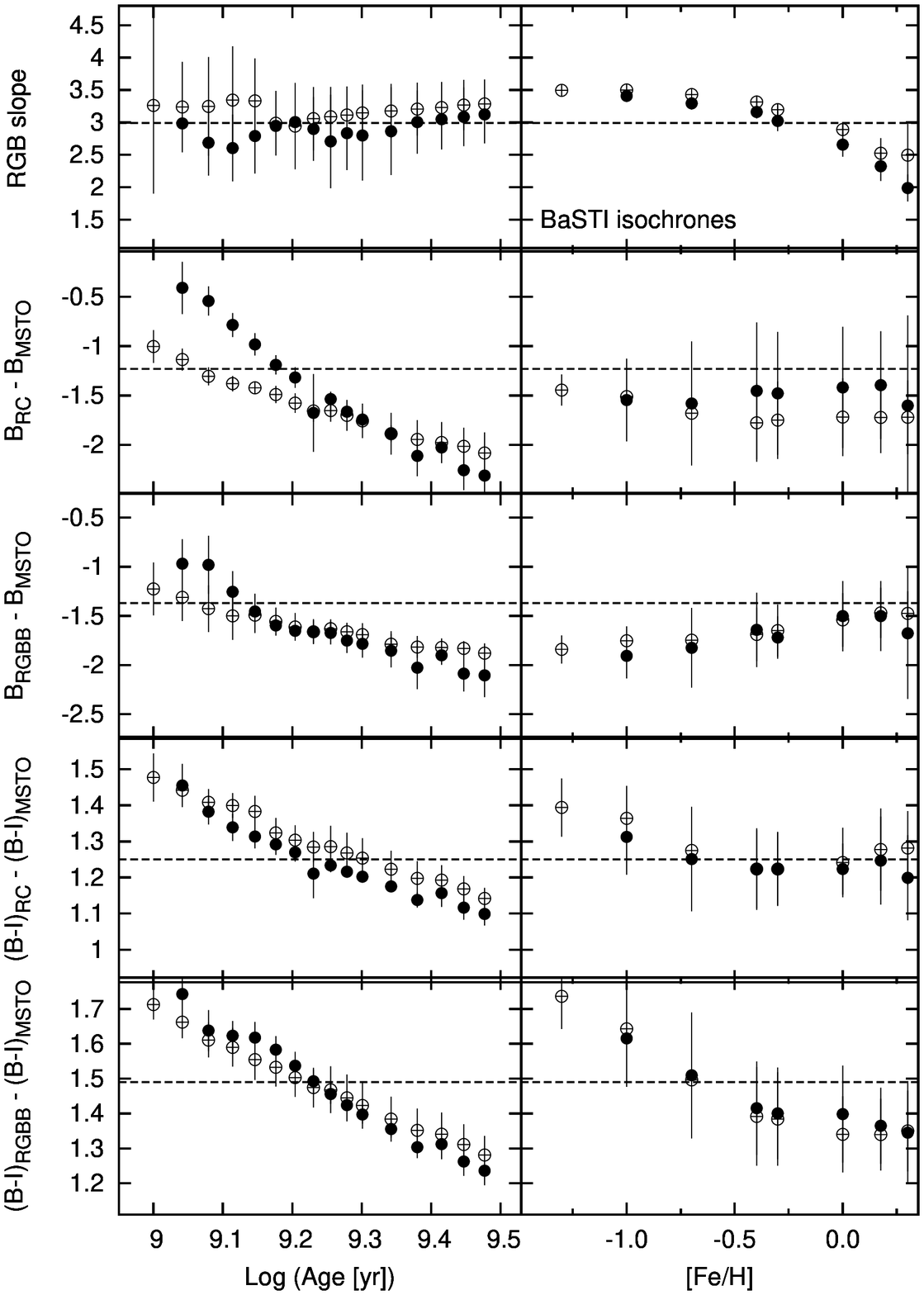}
\caption{Same as Fig.\ \ref{f:agemetplot_girardi}, but now for the Teramo
  isochrones \citep{piet+04}. The filled circles refer to the isochrones with
  convective overshooting, while the open circles refer to those without. 
\label{f:agemetplot_basti}}    
\end{figure*}

The sensitivity of parameters (1), (2), and (3) mentioned above to population
parameters in the age range 1\,--\,3 Gyr was illustrated in Figures 9\,--\,11 
of Paper I for all three isochrone families used here. Since we are using
parameters (1a) and (2a) instead of (1) and (2) for some clusters in this
paper, we now show a comparison of the sensitivity of parameters (1), (1a), (2),
(2a), and (3) to age and [Fe/H] for the Padova and Teramo isochrone
families in Figures \ref{f:agemetplot_girardi} and \ref{f:agemetplot_basti} for
the age range 1\,--\,3 Gyr. 
These plots 
show that while details of the dependences of $\Delta B_{\rm RC}^{\rm MSTO}$
and $\Delta (B-I)_{\rm RC}^{\rm MSTO}$ on age and \FeH\ are  
different from those of $\Delta B_{\rm RGBB}^{\rm MSTO}$ and $\Delta
(B-I)_{\rm RGBB}^{\rm MSTO}$, both sets of parameters do yield mutually
consistent results when compared with the observed values. As already
mentioned in Paper I, the RGB slope is highly sensitive to metallicity and
almost independent of age in the range studied here. 

We then select all isochrones (within each family) for which the values of
the three parameters mentioned above lie within 2 $\sigma$ of the measurement
uncertainty of those parameters on the CMDs. 
This yielded 6\,--\,15 isochrones depending on the isochrone
family. For these isochrones, we then find the best-fit values for distance
modulus $(m-M)_0$ and foreground reddening $A_V$ by means of a least squares
fitting program. For the filter-dependent reddening we use
$A_{\rm F435W} = 1.351\, A_V$, $A_{\rm F555W} = 1.026\, A_V$, and $A_{\rm F814W} =
0.586\, A_V$ (cf.\ Paper~I). 

Finally, the isochrones were overplotted onto the
CMDs for visual examination. As mentioned in Paper I for NGC~1846, 
this revealed that there was a small but 
systematic offset in \FeH\ between best-fit isochrones for \B\
vs.\ \BI\, and \V\ vs.\ \VI\ in the sense that the derived value of 
\FeH\ was always higher for the isochrone fit to \V\ vs.\ \VI\ 
than to \B\ vs.\ \BI. This effect was most significant for the Padova and
Teramo isochrones, and we have suggested (see Paper~I) that the cause is
related to the fact that those two isochrone families 
  derive their
  $T_{\rm eff}$ -- color relations from the ATLAS9 stellar atmosphere models
  of R.\ L.\ Kurucz \citep[e.g.,][]{caskur03} which have been shown to contain
  more flux in the range $\sim$\,5000\,--\,6500 \AA\ (including much of the
  $V$ band) than empirical star spectra from the \citet{pick98} library at the
  same stellar type \citep{mara+08}. Model SEDs therefore have bluer 
  $V\!-\!I$ colors than observed for RGB stars, consistent with what we see. 
Conversely, the Dartmouth isochrones are based on the {\sc Phoenix}
model atmospheres \citep[e.g.,][]{haus+99} which include hundreds of millions
more molecular transitions than the ATLAS9 models and hence a more
accurate opacity modeling. 
Because of this effect, we focus on the \B\ vs.\ \BI\ CMDs for
deriving population parameters.

\begin{figure*}[tp]
\centerline{
\psfig{figure=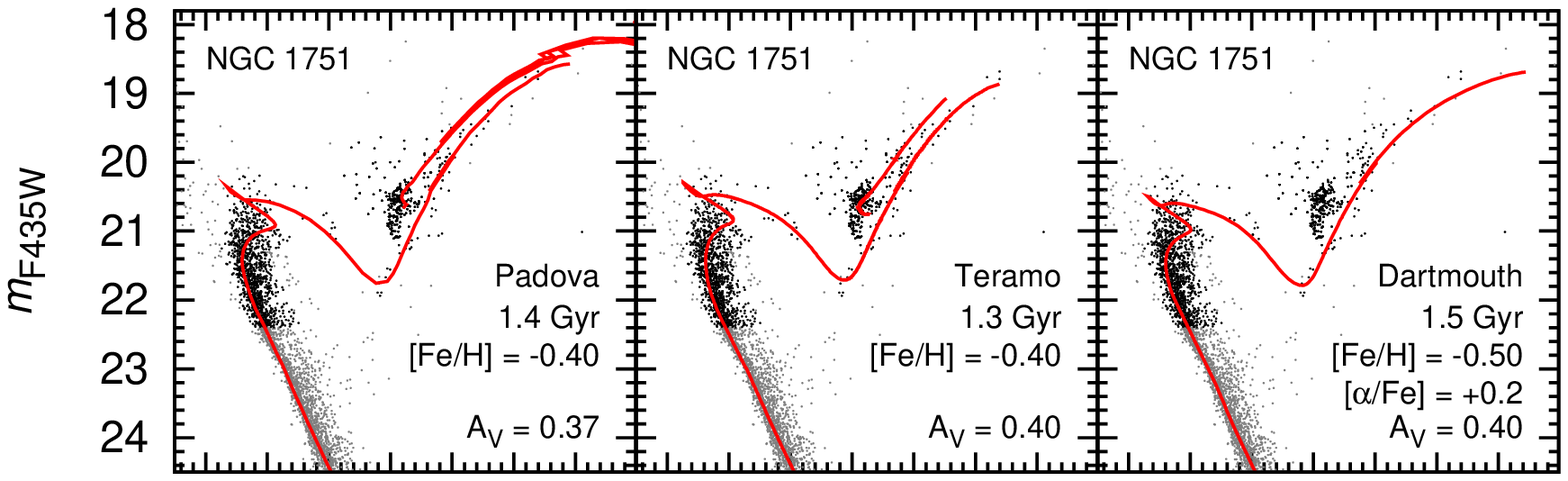,width=16cm}
}
\centerline{
\psfig{figure=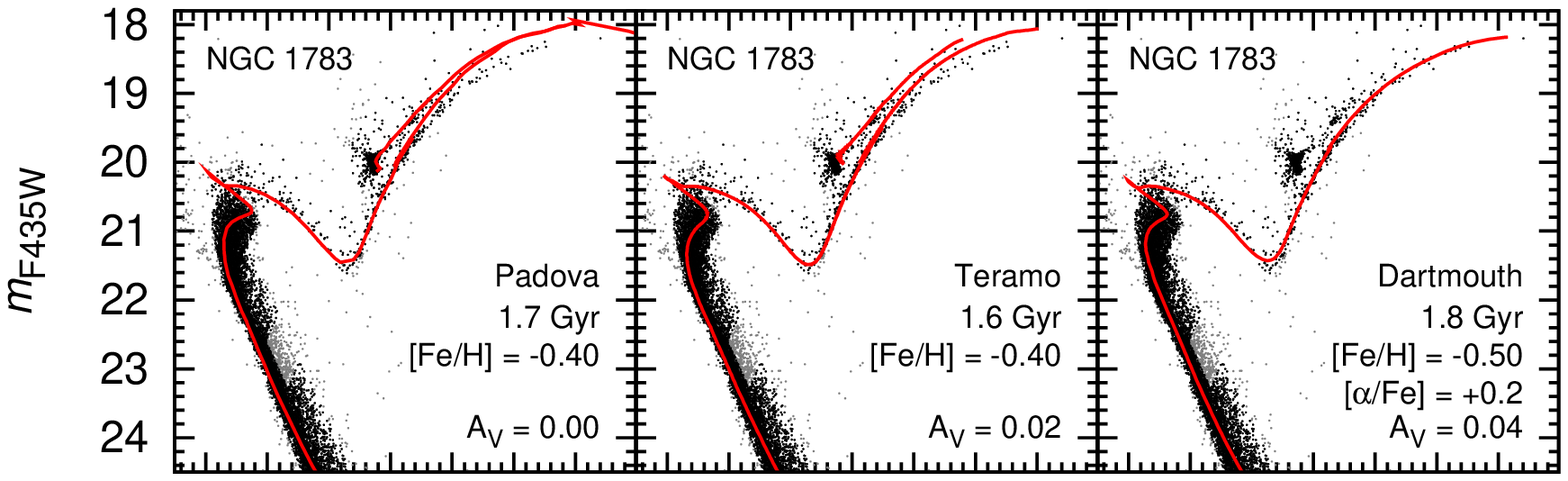,width=16cm}
}
\centerline{
\psfig{figure=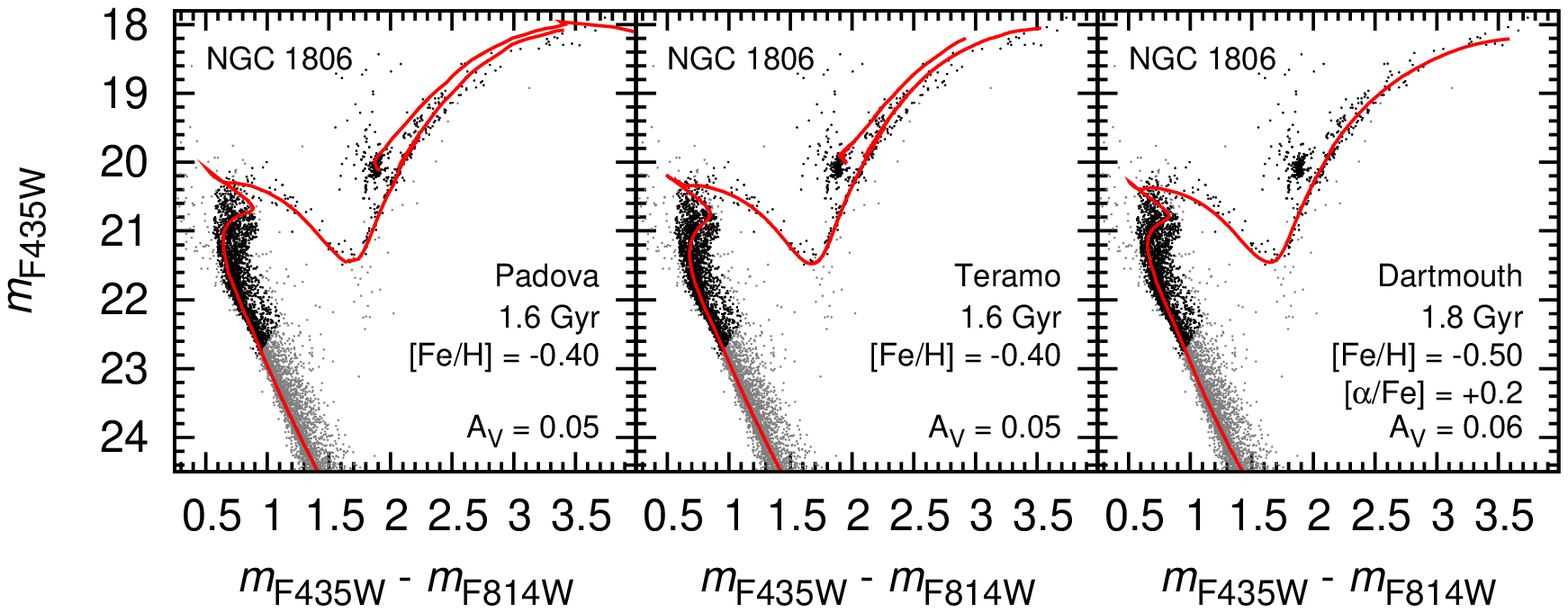,width=16cm}
}
\caption{Best-fit isochrones (solid lines) are superposed onto the CMDs of NGC
  1751, NGC 1783, and NGC 1806 (the CMDs are the same as those shown in 
Fig.\ \ref{f:fullCMDs_1}). The left-hand panels show Padova
  isochrones, the middle panels show Teramo isochrones, and the right-hand
  panels show Dartmouth isochrones. 
Black dots indicate stars in areas of the
  CMD contaminated by field stars by less than 20\%, while grey dots
  indicate stars in other areas. 
\label{f:isofits1}}
\end{figure*}

\begin{figure*}[tp]
\centerline{
\psfig{figure=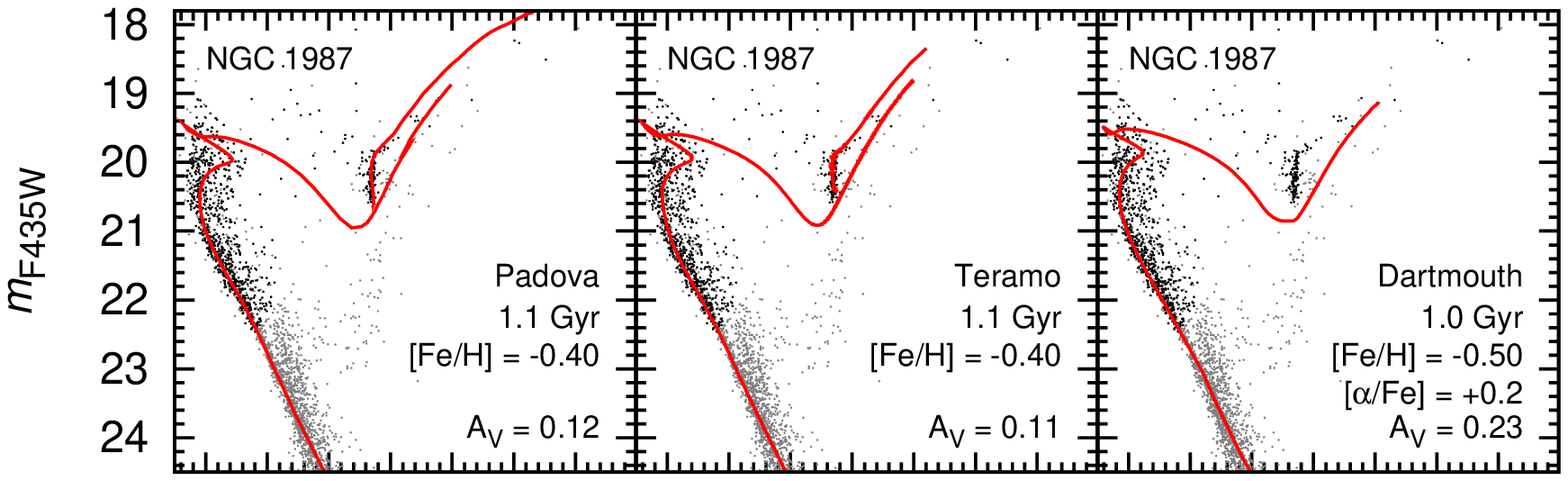,width=16cm}
}
\centerline{
\psfig{figure=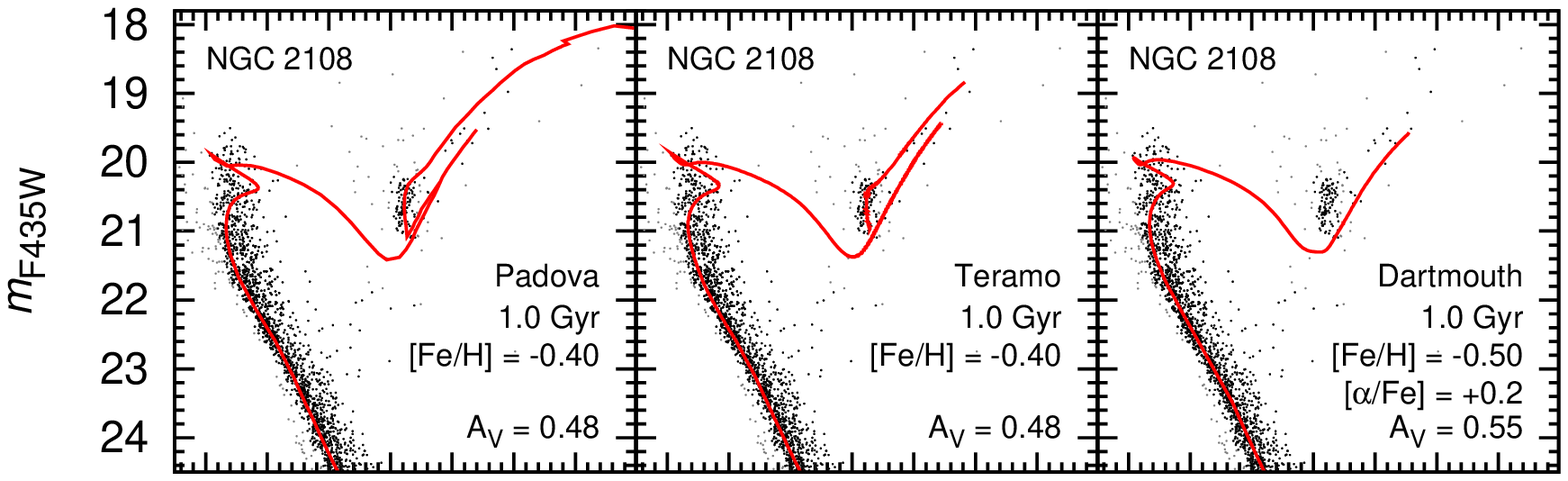,width=16cm}
}
\centerline{
\psfig{figure=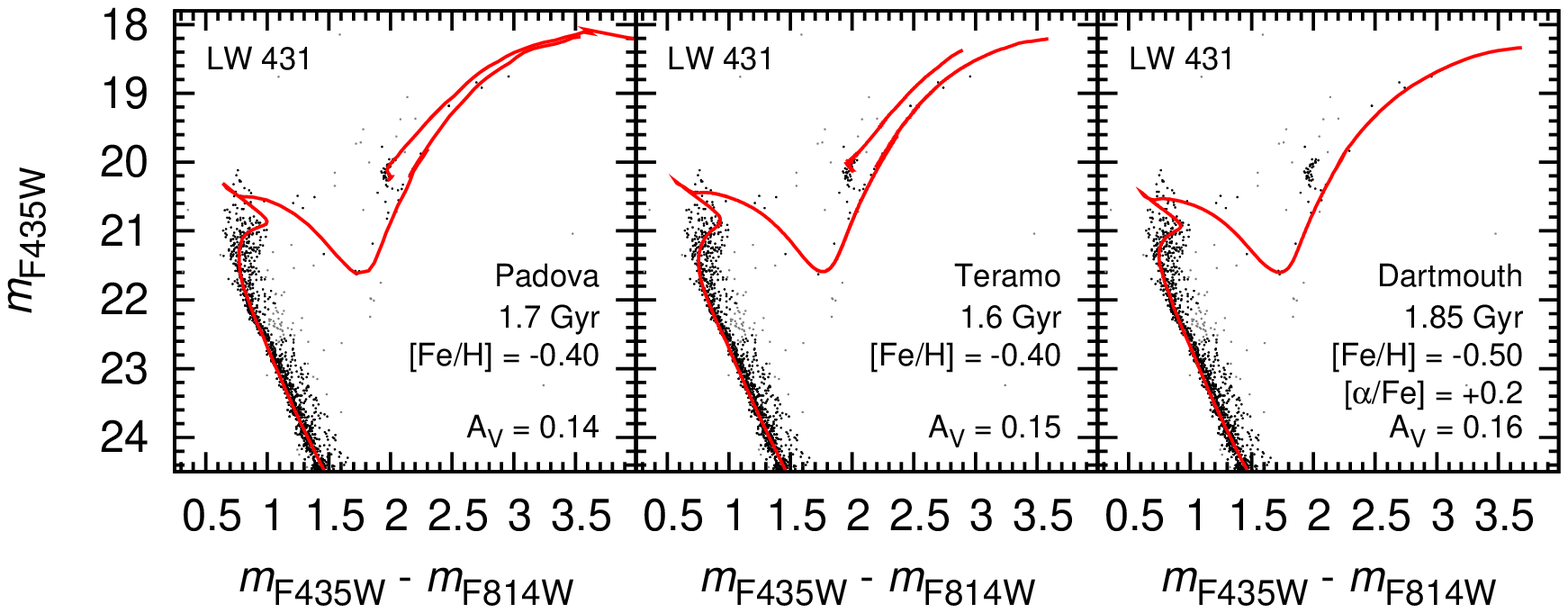,width=16cm}
}
\caption{Same as Fig.\ \ref{f:isofits1}, but now for NGC 1987, NGC 2108, 
  and LW 431. 
\label{f:isofits2}}
\end{figure*}

The best-fit isochrones and their population parameters 
are listed for each isochrone family in Table~\ref{t:bestisotab1}
(in the back of the paper, after the References) and shown in
Figures \ref{f:isofits1}\,--\,\ref{f:isofits2}, superposed onto the
CMDs. The best-fit isochrones of each model family generally match the various
stellar sequences well\footnote{Dartmouth isochrone fits could not be
  performed as described for the clusters NGC~1987, NGC~2108, and LW~431 due
  to the lack of a clear RGB bump and the fact that the Dartmouth isochrones
  do not extend to the RC. The Dartmouth fits shown in Figure~\ref{f:isofits2}
  are instead ``by eye'' fits to the MS and MSTO.}. 
However, one significant difference among the isochrones is seen on the upper
RGB 
  for clusters NGC~1783 and NGC~1806 (the same was seen for NGC~1846,
  see Paper~I). The best-fit Dartmouth isochrones typically provide a better fit to the 
upper RGB than the Padova and Teramo isochrones, both of which appear bluer
than the observed stars. This difference is briefly discussed in 
\S~\ref{s:aFe} below.  

In order to find the spread in age that can explain the extended 
morphology of the MSTO regions of the clusters in our sample, we first note
that the width of the RGB in all clusters is consistent with
photometric errors and that the slope of the RGB in each cluster does not show
evidence for an intrinsic spread in [Fe/H]. Hence we 
fix [Fe/H], $A_V$, and $(m-M)_0$ and vary the isochrone age using steps of
0.05 Gyr until one reaches the extremes of the eMSTO region populated by the
cluster stars. These age spreads are listed in Table~\ref{t:bestisotab2},
along with the 
final adopted parameters for the clusters in our sample, 
including their uncertainties that reflect both errors associated with the
isochrone fitting and systematic uncertainties related to the use of the
different stellar models. The determination of the adopted population
parameters and their uncertainties are discussed further in \S\
\ref{s:syserrors} below. 

\begin{table*}[tbh]
\begin{center}
\footnotesize
%\scriptsize
\caption{Adopted population parameters of the star clusters studied in this
  paper.}
 \label{t:bestisotab2}
\begin{tabular}{@{}lcccccc@{}}
\multicolumn{3}{c}{~} \\ [-2.5ex]   
 \tableline \tableline
\multicolumn{3}{c}{~} \\ [-2.ex]                                                
\multicolumn{1}{c}{Cluster} & Age & Age Range & [Fe/H]\tablenotemark{a} &
$(m-M)_0$ & $A_V$ & log ${\cal{M}}_{\rm cl}$ \\
\multicolumn{1}{c}{(1)}     & (2) & (3)       & (4) &   (5)     & (6) & (7) 
 \\ [0.5ex] \tableline  
\multicolumn{3}{c}{~} \\ [-2.ex]              
NGC 1751 & $1.40 \pm 0.10$ & 1.15\,--\,1.65 & 
  $-0.50 \pm 0.10$ & $18.50 \pm 0.03$ & $0.40 \pm 0.01$ & $4.82 \pm 0.09$ \\
NGC 1783 & $1.70 \pm 0.10$ & 1.50\,--\,1.90 &
  $-0.50 \pm 0.10$ & $18.46 \pm 0.04$ & $0.02 \pm 0.02$ & $5.25 \pm 0.09$ \\
NGC 1806 & $1.67 \pm 0.10$ & 1.57\,--\,1.92 & 
  $-0.50 \pm 0.10$ & $18.46 \pm 0.04$ & $0.05 \pm 0.01$ & $5.03 \pm 0.09$ \\
NGC 1846 & $1.73 \pm 0.10$ & 1.53\,--\,1.93 &
  $-0.50 \pm 0.10$ & $18.45 \pm 0.05$ & $0.08 \pm 0.01$ & $5.17 \pm 0.09$ \\
NGC 1987 & $1.05 \pm 0.05$ & 0.95\,--\,1.20 &
  $-0.50 \pm 0.10$ & $18.38 \pm 0.02$ & $0.16 \pm 0.04$ & $4.49 \pm 0.09$ \\
NGC 2108 & $1.00 \pm 0.05$ & 0.90\,--\,1.10 &
  $-0.50 \pm 0.10$ & $18.45 \pm 0.03$ & $0.50 \pm 0.03$ & $4.41 \pm 0.09$ \\
  LW 431 & $1.73 \pm 0.10$ & 1.53\,--\,1.93 &
  $-0.50 \pm 0.10$ & $18.43 \pm 0.03$ & $0.15 \pm 0.01$ & $4.00 \pm 0.09$ \\ [0.5ex] \tableline
\multicolumn{3}{c}{~} \\ [-2.5ex]              
\end{tabular}
\tablecomments{Column (1): Name of star cluster. (2): Adopted age in Gyr. (3):
  Age range in Gyr associated with the width of the observed MSTO region. (4):
  Adopted [Fe/H] in dex. (5): Adopted distance modulus in mag. (6): Adopted
  foreground $V$-band reddening in mag. (7) Logarithm of photometric mass (in \Msun).}  
\tablenotetext{a}{[$\alpha$/Fe] = +0.2 $\pm$ 0.1 is adopted for all clusters.}
\end{center}
\end{table*}

Table~\ref{t:bestisotab2} also includes present-day cluster masses which are
estimated from the total $V$ magnitudes listed in Table~\ref{t:sample} and the
$A_V$, $(m-M)_0$, [Fe/H], and mean age listed in
Table~\ref{t:bestisotab2}.
The masses use the ${\cal{M}}/L_V$ predicted by the SSP models of \citet{bc03},
assuming a \citet{salp55} IMF. 
The latter models were recently found to provide the best fit (among popular
SSP models) to observed integrated-light photometry of LMC clusters with ages and
metallicities measured from CMDs and spectroscopy of individual RGB stars in
the 1\,--\,2 Gyr age range \citep{pess+08}.

\subsection{Influence of [$\alpha$/Fe] Abundance Ratio} \label{s:aFe}

\begin{figure*}[ptb]
\centerline{
\psfig{figure=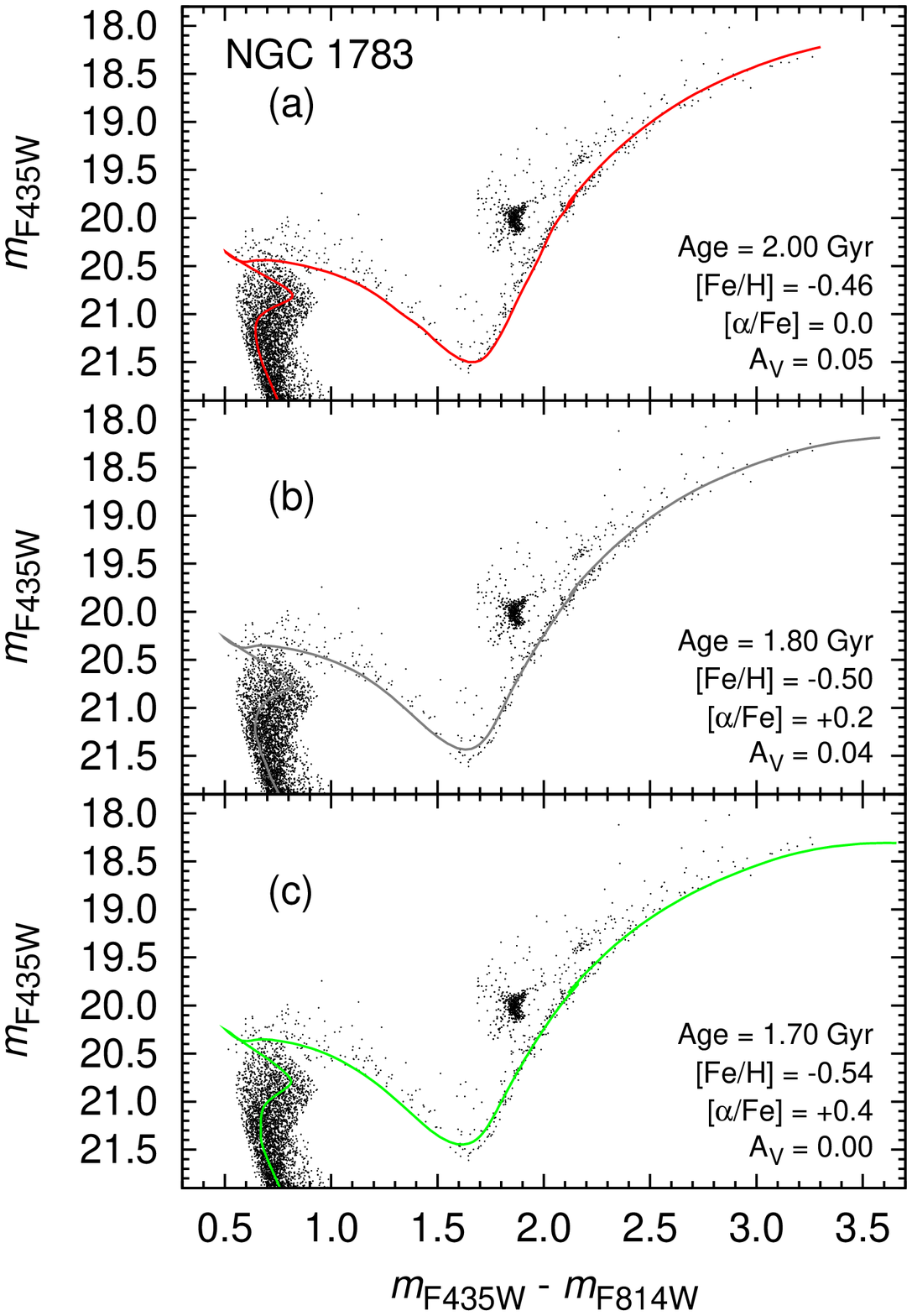,width=6.5cm}
\hspace*{1mm}
\psfig{figure=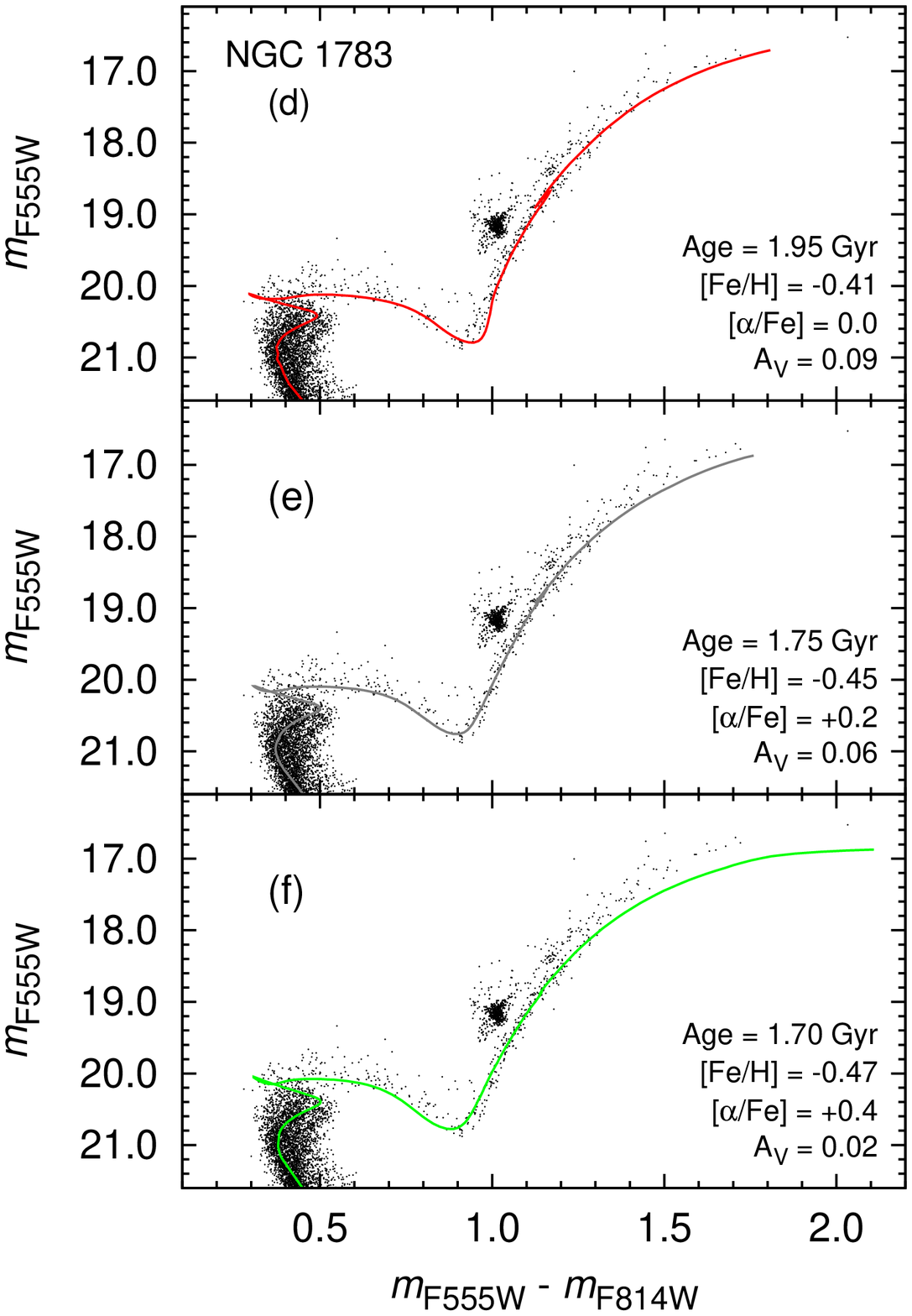,width=6.5cm}
}
\caption{Illustration of the effect of the \afe\ ratio on the
  isochrone morphology along the RGB as well as on the derived age and \FeH\
  for NGC~1783. Panels (a) through (c) show the best-fit Dartmouth
  isochrones for \afe\ = 0.0, +0.2, and +0.4 respectively in \B\ vs.\ \BI. 
  Panels (d) through (f) do the same in \V\ vs.\ \VI.
  The legends list the best-fit population properties in each case.
  Note the effect of the \afe\ ratio on the slope and curvature of
  the RGB, as well on the resulting age and \FeH. 
\label{f:afeplot1}}
\end{figure*}

\begin{figure*}[ptb]
\centerline{
\psfig{figure=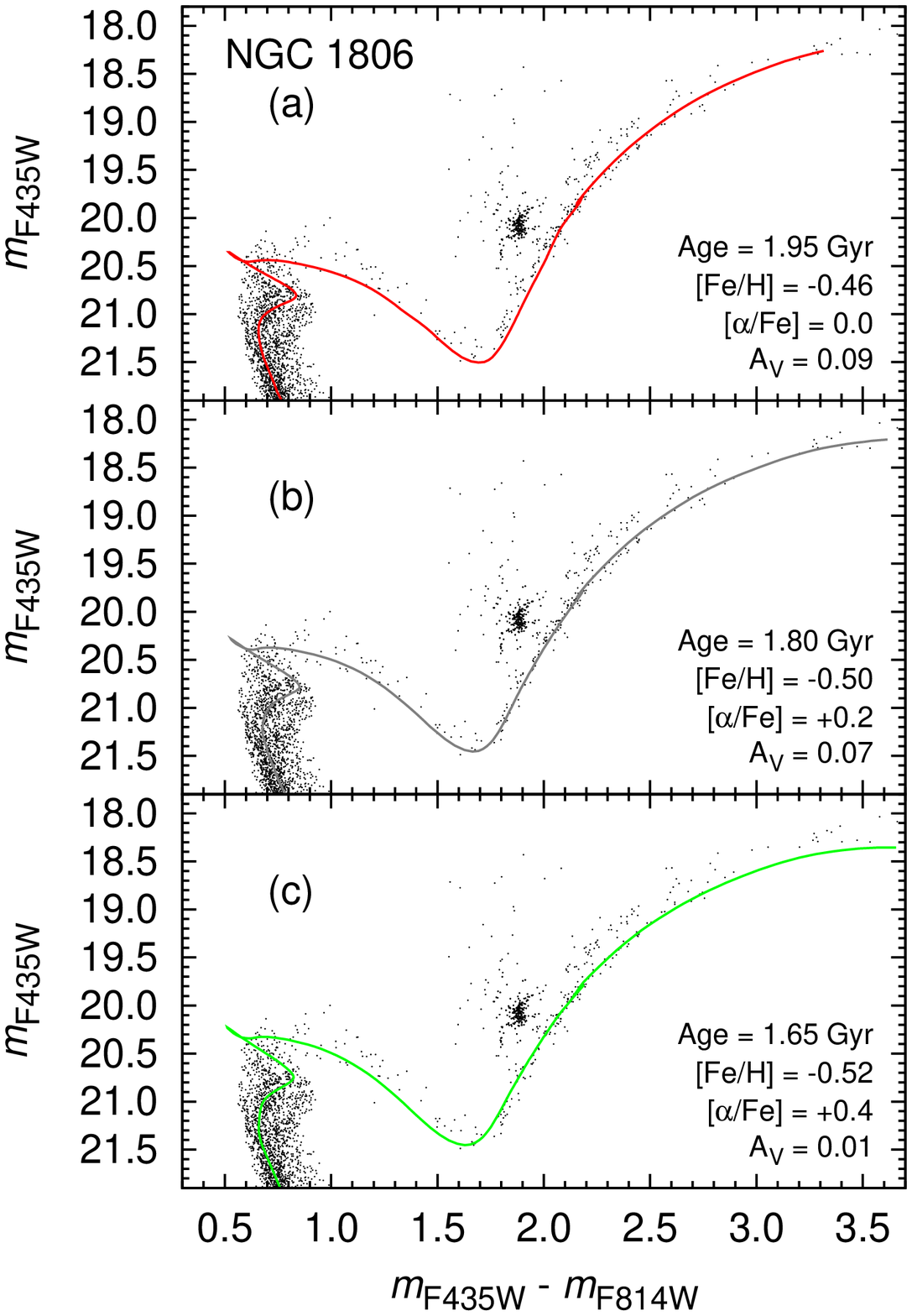,width=6.5cm}
\hspace*{1mm}
\psfig{figure=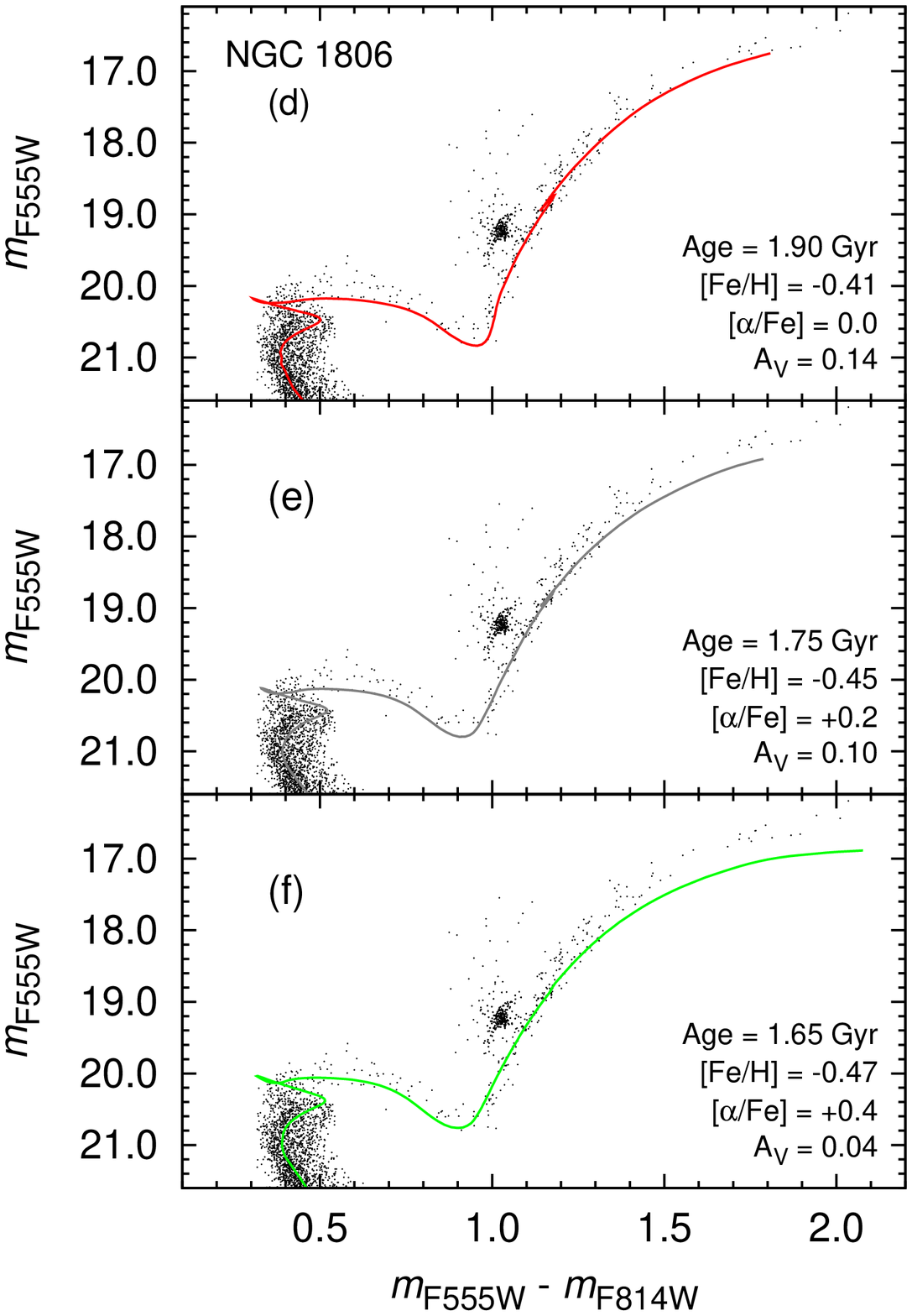,width=6.5cm}
}
\caption{Same as Fig.\ \ref{f:afeplot1}, but now for NGC~1806.
\label{f:afeplot2}}
\end{figure*}

We assess the influence that non-solar \afe\ abundances would have on the derived age and
metallicity by comparing our CMD with Dartmouth isochrones for different values  
of \afe. The result is illustrated in Figures \ref{f:afeplot1} and 
\ref{f:afeplot2}, which show the
best-fit Dartmouth isochrones for \afe\ = 0.0, +0.2, and +0.4 superposed
onto the \B\ vs.\ \BI\ 
 and \V\ vs.\ \VI\ 
CMDs of NGC~1783 and
NGC~1806, the two star clusters in 
our sample (in addition to NGC~1846, cf.\ Paper I) that have RGB sequences
sampled well enough to allow this exercise. 
Note that all three isochrones fit the MSTO and
RGB bump locations well, which is likely (at least partly) due to the
fitting method we used (see above in \S\ \ref{s:isofits}). However,
the detailed fits along the RGB differ significantly from one \afe\
value to another. 
 This is best seen in the \B\ vs.\ \BI\ CMD. 
In particular, there is a relation between the value of  \afe\ and the curvature
of the RGB in the sense that larger \afe\ yields stronger curvature for the RGB. 

As already mentioned in Paper I for NGC~1846, larger values of \afe\ result in
younger fitted ages.
Hence there is a degeneracy between age and \afe\
if one does not take the detailed morphology of the RGB into account. From 
the results for NGC~1783, NGC~1806, and NGC~1846, the amplitude of this effect
is a decrease of 9.3\% ($\pm$ 1.0\%) in age for an increase of \afe\ of 0.2 dex.  

As Figures \ref{f:afeplot1} and \ref{f:afeplot2} show, the best fit to the full
RGB is achieved using the isochrone with \afe\ = +0.2 for all clusters for
which the extent and sampling of the RGB is sufficient to permit this
comparison. Since the fit of the  Dartmouth isochrones with \afe\ =
+0.2 dex to the RGBs is clearly better than any isochrone that uses
solar abundance ratios in any of the three families, we adopt \afe\ =
+0.2 for NGC~1751, NGC~1783, NGC~1806, 
NGC~1846, and LW\,431.  
  Future spectroscopic determinations of \afe\ in RGB stars of these
  clusters and the surrounding field population would be very useful
  to confirm or deny the trends found here from photometry.  

\subsection{Adopted Population Properties and their Systematic
  Uncertainties} \label{s:syserrors}    

We evaluate ``mean'' ages of the star clusters in our 
sample as follows. As to the clusters for which we adopt \afe\ =
+0.2 (i.e., all clusters except NGC~1987 and NGC~2108, the two youngest ones), we
first consider the best-fit ages found for these clusters using the Dartmouth isochrones 
that employ \afe\ = 0.0. Mean values and standard
deviations of the ages found from all three sets of isochrone families that
include treatment of convective overshooting are then derived (at \afe\ =
0.0). Finally, those mean ages are converted to the equivalent for \afe\ = +0.2 
using the relation between age and \afe\ mentioned in \S\,\ref{s:aFe} above.  
As to NGC~1987 and NGC~2108, which are too young to exhibit a RGB bump, the isochrone
fitting method employed here (see \S\,\ref{s:fitmeth}) does not work with Dartmouth
isochrones. Hence we average the results from the Padova and Teramo
isochrone families for those clusters. We note that the ``by eye'' Dartmouth
isochrone fits shown in Figure~\ref{f:isofits2} for those clusters do have 
ages and [Fe/H] values consistent with the Padova and Teramo isochrone fits. 
 
We quantify systematic uncertainties in derived age, [Fe/H], 
distance, and reddening by comparing our best-fit results from 
each set of isochrones that includes treatment of convective overshooting,  
as compiled in Table~\ref{t:bestisotab1}.
These results yield systematic uncertainties of:
$\pm$\,7\% in age, $\pm$\,0.1 dex in [Fe/H], $\pm$\,0.05 mag in $(m-M)_0$ ($\simeq$\,5\% in
linear distance), and $\pm$\,0.02 mag in $A_V$ ($\simeq$\,15\%). 
We suggest that these values represent typical systematic uncertainties
associated with the determination of  population parameters of
intermediate-age star clusters from \B\ vs.\ \BI\ CMD fitting by isochrones
of any given stellar model.  

\section{Constraints on Helium Abundance Spreads} \label{s:Helium}

Massive GCs in our Galaxy such as $\omega$\,Cen, NGC~2808, NGC~1851 and 47~Tuc
have multiple MS and/or SGB sequences, which are typically interpreted
as populations with different Helium abundance
\citep[e.g.,][]{piot+07,milo+08,ande+09}. Note that these Galactic GCs are
much more massive than the LMC clusters in our sample, and were even
more so at their birth since they have undergone $\sim$\,10 Gyr more mass loss
due to disruption processes. Even so, Helium is a natural product of
the same chemical reactions that are thought to be responsible for the
Na-rich and O-poor stars in the Na-O anticorrelations among stars in Galactic GCs
\citep[e.g.,][]{grat+04}. Since the eMSTO phenomenon in the intermediate-age
clusters in our sample may well be causally associated with the Na-O
anticorrelations within Galactic GCs, we evaluate the possibility that enhanced
He may contribute to the eMSTO feature.
We use the Dartmouth isochrones with moderate He enhancement, namely He mass
fraction $Y = 0.33$\footnote{We employ both \afe\ = 0.0 and \afe\ = +0.4
  isochrones in this context; Dartmouth isochrones with \afe\ = +0.2
  are not available for the case of enhanced He.}. 

\begin{figure}[tb]
\centerline{
\psfig{figure=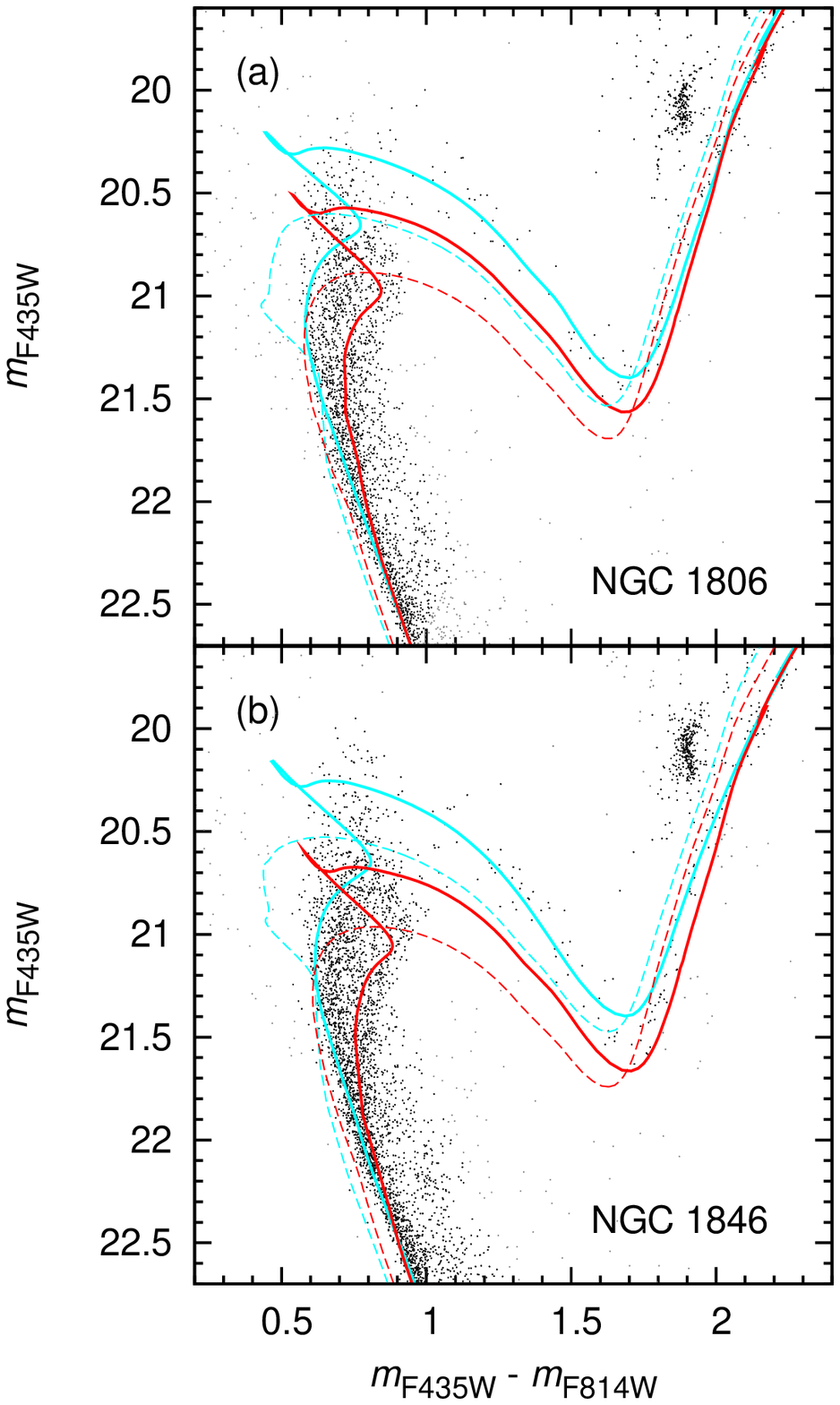,width=8cm}
}
\caption{Predictions from Dartmouth isochrones with primordial (solid
lines) and enhanced (dashed lines) Helium abundance.
The CMDs of NGC~1806 (panel a) and NGC 1846 (panel b) are shown. 
Solid lines show best-fit
  isochrones with \afe\ = 0.0 for the bright end (blue) and faint end (red) of
  the eMSTO region. The dashed lines show isochrones with the same age,
  [Fe/H], and \afe\ as the solid lines of the same color, but now with helium
  mass fraction $Y = 0.33$.  
\label{f:He_fig1}}
\end{figure}

Taking into account that the width of the RGB in our sample is
consistent with small uncertainties in the photometry (of order 0.01 mag in
\BI), we first hold $(m-M)_0$, $A_V$, and [Fe/H] constant at the
values listed in Table~\ref{t:bestisotab1}, and choose ages consistent
with the bright and faint ends of the eMSTO region for each cluster as
listed in Table~\ref{t:bestisotab2}. Figure~\ref{f:He_fig1} shows the
CMDs of NGC~1806  and NGC~1846 along with isochrones for the ages and
[Fe/H] described above, both for $Y = 0.25$ (i.e., primordial He) and $Y =
0.33$ (enhanced He). Note that the He-enriched isochrones do not fit the CMDs
well. In particular, the $Y = 0.33$ isochrones have SGBs and MSes that are
$\sim$\,0.3 mag too faint relative to the data (and the best-fit isochrones
that use primordial He).  
The morphology of the MSTO region of the star clusters is also distinctly
different from that of the isochrones with enhanced He. The latter indicate a
MSTO region which lies roughly $\Delta$\,(\BI)~$\simeq 0.15$ mag on the
blueward (hot) side of the MSTO of the clusters. Finally, the RGB sequences of
the He-enriched isochrones are significantly bluer than the RGB of the clusters. 

\begin{figure}[tb]
\centerline{
\psfig{figure=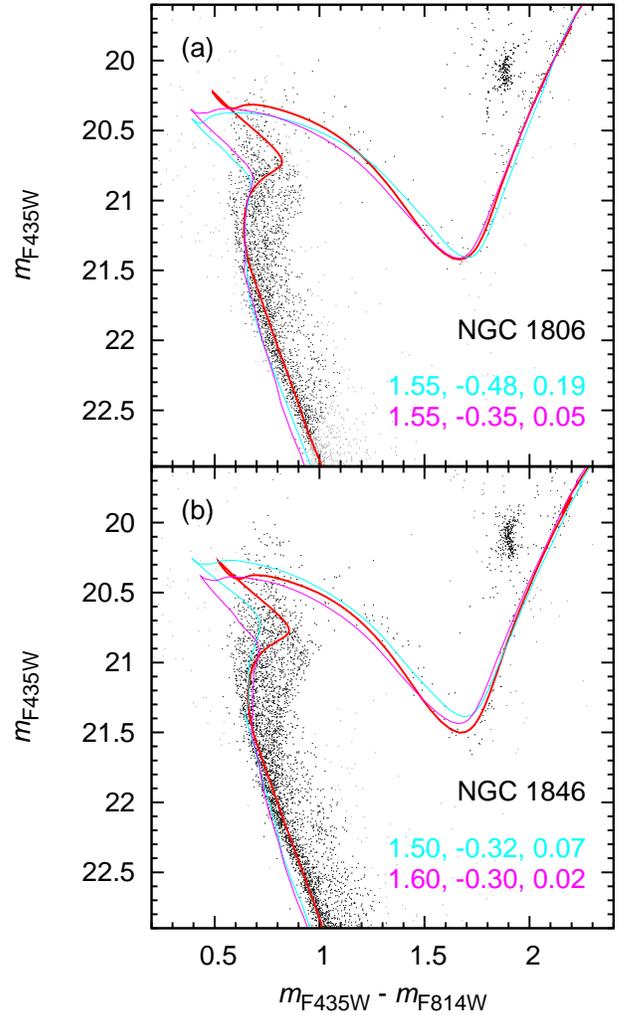,width=8cm}
}
\caption{Similar to Fig.~\ref{f:He_fig1}, but red lines now show best-fit
  Dartmouth isochrones with $Y = 0.25$ (cf.\ Table~\ref{t:bestisotab2}) for the
  overall CMDs of NGC 1806 (panel a) and NGC 1846 (panel b). The blue and
  magenta lines show two of the best-fitting isochrones with $Y = 0.33$ to
  those CMDs. The parameters Age (in Gyr), [Fe/H] (in dex) and $A_V$ (in mag)
  of the latter   two isochrones are shown in the legend (in that order, and
  using the same   color). Note that the best-fit $Y = 0.33$ isochrones always
  have MSes that   are significantly bluer than the cluster MSes.  
\label{f:He_fig2}}
\end{figure}

Considering the alternative (albeit unlikely) possibility that the majority of
the stars in these clusters have enhanced helium abundance, we then leave the
age, [Fe/H], \afe, $(m-M)_0$, and $A_V$ as free parameters and repeat the
isochrone fitting procedure mentioned in \S\,\ref{s:fitmeth} with $Y = 0.33$
isochrones. We used a grid of age and [Fe/H] with steps of $\Delta$\,Age = 0.05
Gyr and $\Delta$\,[Fe/H] = 0.02 dex in this case,
respectively, centered on the values found for primordial
helium. Figure~\ref{f:He_fig2} depicts the two best-fit $Y$ = 0.33 isochrones
to NGC 1806 and NGC~1846, along with the best-fit isochrones listed in
Table~\ref{t:bestisotab1} for comparison purposes. While these He-enhanced
isochrones are a better fit to the RGB, the SGB, and the MSTO region than
those shown in Fig.~\ref{f:He_fig1}, the MSes predicted by the He-enhanced isochrones
are still significantly hotter than those observed for  the clusters (and of the best-fit
isochrones that use primordial He abundance).

We therefore conclude that He enhancement of any subpopulation of
stars in these clusters must be very small. 
Comparing the locations of the MSes of $Y = 0.33$ isochrones
with those of the $Y = 0.25$ isochrones and of the clusters, we estimate 
that $\Delta Y \la 0.02$ in the eMSTO clusters. 
Note that such a small range in He abundance is consistent with the moderate
masses of the eMSTO clusters. Consider for example that He enhancement for a
second-generation population is provided by slow stellar winds of IM-AGB
stars (i.e., AGB stars with $3 \la {\cal{M}}/M_{\odot} \la 8$) which produce
He through the hot-bottom burning process. For that case, \citet{renz08}
estimates that 0.7\% of the initial mass of a stellar population is returned
to the ISM as ``fresh'' He. For a massive eMSTO cluster like NGC~1846 with a 
current mass of $\sim 1.5 \times 10^5$ \Msun\ (cf.\
Table~\ref{t:bestisotab2}), we calculate in Paper III that its mass at an age
of 10 Myr was a factor 2\,--\,3 higher than the current mass, depending on the 
degree of initial mass segregation. For an initial mass of $3 \times 10^5$
\Msun, IM-AGB stars would thus produce 2100 \Msun\ of ``fresh'' He (neglecting 
stochastic effects associated with sampling the mass function). Under the
assumptions that the second generation of stars {\it (i)\/} formed with a star
formation efficiency $\epsilon_{\rm SF} = 0.5$ and {\it (ii)\/} that it currently
represents 65\% of the stars in the cluster (see \S\,\ref{s:MCsims} below),
2100 \Msun\ of He would be equivalent to a helium mass fraction increase of
$0.5 \times 2100/(0.65 \times 1.5 \times 10^5)$ = 0.01 relative to the first
generation. This estimate is consistent with the estimate of the upper limit
on He enhancement mentioned above.

\section{Morphology of the MSTO Regions} \label{s:MSTOmorph}

The eMSTO regions in the clusters in our sample are likely due to the
existence of more than one ``simple'' stellar population. To date, the
presence of eMSTO regions have been interpreted as {\it (i)\/} bimodal
age distributions \citep{mack+08,milo+09}, {\it (ii)\/} age spreads of
200\,--\,500 Myr \citep{goud+09,mack+08,milo+09,gira+09,rube+10,rube+11}, and
{\it (iii)\/} spreads in rotation velocity among turn-off stars \citep{basdem09}. 
The latter study is the only one that does not
invoke a spread in age, or star formation history, as the cause.

\citet{basdem09} claimed that a distribution of rotation velocities
$v_{\rm rot}$ ranging up to $\sim$\,70\% of the critical break-up velocity for
stars of $1.2 \la {\cal{M}}/\mbox{M}_{\odot} \la 1.7$, can mimic the observed
morphologies of eMSTO regions in intermediate-age clusters.
Basically, increasing $v_{\rm rot}$ yields cooler isochrones near the MSTO,
mimicing older ages. However, the study of \citet{gira+11} considered 
effects of rotation and convective core overshoot on isochrones using newly
calculated evolutionary tracks for $v_{\rm rot} = 150$ \kms\ (which is close to
the peak of the observed distribution of rotation rates of such stars), and
found that the effects of both rotation and convective core overshoot cannot
account for the observed widths of the eMSTO region for that
particular value of $v_{\rm rot}$. Furthermore, the $\sim$\,1.4 Gyr old star
clusters NGC~419 and NGC~1751 have been found to host extended red
clumps which require a range of $\sim$\,0.2 \Msun\ in the mass of the
H-exhausted core. Rotation is unlikely to be able to produce this effect
whereas a range in age does so naturally \citep{gira+09,rube+11}.  

While further studies of the effects of stellar rotation on isochrones (e.g.,
using a suitable {\it range\/} of $v_{\rm rot}$ values) should be pursued before
dismissing rotation as a significant cause of eMSTO
regions in intermediate-age star clusters, we focus here on age spreads
which currently seem to be the more likely cause of the eMSTO regions in the
intermediate-age clusters studied here.
We use simulations to investigate the impact that several different parameters
have on the morphology of the MSTO region, e.g.,  formation history, binaries,
etc., and then compare the results with the observations.

\subsection{Monte Carlo Simulations} \label{s:MCsims}

We first simulate cluster CMDs where stars formed in two distinct epochs with
an age difference $\Delta \tau$.
Each simulated cluster CMD is created by populating Dartmouth isochrones with stars
randomly drawn from a Salpeter IMF between 0.1 M$_{\odot}$ and the RGB-tip
mass.
A pair of ages is chosen for each synthetic CMD from an age grid that
encompasses the age interval around the best-fit age implied by the range
given in column (3) of Table~\ref{t:bestisotab2}. The grid is populated using
an age increment of 50 Myr, which is similar to the age of the youngest (most
massive) types of stars that have been put forward as plausible donors of
material from which a new stellar generation can be formed (i.e., the FRMS and
massive binary stars). 
For each pair of ages, we vary the the relative (mass) fraction
in the younger population from 0.80 to 0.20, in 0.05 increments.
The total number of simulated stars is normalized to the observed number of
stars brighter than the 50\% completeness limit. 
We add an unresolved binary companion to a fraction (see below) of the stars,
drawn from the same mass function (i.e., using a flat primary-to-secondary
mass ratio distribution).
Finally, we add photometric errors to the artificial stars,
modeled after the actual distribution of photometric
uncertainties.

We use the width of the upper main sequence, i.e. the part brighter than 
the turn-off of the field stellar population and fainter than the MSTO region
of the clusters, to determine the binary star fraction in our sample clusters. 
We estimate the internal systematic
uncertainty in binary fraction to $\pm 5\%$ and defer a more detailed
discussion of binary parameter degeneracies, in particular binarity vs.\ mass
fractions of stellar generations,
to a future paper. For the purposes of this work the results don't change
significantly within $\sim\!10\%$ of the binary fraction. 

In order to compare the observed and simulated MSTO regions, we use
a ``pseudo-age'' distribution.
The pseudo-age distribution is determined by constructing a 
parallelogram in the CMD with: 
{\it (i)\/} One axis approximately parallel to the isochrones, 
{\it (ii)\/} the other axis approximately perpendicular to
the isochrones, and 
{\it (iii)\/} located in a region of the MSTO
where the split between the isochrones is relatively evident. 
The (\BI, \B) coordinates of the stars in the CMD
are then transformed into the reference
coordinate frame defined by the two axes of the parallelogram, 
and then considering the distributions of the coordinates of the
stars in the direction perpendicular to the isochrones. To translate the
latter coordinate to age, the same procedure is done for the Dartmouth
isochrones for an age range that covers the observed extent of the MSTO
region of the cluster in question, using an age increment of 0.05 Gyr. The
relationship between age and the coordinate in the direction
perpendicular to the isochrones is then determined using a polynomial 
least-squares fit. Since binary stars influence the distribution of stars
in the MSTO region to some (albeit small) extent, we call the resulting
age parameter ``pseudo-age''.  

\begin{figure*}[tb]
\centerline{
\psfig{figure=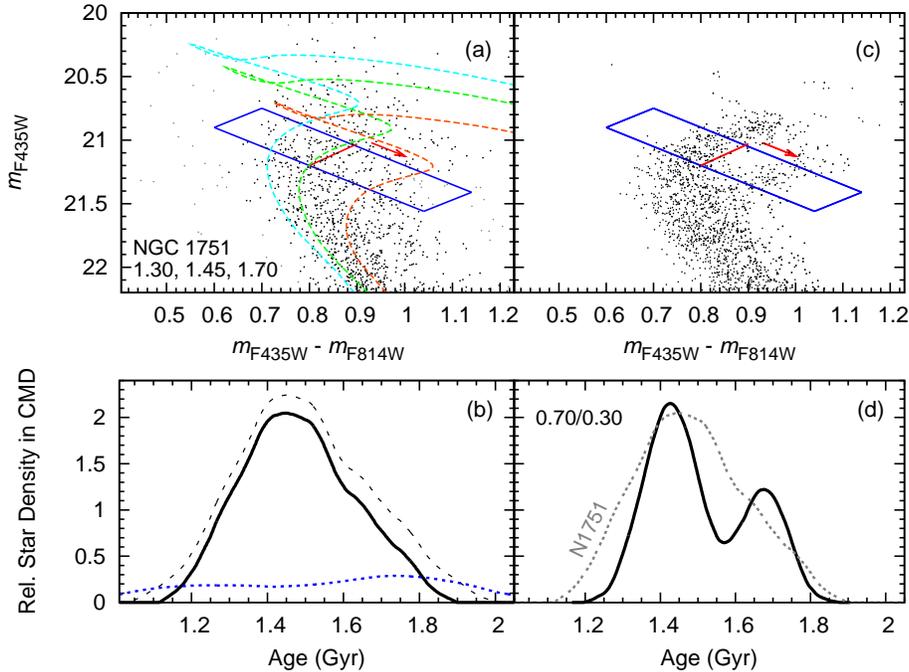,width=12.cm,angle=-90}
}
\caption{{\it Panel (a)}: Enlargement of the CMD of NGC~1751 shown in the
  middle panel of Fig.\ \ref{f:fullCMDs_1}, focusing on the MSTO region.
  The blue, green, and orange curves represent Dartmouth
  isochrones that fit the upper end, middle, and lower end of the
  broad MSTO region, respectively. Their ages (in Gyr) are mentioned
  at the lower left corner of the panel.   
  {\it Panel (b)}: ``Pseudo-age'' distributions of the cluster, using stars
  in the parallelograms shown in panels (a) and derived using a non-parametric
  density estimator (see text in \S\,\ref{s:MSTOmorph} for details). The black
  dashed line represents all stars, the blue dashed line represents stars in
  the background area scaled to the area of the ACS image used for the CMD in
  panel (a), and the solid line represents ``all stars minus background''.  
  The red arrow in panel (a) indicates the positive direction of the
  X axis of panel (b).
  {\it Panels (c) and (d)}: Same as panels (a) and (b), respectively, but now
  for simulations of two SSPs including binary stars whose properties are
  listed in Table~\ref{t:KStests} and described in \S\,\ref{s:MSTOmorph}. The
  legend on the top left indicates the mass fractions of the younger vs.\ the
  older SSP used in the simulation. The grey short-dashed line in panel (d) is
  a copy of the solid line in panel (b) to allow a direct comparison. 
\label{f:acrossMSTOplot1}}
\end{figure*}

The procedure mentioned above is illustrated in
Figures~\ref{f:acrossMSTOplot1}\,--\,\ref{f:acrossMSTOplot4}. The top panels
show the simulated and observed CMDs and the parallelogram mentioned above (in
blue, with the reference axis along the isochrones in red).
The bottom panels show the corresponding pseudo-age distributions.
These were calculated using the non-parametric Epanechnikov-kernel probability density function
\citep{silv86} for all objects in the parallelograms, in order to avoid
potential biases that can arise if fixed bin widths are used. In the case of
the observed CMD (i.e., panels a and b), this was done both for stars
within a King core radius from the cluster center and for the ``background
region'' for which the CMD was shown in the right panels of
Fig.~\ref{f:fullCMDs_1}. The panels (d) also list the 
fraction of stars in the two distinct SSPs (shown in black dots on panels c).
The intrinsic probability density function of the pseudo-age distribution of
the cluster was then derived by statistical subtraction of the background
region (see panels b). The best-fit pair of two-SSP simulations was selected
by conducting two-sample Kolmogorov-Smirnov (K-S) tests of the probability
density functions of the simulations against those of the cluster data and
picking the simulation with the highest $p$ value. The properties of the
selected simulations, including the best-fit binary fractions and the $p$
values for each star cluster are listed in Table~\ref{t:KStests}.  

\begin{figure*}[tbp]
\centerline{
\includegraphics[angle=-90,width=12.cm]{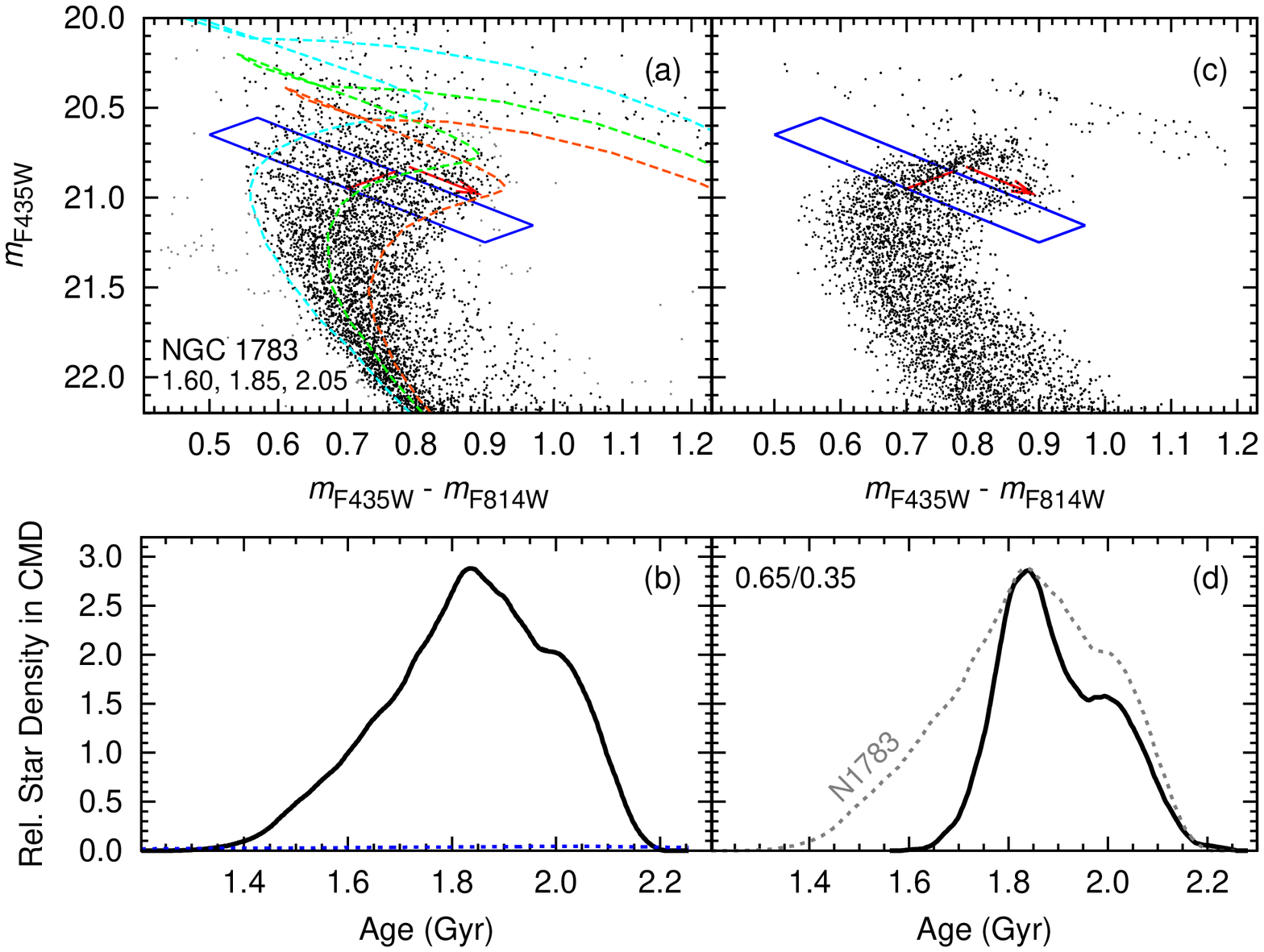}
}
\vspace*{2mm}
\centerline{
\includegraphics[angle=-90,width=12.cm]{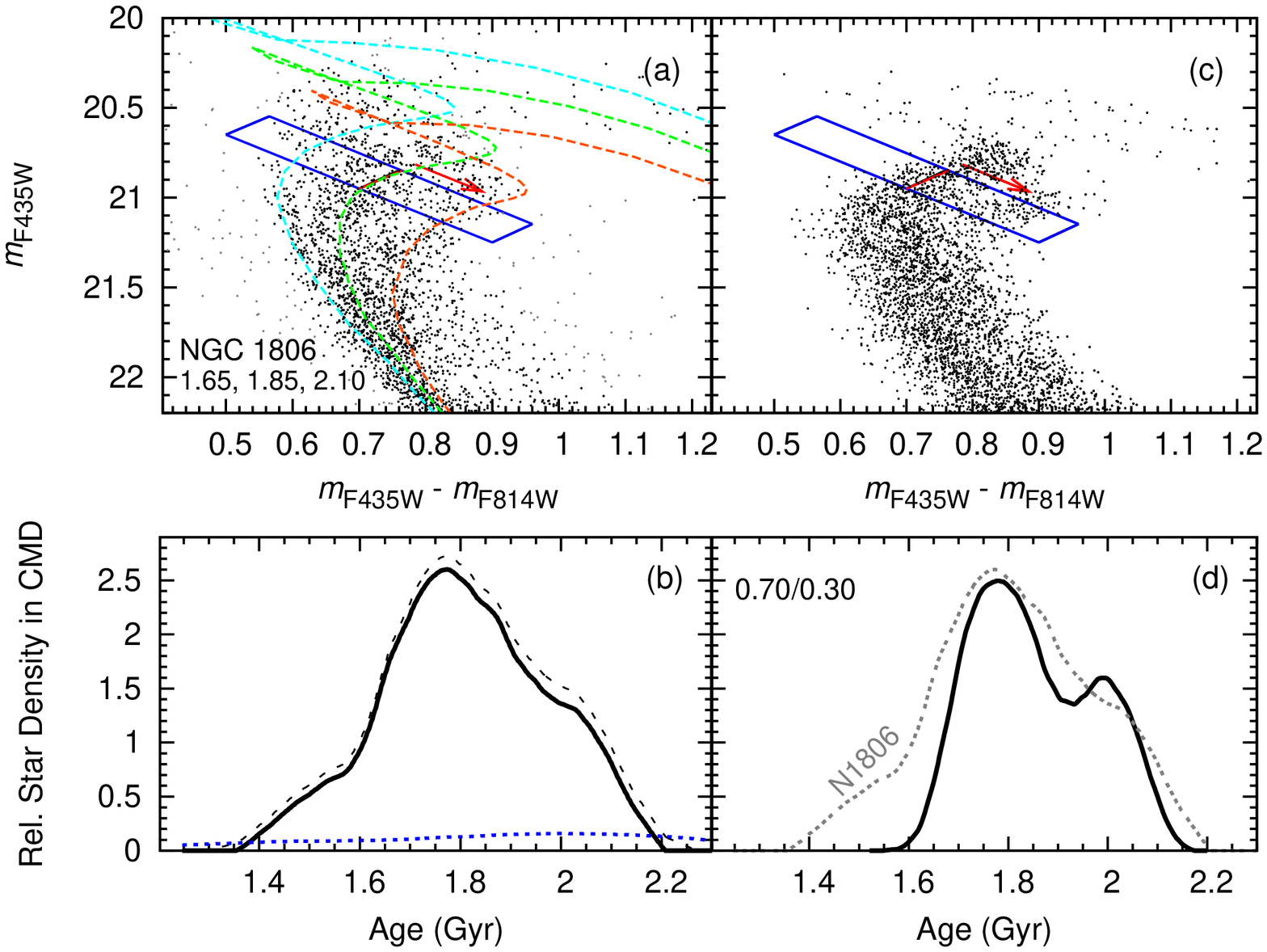}
}
\caption{Same as Fig.~\ref{f:acrossMSTOplot1}, but now for NGC~1783
  and NGC~1806. 
\label{f:acrossMSTOplot2}}
\end{figure*}

\begin{figure*}[tbp]
\centerline{
\psfig{figure=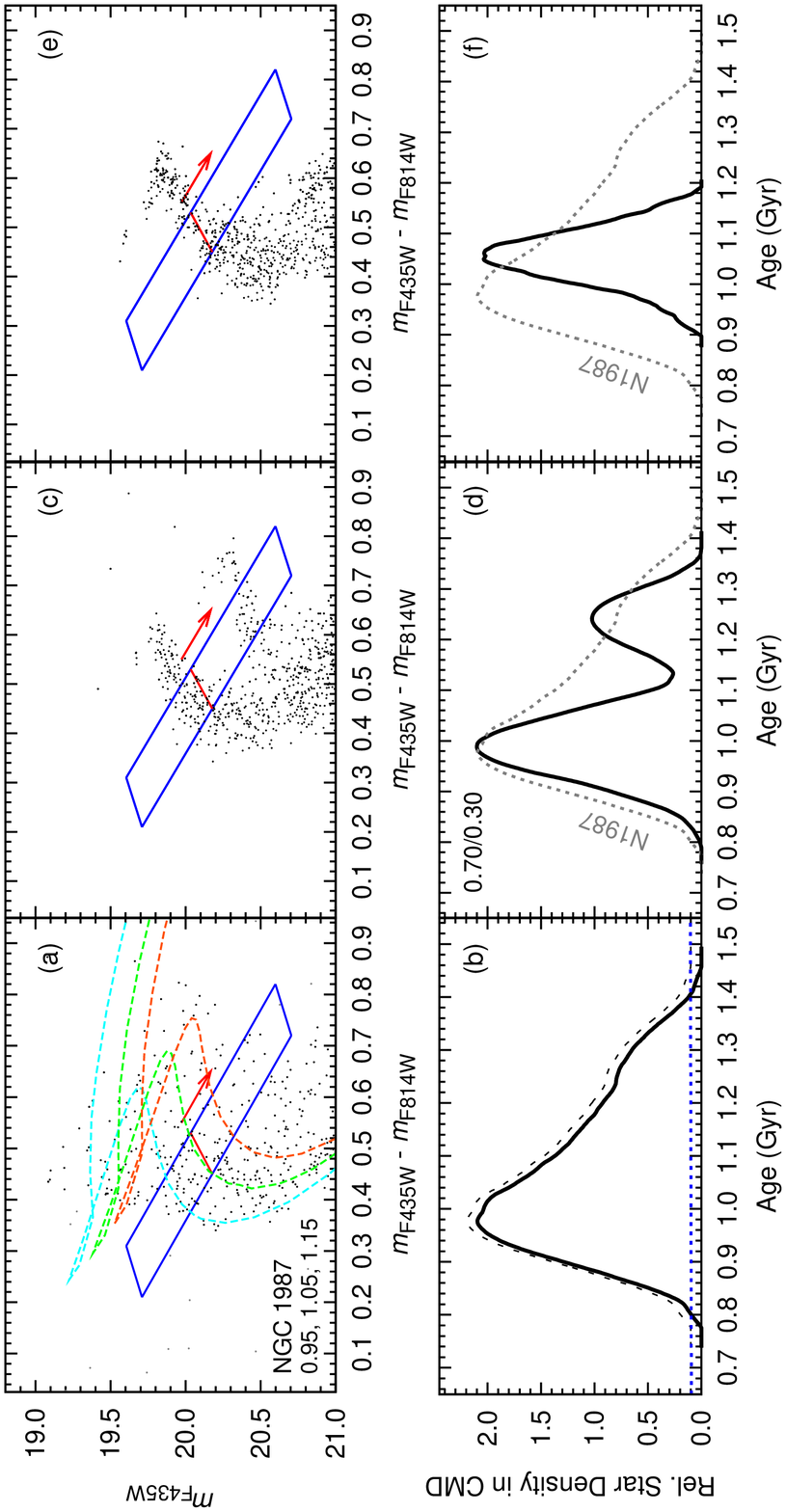,width=16.5cm,angle=-90}
}
\vspace*{2mm}
\centerline{
\psfig{figure=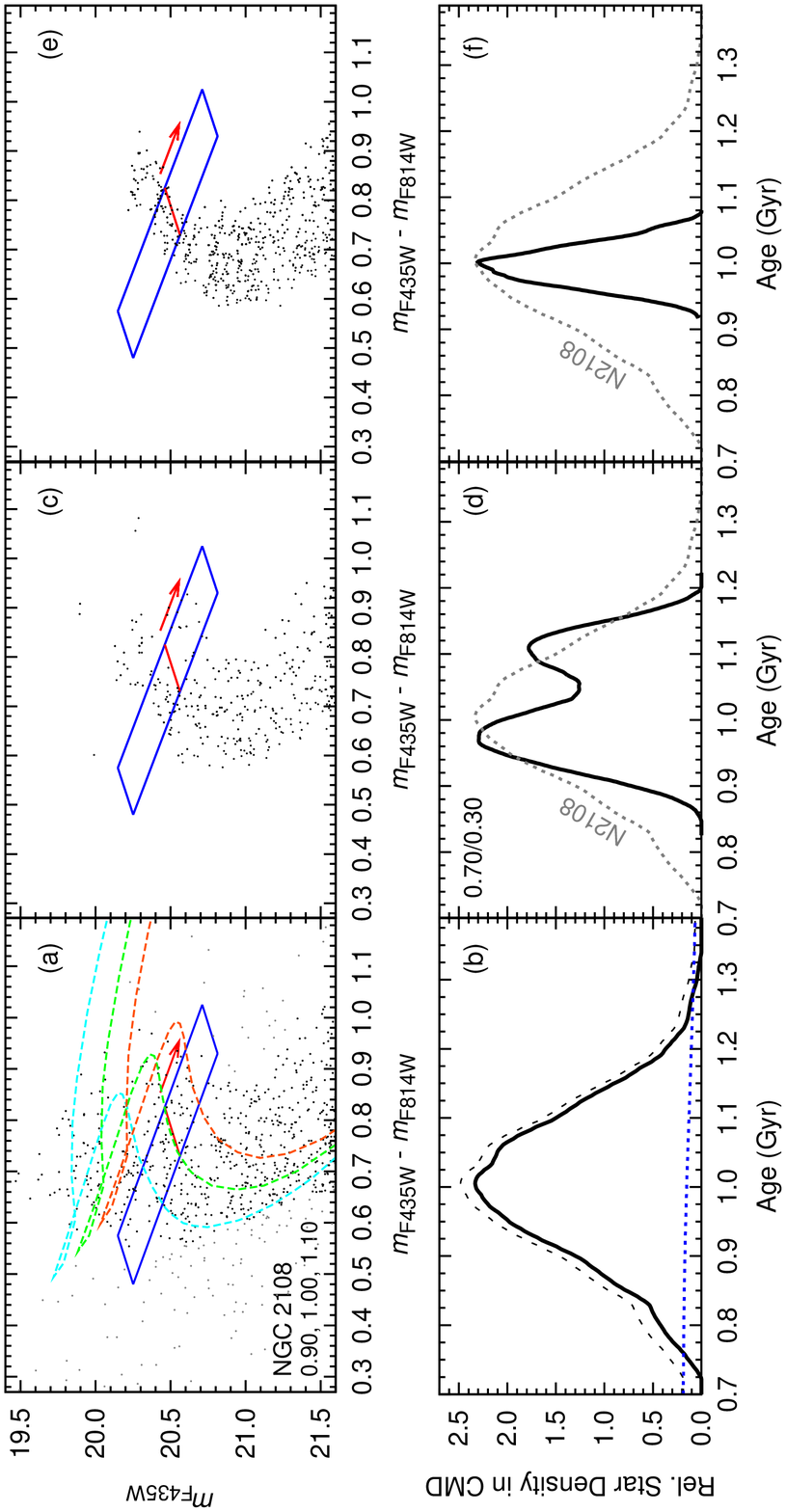,width=16.5cm,angle=-90}
}
\caption{Similar to Figure~\ref{f:acrossMSTOplot1}, but now for NGC 1987 and
  NGC 2108, two of the three lowest-mass clusters 
  in our sample. The extra panels (e) and (f) are the same as panels (c) and
  (d), respectively, but now for simulations of a {\it single\/} SSP including 
  binary stars as described in \S\,\ref{s:MSTOmorph}. 
\label{f:acrossMSTOplot3}}
\end{figure*}

\begin{figure*}[tb]
\centerline{
\psfig{figure=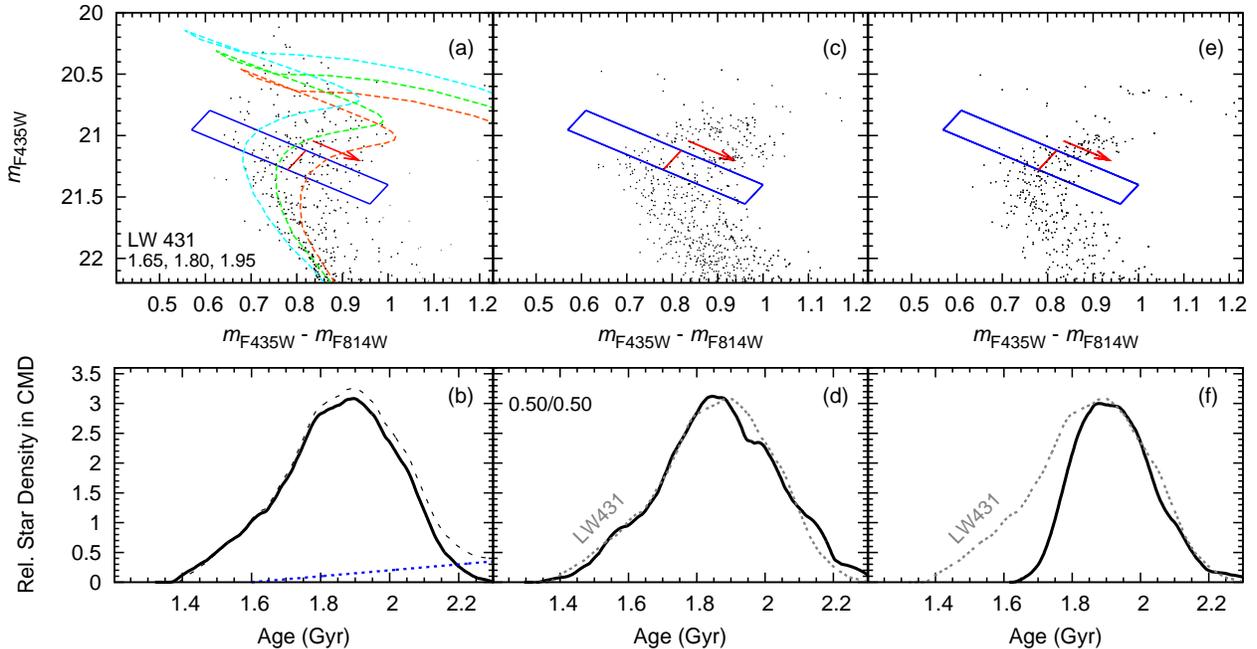,width=16.5cm,angle=-90}
}
\caption{Same as Figure~\ref{f:acrossMSTOplot3}, but now for LW\,431. 
\label{f:acrossMSTOplot4}}
\end{figure*}

\begin{table}[tbh]
\begin{center}
\footnotesize
%\scriptsize
\caption{Properties of best-fit two-SSP simulations of the star clusters in
  our sample.}
\label{t:KStests}
% name, age1, age2, frac_age1, D, $p$
\begin{tabular}{@{}lcccccc@{}}
\multicolumn{3}{c}{~} \\ [-2.5ex]   
 \tableline \tableline
\multicolumn{3}{c}{~} \\ [-2.ex]                                                
\multicolumn{1}{c}{Cluster} & Age$_1$& Age$_2$ & $f_{\rm Y}$ & $f_{\rm binary}$ & 
 $p_{\rm KS}$ \\
\multicolumn{1}{c}{(1)}     & (2)     & (3)     & (4) &   (5)            & 
 (6) \\ [0.5ex] \tableline  
\multicolumn{3}{c}{~} \\ [-2.ex]              
NGC 1751 & 1.40 & 1.70 & 0.65 & 0.30 & 0.20 \\
NGC 1783 & 1.80 & 2.00 & 0.65 & 0.25 & 0.00 \\
NGC 1806 & 1.75 & 2.00 & 0.70 & 0.25 & 0.14 \\
NGC 1846\rlap{$^{\rm a}$} & 1.75 & 2.05 & 0.65 & 0.15 & 0.11 \\
NGC 1987 & 1.00 & 1.25 & 0.70 & 0.35 & 0.48 \\
         & 1.05 & 1.05 & 1.00 & 0.35 & 0.02 \\
NGC 2108 & 0.95 & 1.10 & 0.70 & 0.45 & 0.42 \\
         & 1.00 & 1.00 & 1.00 & 0.45 & 0.03 \\
  LW 431 & 1.80 & 2.05 & 0.50 & 0.20 & 0.90 \\
         & 1.90 & 1.90 & 1.00 & 0.20 & 0.15 \\ [0.5ex] \tableline
\multicolumn{3}{c}{~} \\ [-2.5ex]              
\end{tabular}
\tablecomments{Column (1): Name of star cluster. (2): Age of youngest
  SSP in Gyr. (3): Age of oldest SSP in Gyr. (4): Best-fit mass fraction of
  youngest SSP. (5): Best-fit binary fraction. (6): $p$ value of two-sided K-S 
  test of the simulated data in the parallelograms shown in
  Figures~\ref{f:acrossMSTOplot1}\,--\,\ref{f:acrossMSTOplot4} against the
  observed data. See \S\,\ref{s:MSTOmorph} for details.}
\tablenotetext{a}{Values taken from Paper I.}
\end{center}
\end{table}

  During the refereeing process of this manuscript, a paper by
  \citet{yang+11} appeared, reporting on simulations of the impact of
  {\it interactive\/} binaries to eMSTO morphologies of intermediate-age star
  clusters. Their calculations, which assume that all cluster stars are
  members of binary systems, show that the presence of interactive binaries
  would cause {\it (i)\/} a slight extension of the MSTO region towards the
  blue and {\it (ii)\/} a ``fan'' of low stellar density in the CMD towards
  brighter magnitudes and bluer colors than the MSTO. The distribution of this
  ``fan'' on the CMD shows similarities to the presence of younger stars,
  although most ``fan'' stars are fainter than the MSTO region of younger
  isochrones. We believe that the actual impact of such interactive binaries
  to the eMSTO morphologies of the star clusters in our sample is
  insignificant for two main reasons. First, the extension of the MSTO region
  mentioned in point {\it (i)\/} above involves color changes that are
  significantly smaller than the observed color ranges encompassed by the
  eMSTOs shown in the current paper. Second, a comparison between the middle
  and right-hand panels in our Figures~\ref{f:fullCMDs_1} and \ref{f:fullCMDs_2}
  reveals that the stars in our CMDs that could be interactive binary stars
  in the ``fan'' mentioned in point {\it (ii)\/} above are most likely LMC
  field stars rather than cluster stars, since they are at least as
  abundant in the ``field'' CMD as in the ``cluster'' CMD.

\subsection{Ability to Resolve Populations of Different Ages} \label{s:ageres}

\begin{figure}[tb]
\centerline{
\psfig{figure=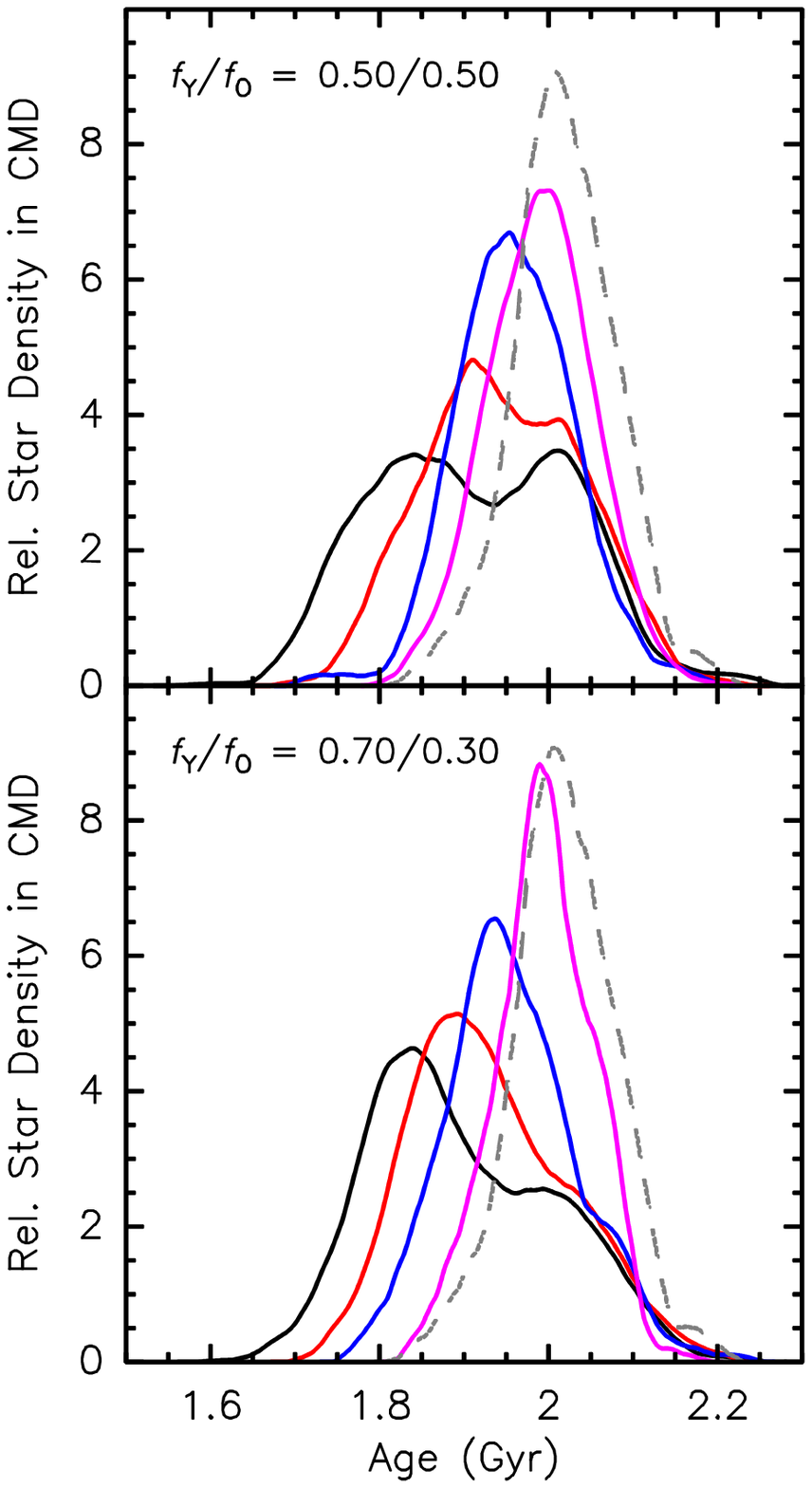,width=7.cm}
}
\caption{Illustration of age resolution of our ``pseudo-age''
  distributions. {\it Top panel}: Pseudo-age distributions of simulations of two 
  SSPs including binary stars as described in \S\,\ref{s:MSTOmorph}.2, using stars
  in the parallelograms shown in panels (a) and (c) of 
  Figure~\ref{f:acrossMSTOplot2} for the case of NGC~1783. The two SSPs have
  equal mass fractions in this panel. The age of the older population is 2.0
  Gyr, and $\Delta$\,age = 0 Myr (dashed line), 50 Myr (magenta line), 100
  Myr (blue line), 150 Myr (red line), and 200 Myr (black line). {\it Bottom
    panel}: Same as top panel, but now for the case where 70\% of the stars
  are in the younger generation. Note that age differences between successive
  stellar generations of 100\,--\,150 Myr or more are discernable using this
  method at an age of 2 Gyr.   
\label{f:ageres}}
\end{figure}

How well does the method described above allow us to
resolve two (or more) discrete star formation events separated in time by
$\Delta \tau$?
To address this question we use our simulations for NGC 1783, the cluster with
the highest present-day mass in our sample. We consider the simulations that
involved an age of the first generation of 2.00 Gyr with $\Delta \tau$ values
of 0 (i.e., a single SSP), 50, 100, 150 and 200 Myr, and two different mass 
fractions, namely $f_{\rm Y}$ = 0.70 and $f_{\rm Y}$
= 0.50 for the younger generation. 
These mass fractions were chosen to bracket the values found for the
clusters in our sample (cf.\ Table~\ref{t:KStests}). Results of these
simulations are shown in Figure~\ref{f:ageres}. For $f_{\rm Y}$ = 0.50, the
bimodality of the pseudo-age distribution shows up clearly for $\Delta
\tau \ga 150$ Myr, while $\Delta \tau \ga 100$ Myr yield a
significantly broader distribution than a single SSP. For $f_{\rm Y}$ = 0.70,
the ``hump'' on the right side of the pseudo-age distribution (due to the
older generation) is already readily recognizable at $\Delta \tau \ga 100$
Myr. Hence our method can recognize age bimodality in CMDs
if $\Delta \tau \ga 100-150$ Myr at an age of 2.0 Gyr, i.e., $0.05 \la (\Delta
\tau / \tau) \la 0.08$, with the exact value depending on the mass fractions
of the different generations\footnote{Obviously, the age resolution
depends on the quality of the data, and needs to be
evaluated individually for each dataset.}.  

\subsection{Comparison With Observations}

As Figures~\ref{f:acrossMSTOplot1}\,--\,\ref{f:acrossMSTOplot4} show, the
distribution of observed stars in the parallelogram typically peaks near the
``young'' end of the age range and then declines more or less 
uniformly towards older as well as younger ages. A comparison of panels (b) with (d)
for a given star cluster shows that the distribution of the observed stars is typically more
continuous than those of the best-fit two-SSP simulations. 
Two-sample K-S tests confirm this visual impression: The $p$ values of the K-S
tests to compare the distributions of the simulated CMDs of two SSPs with the
best-fit mass fractions of the younger population against the data 
do not exceed 20\% for the clusters (see Table~\ref{t:KStests}). 
NGC 1987, NGC 2108, and LW 431, the three lowest-mass clusters in our 
sample, do not follow this rule (see
Fig.~\ref{f:acrossMSTOplot3}\,--\,\ref{f:acrossMSTOplot4}); these  
three cases are briefly discussed below. 
Therefore, a main conclusion of this paper is that the more 
massive clusters in our sample are better explained by a population with a
{\it distribution of ages\/} rather than by two discrete SSPs.
  This conclusion differs from that of \citet{mack+08} whose analysis
  favored a bimodal distribution of ages for the clusters NGC~1806 and
  NGC~1846 and that of \citet{milo+09}, whose analysis did so for the
  clusters NGC~1751, NGC~1783, NGC~1806 and NGC~1846.
The smooth and extended nature of the pseudo-age distributions seems to
suggest that star formation did not actually occur in discrete events. This is
further discussed in \S\,\ref{s:disc} below. 

As to the low-mass clusters NGC 1987, NGC 2108 and LW 431, panels (c) and (d)
in Figures~\ref{f:acrossMSTOplot3} and  \ref{f:acrossMSTOplot4} compare
pseudo-age distributions for the observations and the simulated bimodal
formation histories. 
The K-S test
comparing these distributions indicate that a bimodal age distribution result
in formally acceptable fits (cf.\ Table~\ref{t:KStests}).
However, these results do not preclude a continuous distribution of ages
within these clusters.
We also compared the pseudo-age distributions for these three
clusters with those resulting from a single-age population,
as shown in panels (e) and (f) of
Figures~\ref{f:acrossMSTOplot3} and \ref{f:acrossMSTOplot4}.
The observed distributions are significantly broader than a single-age
population, supporting our general conclusion that more than one ``simple''
population is required to explain the morphology of the MSTO regions of these
low-mass clusters as well.

\section{Implications Regarding The Nature of Multiple Populations in Star
  Clusters} \label{s:disc} 

The results of the experiments described above have important implications
regarding the origin of multiple populations in star clusters in general,
including the situation seen in many (ancient) globular clusters in our
Galaxy. 
The following discussion is relevant for star clusters 
with initial masses that were too low to retain gas expelled by energetic
supernova (SN) explosions and/or to capture significant numbers of field stars from
their host (dwarf) galaxies. In practice this restriction corresponds roughly to 
masses $\la 2 \times 10^6$ \Msun\ \citep[cf.][]{basgoo06,fell+06} at an age of
13 Gyr. The clusters in our sample are indeed less massive than that upper limit. 

A main result of this work is that the clusters in our sample,
particularly the four most massive ones, appear to have formed stars more or
less continously for $200-500$~Myr, and are inconsistent with a bimodal age
distribution. This result rules out scenarios
where two star clusters have merged together or where
a star cluster has merged with a giant molecular cloud \citep{bekmac09},
since these would lead to strongly bimodal age distributions.

We believe our results 
constitute support for the ``in situ'' scenario \citep[e.g.,][]{derc+08,renz08} in
which star clusters with masses high enough to retain ejecta in {\it slow\/}
winds of stars (as opposed to SN ejecta) of the first generation gather this
material in their central regions where secondary star formation can occur. 
Dynamical evidence to further support this scenario for the clusters in our
sample is presented in Paper III. In the context of this scenario, the
hitherto suggested source(s) of the ejecta are FRMS \citep{decr+07}, massive
binary stars \citep{demink+09}, and IM-AGB stars
\citep[e.g.,][]{danven07}. Note that the ejecta from FRMS and massive binary
stars are produced on time scales that are 
significantly shorter than those from IM-AGB stars ($10 - 30$~Myr versus $50 -
300$~Myr, respectively; see e.g.\ \citealt{decr+07,grat+04,vendan08}). Since
our simulations in \S\,\ref{s:MSTOmorph} show that two populations with ages
separated by 100\,--\,150 Myr or more would result in observable bimodality in
our MSTO photometry, the combination of the observed age ranges of
200\,--\,500 Myr and the absence of clear bimodality in the pseudo-age
distributions seems to suggest at face value that FRMS and/or massive binary
stars could well be significant contributors to the enriched material used for
the secondary stellar population(s). However, IM-AGB stars also seem likely
significant contributors, since the (pseudo-)age distributions of the star
clusters 
in our sample do typically peak at the younger end of the age range
(especially for the more massive clusters in our sample). 

Finally, the slow winds from all suggested stellar types would lead to chemical
enrichment of light elements in the second generation of stars due to
products from the CNO and ON cycles. 
In the context of the ``in situ'' scenario, one would expect to see significant
and correlated variations in light element abundances (e.g., N, O, Na) among the
stars in the star clusters with relatively high masses in our sample, likely
in a way similar to the Na-O anticorrelation found in  Galactic GCs \citep[see
also][]{conr11,kell+11}. If instead the main cause of the eMSTO regions is not an
age spread, one would {\it not\/} expect to see any variations in
light-element abundances, even in the most massive eMSTO clusters.  
Measuring the chemical composition of RGB stars in LMC clusters
with eMSTO regions is feasible (albeit challenging) with current spectrographs on
8-10m-class telescopes \citep[e.g.,][]{mucc+08} and should provide additional
relevant evidence to help decipher the most likely scenario responsible for
populating the eMSTO regions. 

\section{Summary and Conclusions} \label{s:conc}

We have used deep {\it BVI} photometry from {\it HST/ACS} images to construct
CMDs of 7 intermediate-age star clusters in the LMC.  We have used the
{\it ePSF\/} fitting technique developed by J.\ Anderson, 
which returns high-accuracy photometry of cluster stars extending some 5
magnitudes below the main sequence turnoff for all our target clusters. 
We fit isochrones from the Padova, Teramo, and Dartmouth groups in order
to determine the best-fit stellar population parameters for the clusters in our
sample. All three sets give a reasonably good fit to the CMDs, although there
are significant differences between the observations and predictions in the shape 
of the RGB. The overall best fit to the entire CMD is typically achieved using
the Dartmouth isochrones, but this distinction can formally only be made for
star clusters massive enough to have a well-populated RGB. 

We use the results from the isochrone fitting to quantify typical systematic
errors of fitted population parameters for intermediate-age star
clusters in the LMC introduced by using any one family of
isochrones, and the assumption of solar \afe\ ratio. These systematic errors are
typically of order 15\% for any given stellar population parameter. 

The CMDs for the clusters in our sample show a number of interesting features:
{\it (i)\/} a very narrow RGB; % red giant branch; 
{\it (ii)\/} a MSTO %main sequence turnoff (MSTO) 
region that is clearly more extended than the (fainter) single-star main
sequence;  {\it (iii)\/} an obvious sequence of unresolved binary stars,
somewhat brighter than the single star main sequence.
We have tested the role played by binary populations on the morphology of the
MSTO region in the CMD via Monte-Carlo simulations of multiple stellar generations.
Our multi-SSP models include 
a realistic treatment of photometric uncertainties and incompleteness, and
employ a flat distribution of primary-to-secondary stellar mass ratios for
binary stars.  A quantitative comparison of the distribution of the stars in
the MSTO region with those in simulations that incorporate two SSPs with 
age differences consistent with the results of the isochrone fitting shows, for
{\it all\/} clusters in our sample, that {\it (i)\/} their MSTO 
region is significantly more extended than can be
explained by a single SSP; {\it (ii)\/} the MSTO regions 
are statistically better described by a population with a {\it distribution of
  ages\/} rather than by two discrete SSPs as suggested by several others. 
We cannot however, formally rule out a bimodal age distribution
for the three lowest mass clusters in our sample.

We also conclude that any He
enhancement of a second generation of stars must be very small 
($\Delta Y \la 0.02$) for the clusters in our sample, based on a 
comparison of the cluster CMDs with isochrones with enhanced helium
mass fractions as well as dynamical arguments.

Viable sources of the material used to form secondary generations of stars in
star clusters include fast-rotating massive stars, massive
binary stars, and intermediate-mass AGB stars. Further studies of element
abundance ratios from high-resolution spectroscopy of individual cluster stars
should be very useful in further constraining the nature of the eMSTO 
regions in massive intermediate-age star clusters in the LMC. In particular,
if these clusters indeed host a range of stellar ages, one would expect to see 
correlated variations in light element abundances (e.g., N, O, Na) among the
stars in the more massive star clusters in our sample, likely in a way similar
to the Na-O anticorrelation found in Galactic GCs of similar (initial)
mass. If instead the main cause of the eMSTO regions is not an age spread, one
would {\it not\/} expect to see any variations in light-element abundances,
even in the most massive eMSTO clusters.

\paragraph*{Acknowledgments.}~We are grateful to Jay Anderson for his
support and help in using his {\it ePSF}-related programs, and we acknowledge
stimulating discussions with Aaron Dotter, Leo Girardi, and Selma de Mink.
  We gratefully acknowledge the useful comments and suggestions of the
  anonymous referee. 
T. H. P.\ acknowledges support by the FONDAP Center for Astrophysics 15010003
and BASAL Center for Astrophysics and Associated Technologies PFB-06,
Conicyt, Chile. He also gratefully acknowledges past support from the
National Research Council of  Canada in the form of a Plaskett
Research Fellowship, during which part of this research was
conducted. 
R. C.\ acknowledges support from the National
Science Foundation through CAREER award 0847467.
This research was supported in part by the National Science Foundation under 
Grant No.\ PHY05-51164.
Support for {\it HST\/} Program GO-10595 was provided by
NASA through a grant from the Space Telescope Science Institute, 
which is operated by the Association of Universities for Research in
Astronomy, Inc., under NASA contract NAS5--26555. We acknowledge the use of
the $R$ Language for Statistical Computing, see http://www.R-project.org.

\setcounter{table}{1}

\begin{turnpage}
\begin{deluxetable*}{@{}lccccccccc@{}}
%\rotate
\tablewidth{0pt}
%\tabletypesize{\scriptsize}
\tablecolumns{10}
\tablecaption{Best-fit population parameters of the star clusters studied in this
  paper as derived from different isochrone families. 
\label{t:bestisotab1}}
% name, (age, age range, [Fe/H], m-M, A_V)*3, Age, Age range, [Fe/H], (m-M), A_V
\tablehead{
\colhead{} & \multicolumn{3}{|c|}{Padova fits\tablenotemark{a}} & 
 \multicolumn{3}{c|}{Teramo fits\tablenotemark{a}} & 
 \multicolumn{3}{c}{Dartmouth fits\tablenotemark{b}} \\ 
%\cline{2-4}  \cline{5-7} \cline{8-10} \\ 
 \colhead{Cluster} & \colhead{Age} & %\colhead{[Fe/H]} & 
  \colhead{$(m-M)_0$} & \colhead{$A_V$} & \colhead{Age} & %\colhead{[Fe/H]} & 
  \colhead{$(m-M)_0$} & \colhead{$A_V$} & \colhead{Age} & %\colhead{[Fe/H]} & 
  \colhead{$(m-M)_0$} & \colhead{$A_V$} \\
}
%\multicolumn{1}{c}{(1)}     & \multicolumn{1}{c}{(2)}        & (3) & (4) & 
% (5) \\ [0.5ex] \tableline
\startdata                                  
NGC 1751 & 
 $1.40 \pm 0.05$ & %$-0.40 \pm 0.05$ &
  $18.50 \pm 0.02$ & $0.38 \pm 0.02$ & 
 $1.30 \pm 0.05$ & %$-0.40 \pm 0.05$ & 
  $18.55 \pm 0.02$ & $0.40 \pm 0.02$ & 
 $1.50 \pm 0.02$ & %$-0.50 \pm 0.05$\tablenotemark{d} & 
  $18.48 \pm 0.02$ & $0.40 \pm 0.02$ \\
NGC 1783 & 
 $1.70 \pm 0.05$ & %$-0.40 \pm 0.05$ & 
  $18.49 \pm 0.03$ & $0.00 \pm 0.02$ & 
 $1.60 \pm 0.05$ & %$-0.40 \pm 0.05$ & 
  $18.50 \pm 0.02$ & $0.02 \pm 0.02$ & 
 $1.80 \pm 0.02$ & %$-0.50 \pm 0.05$\tablenotemark{d} & 
  $18.40 \pm 0.02$ & $0.04 \pm 0.02$ \\
NGC 1806 & 
 $1.60 \pm 0.05$ & %$-0.40 \pm 0.05$ & 
  $18.50 \pm 0.03$ & $0.05 \pm 0.03$ & 
 $1.60 \pm 0.05$ & %$-0.40 \pm 0.05$ & 
  $18.45 \pm 0.02$ & $0.05 \pm 0.02$ & 
 $1.80 \pm 0.02$ & %$-0.50 \pm 0.05$\tablenotemark{d} & 
  $18.40 \pm 0.02$ & $0.07 \pm 0.02$ \\
NGC 1846 & 
 $1.70 \pm 0.05$ & %$-0.40 \pm 0.05$ &
  $18.42 \pm 0.03$ & $0.07 \pm 0.02$ & 
 $1.60 \pm 0.05$ & %$-0.40 \pm 0.05$ & 
  $18.50 \pm 0.02$ & $0.08 \pm 0.01$ & 
 $1.90 \pm 0.03$ & %$-0.50 \pm 0.05$\tablenotemark{d} & 
  $18.41 \pm 0.02$ & $0.09 \pm 0.02$ \\
NGC 1987 & 
 $1.10 \pm 0.05$ & %$-0.40 \pm 0.05$ & 
  $18.37 \pm 0.03$ & $0.12 \pm 0.02$ & 
 $1.10 \pm 0.05$ & %$-0.40 \pm 0.05$ & 
  $18.38 \pm 0.02$ & $0.11 \pm 0.02$ & 
 $1.00 \pm 0.05$ & %$-0.50 \pm 0.05$\tablenotemark{d} & 
  $18.38 \pm 0.02$ & $0.23 \pm 0.03$ \\
NGC 2108 & 
 $1.00 \pm 0.05$ & %$-0.40 \pm 0.05$ & 
  $18.45 \pm 0.02$ & $0.48 \pm 0.02$ & 
 $1.00 \pm 0.05$ & %$-0.40 \pm 0.05$ & 
  $18.48 \pm 0.02$ & $0.48 \pm 0.01$ & 
 $1.00 \pm 0.05$ & %$-0.50 \pm 0.05$\tablenotemark{d} & 
  $18.40 \pm 0.02$ & $0.55 \pm 0.02$ \\
  LW 431 & 
 $1.70 \pm 0.05$ & %$-0.40 \pm 0.05$ & 
  $18.45 \pm 0.03$ & $0.14 \pm 0.02$ & 
 $1.60 \pm 0.05$ & %$-0.40 \pm 0.05$ & 
  $18.46 \pm 0.02$ & $0.15 \pm 0.02$ & 
 $1.90 \pm 0.05$ & %$-0.50 \pm 0.05$\tablenotemark{d} & 
  $18.38 \pm 0.02$ & $0.16 \pm 0.02$ 
%\multicolumn{3}{c}{~} \\ [-2.5ex]              
%\end{tabular}
\enddata
\tablecomments{Column (1): Name of star cluster. (2): Adopted age in Gyr. (3):
  Age range in Gyr associated with the width of the observed MSTO region. (4):
  Adopted [Fe/H] in dex. (5): Adopted distance modulus in mag. (6): Adopted
  foreground $V$-band reddening in mag.}  
\tablenotetext{a}{For the Padova and Teramo isochrones, the best-fit [Fe/H] was $-0.40
  \pm 0.05$ for all clusters.}
\tablenotetext{b}{For the Dartmouth isochrones, the best-fit [Fe/H] and
  [$\alpha$/Fe] were $-0.50 \pm 0.05$ and 
  $+0.2 \pm 0.1$, respectively, for all clusters.}
\end{deluxetable*}
\end{turnpage}

%%%% ENDING DOCUMENT HERE %%%%


\begin{thebibliography}{}
\bibitem[Anderson \& King(2000)]{andkin00}
Anderson, J., \& King, I. R. 2000, \pasp,  112,  1360
\bibitem[Anderson \& King(2006)]{andkin06}
Anderson, J., \& King, I. R. 2006, ``PSFs, Photometry, and Astrometry for the
 ACS/WFC'', ACS Instrument Science Report 2006-01 (Baltimore: STScI)
\bibitem[Anderson et al.(2008a)]{ande+08a}
Anderson, J., Sarajedini, A., Bedin, L. R., King, I. R., Piotto, G., Reid, 
 I. N., Siegel, M., Majewski, S. R., et al.\ 2008, \aj, 135, 2055
\bibitem[Anderson et al.(2008b)]{ande+08b}
Anderson, J., King, I. R., Richer, H. B., Fahlman, G. G., Hansen, B. M. S.,
 Hurley, J., Kalirai, J. S., Rich, R. M., et al.\ 2008, \aj, 135, 2114
\bibitem[Anderson et al.(2009)]{ande+09}
Anderson, J., Piotto, G., King, I. R., Bedin, L. R., \& Guthathakurta, P.\
 2009, \apjl, 697, L58
\bibitem[Bastian \& Goodman(2006)]{basgoo06}
Bastian, N., \& Goodman, S. P. 2006, \mnras, 369, L9
\bibitem[Bastian \& de Mink(2009)]{basdem09}
Bastian, N., \& de Mink, S. E.\ 2009, \mnras, 398, L11
\bibitem[Bedin et al.(2004)]{bedi+04}
Bedin, L.~R., Piotto, G., Anderson, J., Cassisi, S., King, I.~R., Momany, Y., 
 \& Carraro, G.\ 2004, \apjl, 605, L125 
\bibitem[Bekki \& Mackey(2009)]{bekmac09}
Bekki, K., \& Mackey, A. D.\ 2009, \mnras, 394, 124
\bibitem[Bertelli et al.(2003)]{bert+03} 
Bertelli, G., Nasi, E., Girardi, L., Chiosi, C. Zoccali, M., \& Gallart, C.\ 
 2003, \aj, 125, 770
\bibitem[Bica et al.(1996)]{bica+96}
Bica, E., Clar\'{\i}a, J. J., Dottori, H., Santos, J. F. C., \& Piatti,
 A. E.\ 1996, \apjs, 102, 57
\bibitem[Bruzual \& Charlot(2003)]{bc03} 
Bruzual, G. A., \& Charlot, S., 2003, \mnras, 344, 1000
\bibitem[Cardelli et al.(1989)Cardelli, Clayton, \& Mathis]{card+89}
Cardelli, J. A., Clayton, G. C., \& Mathis, J. S. 1989, \apj, 345, 245
\bibitem[Carretta et al.(2009)]{carr+09}
Carretta, E., et al.\ 2009, \aap, 505, 117
\bibitem[Carretta et al.(2010)]{carr+10}
Carretta, E., et al.\ 2010, \apjl, 714, L7
\bibitem[Castelli \& Kurucz(2003)]{caskur03}
Castelli, F., \& Kurucz, R. L., 2003, Modelling of Stellar
 Atmospheres, ed.\ N.\ Piskunov, W.\ W.\ Weiss, \& D.\ F.\ Gray (San
 Francisco: ASP), A20 
\bibitem[Conroy(2011)]{conr11}
Conroy, C.\ 2011, submitted to ApJ (arXiv:1101.2208v1)
\bibitem[D'Antona \& Ventura(2007)]{danven07}
D'Antona, F., \& Ventura, P.\ 2007, \mnras, 379, 1431
\bibitem[Decressin et al.(2007)]{decr+07}
Decressin, T., Meynet, G., Charbonnel, C., Prantzos, N., \& Ekstr\'om, S.\
 2007, \aap, 464, 1029
\bibitem[de Mink et al.(2009)]{demink+09}
de Mink, S. E., Pols, O. R., Langer, N., \& Izzard, R. G.\ 2009, \aap, 5007, L1
\bibitem[D'Ercole et al.(2010)]{derc+10}
D'Ercole, A., D'Antona, F., Ventura, P., Vesperini, E., \& McMillan, S. L. W.\ 
 2010, \mnras, 407, 854
\bibitem[D'Ercole et al.(2008)]{derc+08}
D'Ercole, A., Vesperini, E., D'Antona, F., McMillan, S. L. W., \& Recchi, S.\
 2008, \mnras, 391, 825
\bibitem[Dotter et al.(2008)]{dott+08}
Dotter, A., Chaboyer, B., Jevremovi\'c, D., Kostov, V., Baron, E., \&
 Ferguson, J. W.\ 2008, \apjs, 178, 89
\bibitem[Elson \& Fall(1985)]{elsfal85}
Elson, R. A. W., \& Fall, S. M.\ 1985, \apj, 299, 211
\bibitem[Elson \& Fall(1988)]{elsfal88}
Elson, R. A. W., \& Fall, S. M.\ 1988, \aj, 96, 1383
\bibitem[Fellhauer et al.(2006)]{fell+06}
Fellhauer, M., Kroupa, P., \& Evans, N. W.\ 2006, \mnras, 372, 338
\bibitem[Fusi Pecci et al.(1990)]{fusi+90}
Fusi Pecci, F., Ferraro, F. R., Crocker, D. A., Rood, R. T., \& Buonanno, R.\ 
 1990, \aap, 238, 95
\bibitem[Girardi et al.(1995)]{gira+95}
Girardi, L., Chiosi, C., Bertelli, G., \& Bressan, A. 1995, \aap, 298, 87
\bibitem[Girardi et al.(2000)]{gira+00}
Girardi, L., Bressan, A., Bertelli, G., \& Chiosi, C. 2000, A\&AS, 141, 371 
\bibitem[Girardi et al.(2008)]{gira+08}
Girardi, L., Dalcanton, J., Williams, B., de Jong, R. S., Gallart, C., et al.\
 2008, \pasp, 120, 583
\bibitem[Girardi et al.(2009)]{gira+09}
Girardi, L., Rubele, S., \& Kerber, L.\ 2009, \mnras, 394, L74
\bibitem[Girardi et al.(2011)]{gira+11}
Girardi, L., Eggenberger, P., \& Miglio, A.\ 2011, \mnras, in press
 (doi: 10.1111/j.1745-3933.2011.01013.x, arXiv:1101.1180v1) 
\bibitem[Glatt et al.(2008)]{glat+08}
Glatt, K.\ et al., 2008, \aj, 135, 1703
\bibitem[Goudfrooij et al.(2006)]{goud+06}
Goudfrooij, P., Gilmore, D., Kissler-Patig, M., \& Maraston, C.\ 2006,
 \mnras, 369, 697
\bibitem[Goudfrooij et al.(2007)]{goud+07}
Goudfrooij, P., Schweizer, F., Gilmore, D., \& Whitmore, B. C. 2007,
 \aj, 133, 2737
\bibitem[Goudfrooij et al.(2009)]{goud+09}
Goudfrooij, P., Puzia, T. H., Kozhurina-Platais, V., \& Chandar, R.\ 2009,
 \aj, 137, 4988 (Paper I)
\bibitem[Goudfrooij et al.(2011b)]{goud+11b}
Goudfrooij, P., Puzia, T. H., Chandar, R., \& Kozhurina-Platais, V.\ 2011, 
 accepted for publication by \apj\ (Paper III)
\bibitem[Gratton et al.(2004)Gratton, Sneden, \& Carretta]{grat+04}
Gratton, R., Sneden, C., \& Carretta, E.\ 2004, \araa, 42, 385
\bibitem[Hauschildt et al.(1999)]{haus+99}
Hauschildt, P. H., Allard, F., Ferguson, J., Baron, E., \& Alexander,
 D.\ 1999b, \apj, 525, 871 
\bibitem[Hilker \& Richtler(2000)]{hilric00}
Hilker, M., \& Richtler, T.\ 2000, \aap, 362, 895
\bibitem[Keller et al.(2011)Keller, Mackey, \& Da Costa]{kell+11}
Keller, S. C., Mackey, A. D., \& Da Costa, G. S.\ 2011, \apj, 731, 22
\bibitem[Kerber et al.(2007)Kerber, Santiago, \& Brocato]{kerb+07}
Kerber, L. O., Santiago, B. X., \& Brocato, E., 2007, \aap, 462, 139
\bibitem[King(1962)]{king62}
King, I. 1962, \aj, 67, 471 
\bibitem[Kozhurina-Platais et al.(2007)Kozhurina-Platais, Goudfrooij, \& Puzia]{kozh+07}
Kozhurina-Platais, V., Goudfrooij, P., \& Puzia, T. H.\ 2007, ACS Instrument
 Science Report 2007-04 (Baltimore: STScI)
\bibitem[Lee et al.(1999)]{lee+99}
Lee, Y., Joo, J., Sohn, Y., Rey, S., Lee, H., \& Walker, A. R.\ 1999, \nat,
 402, 55
\bibitem[Mackey \& Broby Nielsen(2007)]{macbro07}
Mackey, A. D., \& Broby Nielsen, P. 2007, \mnras, 379, 151
\bibitem[Mackey et al.(2008)]{mack+08}
Mackey, A. D., Broby Nielsen, P., Ferguson, A. M. N., \& Richardson,
 J. C. 2008, \apj, 681, L17
\bibitem[Maraston et al.(2008)]{mara+08}
Maraston, C., Str\"omb\"ack, G., Thomas, D., Wake, D. A., \& Nichol, D. 2008,
 \mnras, 394, L107
\bibitem[Marigo et al.(2008)]{marigo+08}
Marigo, P., Girardi, L., Bressan, A., Groenewegen, M. A. T., Silva, L., \&
 Granato, G. L. 2008, \aap, 482. 883
\bibitem[Mucciarelli et al.(2007)]{mucc+07}
Mucciarelli, A., Ferraro, F. R., Origlia, L., \& Fusi Pecci, F. 2007, \aj, 133, 
 2053
\bibitem[Mucciarelli et al.(2008)]{mucc+08}
Mucciarelli, A., Carreta, E., Origlia, L., \& Ferraro, F. R.\ 2008, \aj, 136, 
 375
\bibitem[Milone et al.(2008)]{milo+08}
Milone, A. P., et al.\ 2008, \apj, 673, 241
\bibitem[Milone et al.(2009)]{milo+09}
Milone, A. P., Bedin, L. R., Piotto, G., \& Anderson, J.\ 2009, \aap, 497, 755
\bibitem[Pessev et al.(2008)]{pess+08} 
Pessev, P. M., Goudfrooij, P., Puzia, T. H., \& Chandar, R.\ 2008, \mnras, 385, 1535
\bibitem[Pickles(1998)]{pick98}
Pickles, A. J. 1998, \pasp, 110, 863
\bibitem[Pietrinferni et al.(2004)]{piet+04}
Pietrinferni, A., Cassisi, S., Salaris, M., \& Castelli, F. 2004, \apj, 612,
 168
\bibitem[Pietrinferni et al.(2006)]{piet+06}
Pietrinferni, A., Cassisi, S., Salaris, M., \& Castelli, F. 2006, \apj, 642, 
 797
\bibitem[Piotto et al.(2007)]{piot+07}
Piotto, G., et al.\ 2007, \apjl, 661, L53
\bibitem[Renzini(2008)]{renz08} 
Renzini, A.\ 2008, \mnras, 391, 354
\bibitem[Rubele et al.(2010)]{rube+10}
Rubele, S., Kerber, L., \& Girardi, L.\ 2010, \mnras, 403, 1156
\bibitem[Rubele et al.(2011)]{rube+11}
Rubele, S., Girardi, L., Kozhurina-Platais, V., Goudfrooij, P., \& Kerber, L.\
 2011, \mnras, 413, doi:10.1111/j.1365-2966.2011.18538.x
\bibitem[Salpeter(1955)]{salp55}
Salpeter, E. E.\ 1955, \apj, 121, 161
\bibitem[Sarajedini \& Layden(1995)]{sarlay95}
Sarajedini, A., \& Layden, A. C.\ 1995, \aj, 109, 1086
\bibitem[Sarajedini et al.(2007)]{sara+07}
Sarajedini, A., Barker, M. K., Geisler, D., Harding, P., \& Schommer, R.\
 2007, \aj, 133, 290
\bibitem[Searle et al.(1980)Searle, Wilkinson, \& Bagnuolo]{swb80}
Searle, L., Wilkinson, A., \& Bagnuolo, W. G.\ 1980, \apj, 239, 803
\bibitem[Silverman(1986)]{silv86} 
Silverman, B. W. 1986, in {\it Density Estimation for Statistics and
  Data Analysis}, Chap and Hall/CRC Press, Inc. 
\bibitem[van der Wel et al.(2006)]{vdwel+06}
van der Wel, A., Franx, M., van Dokkum, P. G., Huang, J., Rix, H.-W., \&
 Illingworth, G. D. 2006, \apj, 2006, L21
\bibitem[Ventura \& D'Antona(2008)]{vendan08}
Ventura, P., \& D'Antona, F.\ 2008, \mnras, 385, 2034
\bibitem[Villanova et al.(2007)]{vill+07}
Villanova, S., et al.\ 2008, \apj, 663, 296
\bibitem[Yang et al.(2011)]{yang+11}
Yang, W., Meng, X., Bi, S., Tian, Z., Li, T., \& Liu, K.\ 2011, \apj, 731, L37
\end{thebibliography}
\end{document}